\documentclass[preprint,floatfix] {revtex4} 
\newcommand{\rvec}{\mathrm {\mathbf {r}}} 
\newcommand{\pvec}{\mathrm {\mathbf {p}}} 
\usepackage{graphicx}
\usepackage{subfigure}
\usepackage{xcolor}
\usepackage{amsmath}
\usepackage{enumerate}
\usepackage{tabularx}

\usepackage{color, soul}
\definecolor{darkblue}{rgb}{0,0,0.5}
\setulcolor{darkblue}

\begin{document}

\title{Information-entropic measures in confined isotropic harmonic oscillator}

\author{Neetik Mukherjee}
\altaffiliation{Email: neetik.mukherjee@iiserkol.ac.in.}

\author{Amlan K.~Roy}
\altaffiliation{Corresponding author. Email: akroy@iiserkol.ac.in, akroy6k@gmail.com.}
\affiliation{Department of Chemical Sciences\\
Indian Institute of Science Education and Research (IISER) Kolkata, 
Mohanpur-741246, Nadia, WB, India}

\begin{abstract}
Information based uncertainty measures like R{\'e}nyi entropy (R), Shannon entropy (S) and Onicescu energy (E) (in both position and 
momentum space) are employed to understand the influence of radial confinement in isotropic harmonic oscillator. The transformation 
of Hamiltonian in to a dimensionless form gives an idea of the composite effect of oscillation frequency ($\omega$) and confinement radius ($r_{c}$).
For a given quantum state, accurate results are provided by applying respective \emph{exact} analytical wave function in $r$ space.
The $p$-space wave functions are produced from Fourier transforms of radial functions. Pilot calculations are done taking order of entropic 
moments ($\alpha, \beta$) as $(\frac{3}{5}, 3)$ in $r$ and $p$ spaces. A detailed, systematic analysis is performed for confined harmonic 
oscillator (CHO) with respect to state indices $n_{r},l$, and $r_c$. It has been found that, CHO acts as a bridge between particle 
in a spherical box (PISB) 
and free isotropic harmonic oscillator (IHO). At smaller $r_c$, $E_{\rvec}$ increases and $R_{\rvec}^{\alpha}, S_{\rvec}$ decrease with 
rise of $n_{r}$. At moderate $r_{c}$, there exists an interaction between two competing factors: 
(i) radial confinement (localization) and (ii) accumulation of radial nodes with growth of $n_{r}$ (delocalization). Most of these results 
are reported here for the first time, revealing many new interesting features. 

\vspace{3mm}
{\bf PACS:} 03.65-w, 03.65Ca, 03.65Ta, 03.65.Ge, 03.67-a.

\vspace{3mm}
{\bf Keywords:} R\'enyi entropy, Shannon entropy,  Onicescu energy, Confined isotropic harmonic oscillator, 
Particle in a symmetric box. 

\end{abstract}
\maketitle

\section{introduction}
In recent years, interest in studying spacially confined quantum systems has enhanced significantly. A quantum mechanical 
particle under extreme pressure environment exhibits many fascinating, notable physical and chemical properties 
\cite{michels37,sabin2009,sen2014electronic}. Discovery and development of modern experimental techniques have 
also inspired extensive research activity to explore and study such systems 
\cite{sabin2009,sarsa11,sech11,katriel12,cabrera13,sen2014electronic}. They have potential applications in a 
wide range of problems namely, quantum wells, quantum wires, quantum dots, defects in solids, 
super-lattice structure, as well as nano-sized circuits such as quantum computer, etc. Besides, they have uses in cell-model 
of liquid, high-pressure physics, astrophysics \cite{pang11}, study of impurities in semiconductor materials, matrix isolated 
molecules, endohedral complexes of fullerenes, zeolites cages, helium droplets, nano-bubbles, \cite{sabin2009} etc. 

In last ten years, extensive theoretical works have been published covering a wide variety of confining potentials. Two such prototypical 
systems that have received maximum attention are confined harmonic oscillator (CHO) (in 1D, 2D, 3D, and D dimension) \cite{coll17,aquino97,campoy2002,montgomery07,roy14,
ghosal16} and confined hydrogen atom (CHA) inside a spherical enclosure \cite{coll17,goldman92,aquino95,garza98,laughlin02,burrows06,
aquino07cha,baye08,ciftci09,sen2014electronic,roy15,centeno17}. The (CHO) model within an impenetrable barrier was 
explored quite extensively leading to a host of interesting properties$-$both from physical and mathematical perspective. 
They offer some unique phenomena, especially relating to \emph{simultaneous, incidental and inter-dimensional} 
degeneracy \cite{montgomery07}. A large variety of theoretical methods were employed; a selected set includes perturbation theory, 
Pad\'e approximation, WKB method, Hypervirial theorem, power-series solution, super-symmetric quantum mechanics, Lie algebra, 
Lagrange-mesh method, asymptotic iteration method, generalized pseudo-spectral method, etc. \cite{goldman92,aquino95,garza98,
laughlin02,burrows06,aquino07cha,baye08,ciftci09,roy15} and references therein. \emph{Exact} solutions \cite{burrows06} 
are expressible in terms of Kummer confluent hypergeometric function.

In recent years, significant attention was paid to explore various information measures (IE), namely, Fisher information 
(I), Shannon entropy (S), R{\'e}nyi entropy (R), Onicescu energy (E) and several complexities 
in a multitude of physical, chemical systems, including central potentials. The literature is quite vast. In a quantum system, 
R, called information generating functionals, is closely related to
entropic moments (discussed later), and completely characterize density $\rho(\rvec)$.  It is successfully used to investigate 
and predict certain quantum properties and phenomena like entanglement, communication protocol, correlation de-coherence, measurement, 
localization properties of Rydberg states, molecular reactivity, multi-fractal thermodynamics, production of multi-particle in 
high-energy collision, disordered systems, spin system, quantum-classical correspondence, localization in phase space 
\citep{varga03,renner05, levay05,verstraete06,
bialas06,salcedo09,liu15}, etc. It is interesting to note that, S, E are two particular cases of R \cite{sen12,bbi06}.
S and E quantify the information content in different and complimentary way. Former 
refers to the expectation value of logarithmic probability density function and is a global measure of spread of density. On the other 
hand, E is quantified as the second-order entropic moment \cite{onicescu66}. It becomes minimum for equilibrium and hence 
often termed as disequilibrium. In recent years, S is examined in a number of systems, 
such as, P\"oschl-Teller \cite{sun2013quantum}, 
Rosen-Morse \cite{sun2013quantum1}, pseudo-harmonic \cite{yahya2015}, squared tangent well \cite{dong2014quantum}, 
hyperbolic \cite{valencia2015quantum}, position-dependent mass Schr\"odinger equation \cite{chinphysb,yanez2014quantum}, 
infinite circular well \cite{song2015shannon}, hyperbolic double-well (DW) potential \cite{sun2015shannon}, etc. Recently, 
some of these measures have been found to be quite efficient and useful to explain the oscillation and 
localization-delocalization behavior of a particle in symmetric and asymmetric DW potential \cite{neetik15,neetik16}, as well as 
in a confined 1D quantum harmonic oscillator \cite{ghosal16}. 

IE quantifies the spatial delocalization of single-particle density of a system in several complimentary ways. Arguably, 
these are the most appropriate uncertainty measures, as they do not make any reference to some specific point of the 
resembling Hilbert space. Moreover, these are closely related to some energetic and experimentally measurable quantities 
\cite{gonzalez03,sen12} of a system. In case of $R$ and $S$, some lower bound is available, which do not
depend on quantum number. But, for $I$ both upper and lower bounds have been established, which strictly change with quantum numbers 
\cite{bbi06,bbi75,romera05}.  

A vast majority of IE-related works, mentioned above and elsewhere, deal with a \emph{free or unconfined} systems.
However, such study for \emph{confined} quantum systems is very rare. In last few years, some such results have been published for
symmetric and asymmetrically confined 1-D harmonic oscillator \cite{laguna14,ghosal16} and confined hydrogen atom 
\cite{mukherjee18,mukherjee18a,majumdar17,mukherjee18b}. However, to the best of our knowledge, such investigation for a 3-D CHO system 
has not yet been done. Hence, it would be highly desirable to explore and inspect these quantities for such system in some detail. In this
work, we have pursued a detailed analysis of R, S, E for CHO. Moreover, we have transformed our original Hamiltonian into a 
dimensionless form \cite{patil07} to make the results more general and interesting, from the view point of an experimentalists 
\cite{zawadzki87, buttiker88}. This modification leads to a dimensionless parameter $\left(\eta=\frac{m\omega r_c^{4}}{\hbar^{2}}\right)$,
which depends on the product of $\omega$ and quartic power of $r_c$. Thus, at first, we
analyze the variation of R, S, E for an arbitrary state in CHO for small, intermediate and large regions of $\eta$ in conjugate $r,~p$ 
spaces. Later, we proceed for a detailed exploration of these measures as functions of $r_c$. These are provided for a general state 
having principal and azimuthal quantum numbers $n,~l$,
while keeping magnetic quantum number $m=0$. In $r$ space all the calculations are performed taking exact wave function. However, such 
expressions are unavailable in $p$-space, and hence numerical Fourier transforms require to be carried out. It is important to note that, 
no such literature is available for CHO. This work has been
arranged in the following manner. Section~2, gives the essential points of methodology, then Section~3 provides a details discussion on the results 
of aforesaid measures for CHO, while we conclude with a few remarks in section 4.      
       
\section{Methodology}
The time-independent, non-relativistic wave function for a CHO system, in $r$ space may be expressed as, 
\begin{equation} 
\Psi_{n_r,l,m} (\rvec) = \psi_{n_r, l}(r)  \ Y_{l,m} (\Omega), 
\end{equation}
with $r$ and $\Omega$ illustrating the radial distance and solid angle successively. Here $\psi_{n,l}(r)$ represents the radial part 
and $Y_{l,m}(\Omega)$ identifies spherical harmonics.
The pertinent radial Schr\"odinger equation under the influence of confinement is (atomic unit employed unless mentioned otherwise), 
\begin{equation}
	\left[-\frac{1}{2} \ \frac{d^2}{dr^2} + \frac{l (l+1)} {2r^2} + v(r) +v_c (r) \right] \psi_{n_r,l}(r)=
	\mathcal{E}_{n_r,l}\ \psi_{n_r,l}(r),
\end{equation}
where $v(r)=\frac{1}{2}\omega^{2}r^{2}$. Our required confinement effect is introduced by 
invoking the following potential: $v_c(r) = +\infty$ for $r > r_c$, and $0$ for $r \leq r_c$, where $r_c$ signifies 
radius of confinement.

\emph{Exact} generalized radial wave function for a CHO is mathematically expressed as \cite{montgomery07}, 
\begin{equation}
\psi_{n_{r}, l}(r)= N_{n_{r}, l} \ r^{l} \ _{1}F_{1}\left[\frac{1}{2}\left(l+\frac{3}{2}-\frac{\mathcal{E}_{n_{r},l}}{\omega}\right),
(l+\frac{3}{2}),\omega r^{2}\right] e^{-\frac{\omega}{2}r^{2}}.
\end{equation}
Here, $N_{n_r, l}$ represents normalization constant and $\mathcal{E}_{n_r,l}$ corresponds to the energy of a given state characterized by 
quantum numbers $n_r,l$, whereas $_1F_1\left[a,b,r\right]$ signifies confluent hypergeometric 
function. Allowed energies are computed by applying the boundary condition $\psi_{n_r,\ell} (0)= \psi_{n_r,\ell} \ (r_c)=0$. In this 
work, generalized pseudospectral (GPS) method was used to evaluate $\mathcal{E}_{n_r,l}$ of these states. This method has 
provided highly accurate results for various model and real systems including atoms, molecules, some of which could be found in the 
references \cite{roy08,roy8a,roy15,roy15a}. This is very well documented and therefore omitted here.

The $p$-space wave function is obtained from Fourier transform of $r$-space counterpart,  
\begin{equation}
\begin{aligned}
\psi_{n_r,l}(p) & = & \frac{1}{(2\pi)^{\frac{3}{2}}} \  \int_0^{r_{c}} \int_0^\pi \int_0^{2\pi} \psi_{n,l}(r) \ \Theta(\theta) 
 \Phi(\phi) \ e^{ipr \cos \theta}  r^2 \sin \theta \ \mathrm{d}r \mathrm{d} \theta \mathrm{d} \phi   \\
      & = & \frac{1}{2\pi} \sqrt{\frac{2l+1}{2}} \int_0^{r_{c}} \int_0^\pi \psi_{n_r,l} (r) \  P_{l}^{0}(\cos \theta) \ 
e^{ipr \cos \theta} \ r^2 \sin \theta  \ \mathrm{d}r \mathrm{d} \theta.  
\end{aligned}
\end{equation}
Here $\psi_{n_r,l}(p)$ is not normalized and needs to be normalized. Integrating over $\theta$ and $\phi$ yields,  
\begin{equation}
\psi_{n_r,l}(p)=(-i)^{l} \int_0^{r_{c}} \  \frac{\psi_{n_r,l}(r)}{p} \ f(r,p)\mathrm{d}r,    
\end{equation}
where, $f(r,p)$ depends only on $l$ quantum number. It can be expressed in terms of \emph{Cosine} and \emph{Sine} series. More details about 
$f(r,p)$ could be found in \cite{mukherjee18}.

R{\'e}nyi entropies of order $\lambda (\neq 1)$ are obtained by taking logarithm of $\lambda$-order entropic moment. 
In spherical polar coordinate these can be written as,
\begin{equation}
\begin{aligned} 
R_{\rvec}^{\lambda}  =  & \frac{1}{(1-\lambda)}\left( \ln 2\pi + \ln [\omega^{\lambda}_r] + \ln [\omega^{\lambda}_{(\theta, \phi)}] \right),  \\
R_{\pvec}^{\lambda}  =  & \frac{1}{(1-\lambda)}\left( \ln 2\pi + \ln [\omega^{\lambda}_p] + \ln [\omega^{\lambda}_{(\theta, \phi)}] \right).
\end{aligned} 
\end{equation}          
Here $\omega^{\lambda}_{\tau}$s are entropic moments in $\tau$ ($r$ or $p$ or $\theta$) space with order $\lambda$, having forms,
\begin{equation}
\omega^{\lambda}_r= \int_0^\infty [\rho(r)]^{\lambda} r^2 \mathrm{d}r, \ \ \  
\omega^{\lambda}_p= \int_{0}^\infty [\Pi(p)]^{\lambda} p^2 \mathrm{d}p, \ \ \  
\omega^{\lambda}_{(\theta, \phi)}= \int_0^\pi [\chi(\theta)]^{\lambda} \sin \theta \mathrm{d}\theta. 
\end{equation}          

If $\lambda$ corresponds to $\alpha$, $\beta$ in $r$, $p$ spaces respectively, then for R, they obey the 
condition $\frac{1}{\alpha}+\frac{1}{\beta}=2.$ Then one can define total R{\'e}nyi entropy as $R_{t}^{(\alpha,\beta)}$ \cite{bbi06,sen12},
satisfying the following bounds,
\begin{equation}
\begin{aligned}
R_{t}^{(\alpha,\beta)} & =  \frac{2-\alpha-\beta}{(1-\alpha)(1-\beta)} \ \ln 2\pi+ \frac{1}{(1-\alpha)}  
\left( \ln [\omega^{\alpha}_r]+ \ln [\omega^{\alpha}_{(\theta, \phi)}]\right)
 + \frac{1}{(1-\beta)}\left( \ln [\omega^{\beta}_p]+ \ln [\omega^{\beta}_{(\theta, \phi)}]\right)  \\
& \geq 3 \times \left[ -\frac{1}{2}\left(\frac{1}{1-\alpha}\ln\frac{\alpha}{\pi}+\frac{1}{1-\beta}\ln\frac{\beta}{\pi}\right)\right].
\end{aligned} 
\end{equation}
                            
$S_{\rvec}, S_{\pvec}$ and total Shannon entropy $S_{t}$ are expressed in terms of expectation values of logarithmic probability 
density functions, which for a central potential further simplifies \cite{bbi75} as below, 
\begin{equation}
\begin{aligned} 
S_{\rvec} & =  -\int_{{\mathcal{R}}^3} \rho(\rvec) \ \ln [\rho(\rvec)] \ \mathrm{d} \rvec   = 
2\pi \left(S_{r}+S_{(\theta,\phi)}\right), \\   
S_{\pvec} & =  -\int_{{\mathcal{R}}^3} \Pi(\pvec) \ \ln [\Pi(\pvec)] \ \mathrm{d} \pvec   = 
2\pi \left(S_{p}+S_{(\theta, \phi)}\right), \\ 
S_{t} & = 2\pi \left[S_{r}+S_{p}+2S_{(\theta, \phi)}\right] \ \ \geq 3(1+\ln \pi), 
\end{aligned} 
\end{equation}
where the quantities $S_r, S_p$ and $S_{\theta}$ are defined as \cite{bbi75},   
\begin{equation}
\begin{aligned} 
S_{r} & =  -\int_0^\infty \rho(r) \ \ln [\rho(r)] r^2 \mathrm{d}r, \ \ \ \ \ \ \ \ \ \ \ \ 
S_{p}  =  -\int_{0}^\infty \Pi(p) \ln [\Pi(p)]  \ p^2 \mathrm{d}p, \\
\rho(r) & = |\psi_{n,l}(r)|^{2},  \ \ \ \ \ \ \ \ \ \ \ \ \ \ \ \ \ \ \ \ \ \ \ \ \ \ \ \  \Pi(p) = |\psi_{n,l}(p)|^{2}, \\
S_{(\theta, \phi)} & =   -\int_0^\pi \chi(\theta) \ \ln [\chi(\theta)] \sin \theta \mathrm{d} \theta, \ \ \ \ \ \
\chi(\theta)   =  |\Theta(\theta)|^2.  \\   
\end{aligned} 
\end{equation}
       
By definition, E represents the 2nd order entropic moment \cite{sen12}; therefore choice of $\alpha= \beta =2$ transforms Eq.~(8) into the 
following form,
\begin{equation}
E_{r}= \int_0^\infty [\rho(r)]^{2} r^2 \mathrm{d}r, \ \   
E_{p}= \int_{0}^\infty [\Pi(p)]^{2} p^2 \mathrm{d}p, \ \   
E_{\theta, \phi}= \int_0^\pi [\chi(\theta)]^{2} \sin \theta \mathrm{d}\theta, \ \
E=E_{r}E_{p}E_{\theta, \phi}^{2}. 
\end{equation}
where, $E_{t}$ is the total Onicescu energy. Note that, the restriction $\frac{1}{\alpha}+\frac{1}{\beta}=2$ holds for R only, 
and not on E. Hence in our study of R, $\alpha=\frac{3}{5}$ and $\beta=3$ have been chosen.

\begingroup           
\squeezetable
\begin{table}
\caption{$R_{\rvec^{\prime}}^{\alpha},R_{\pvec^{\prime}}^{\beta}$, $R_{t}^{(\alpha,\beta)}$ for $1s,~1p,~1d$ states in PISB and CHO 
(six selected $\eta$). See text for detail.}
\centering
\begin{ruledtabular}
\begin{tabular}{l|l|l|llllll}
State  & Property & PISB($\eta=0$) & $\eta=0.0001$  &  $\eta=0.0625$   &  $\eta=1.0$  & $\eta=5.0625$  &  $\eta=45.6976$  &  $\eta=104.8576$ \\
\hline
         & $R_{\rvec^{\prime}}^{\alpha}$    &  0.871064 & 0.87106349  &  0.87060118  & 0.86363060 & 0.83292175 &  0.50791919 &  0.09617311  \\
1s       & $R_{\pvec^{\prime}}^{\beta}$     &  5.3391   &  5.339416   &  5.339830    & 5.346081   & 5.373798   &  5.678841   &  6.080846    \\
         & $R_{t}^{(\alpha, \beta)}$ &  6.2101   & 6.210480    &  6.210431    & 6.209712   & 6.206720   &  6.186760   &  6.177019   \\
\hline 
         & $R_{\rvec^{\prime}}^{\alpha}$    &  0.740619 & 0.74061892 & 0.74041416 & 0.73732009 & 0.72353315 & 0.55729149 & 0.27721065      \\
1p       & $R_{\pvec^{\prime}}^{\beta}$     &  5.8987   & 5.898746   & 5.898972   & 5.902381   & 5.917583   & 6.101146   & 6.406884        \\
         & $R_{t}^{(\alpha, \beta)}$ &  6.6393   & 6.639365   & 6.639386   & 6.639701   & 6.641116   & 6.658438   & 6.684095        \\
\hline
         & $R_{\rvec^{\prime}}^{\alpha}$    &  0.789638 & 0.78963611 & 0.78953691 & 0.78803644 & 0.78131257 & 0.69457347  & 0.51883620     \\
1d       & $R_{\pvec^{\prime}}^{\beta}$     &  6.3210   & 6.321199   & 6.321338   & 6.323450   & 6.332867   & 6.449503    & 6.670819       \\
         & $R_{t}^{(\alpha, \beta)}$ &  7.1106   & 7.110835   & 7.110875   & 7.111487   & 7.114179   & 7.144076    & 7.189655   
\end{tabular}
\end{ruledtabular}
\end{table}
\endgroup

\section{Result and Discussion}
At the beginning, it may be convenient to point out a few things about the presented results. The net information measures in conjugate $r$ and $p$ spaces 
may be divided into radial and angular segments. In a given space, the results provided here correspond to net measures including
the angular contributions. One can transform the IHO to a CHO by pressing the radial boundary of former from infinity to a finite 
region. This change in radial environment does not affect the angular boundary conditions. Hence, angular portion of these measures 
remains invariant in $r$, $p$ spaces. Furthermore, they change with $l$, $m$ quantum numbers. Throughout the whole article the 
magnetic quantum number $m$ is set to 0, unless stated otherwise. Since the wave function, energy and position expectation values of CHO were 
presented earlier in some details, we do not discuss them in this work. Our primary focus is on information analysis.

\begin{figure}                         
\begin{minipage}[c]{0.32\textwidth}\centering
\includegraphics[scale=0.55]{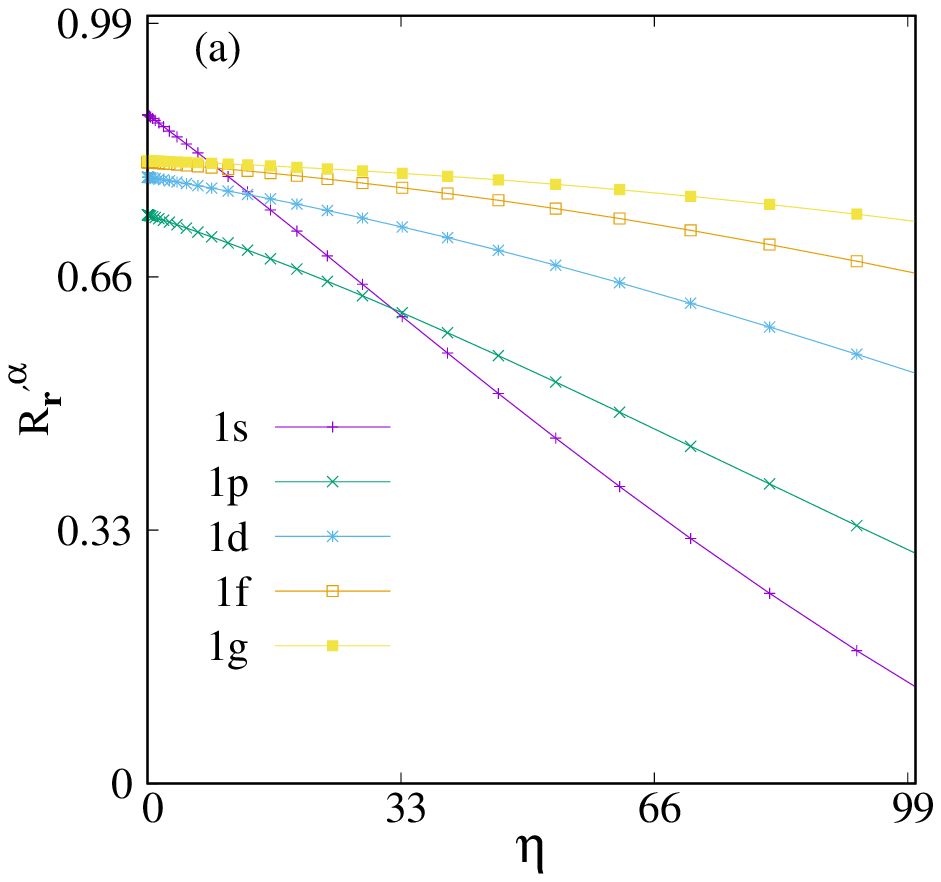}
\end{minipage}%
\hspace{0.02in}
\begin{minipage}[c]{0.32\textwidth}\centering
\includegraphics[scale=0.55]{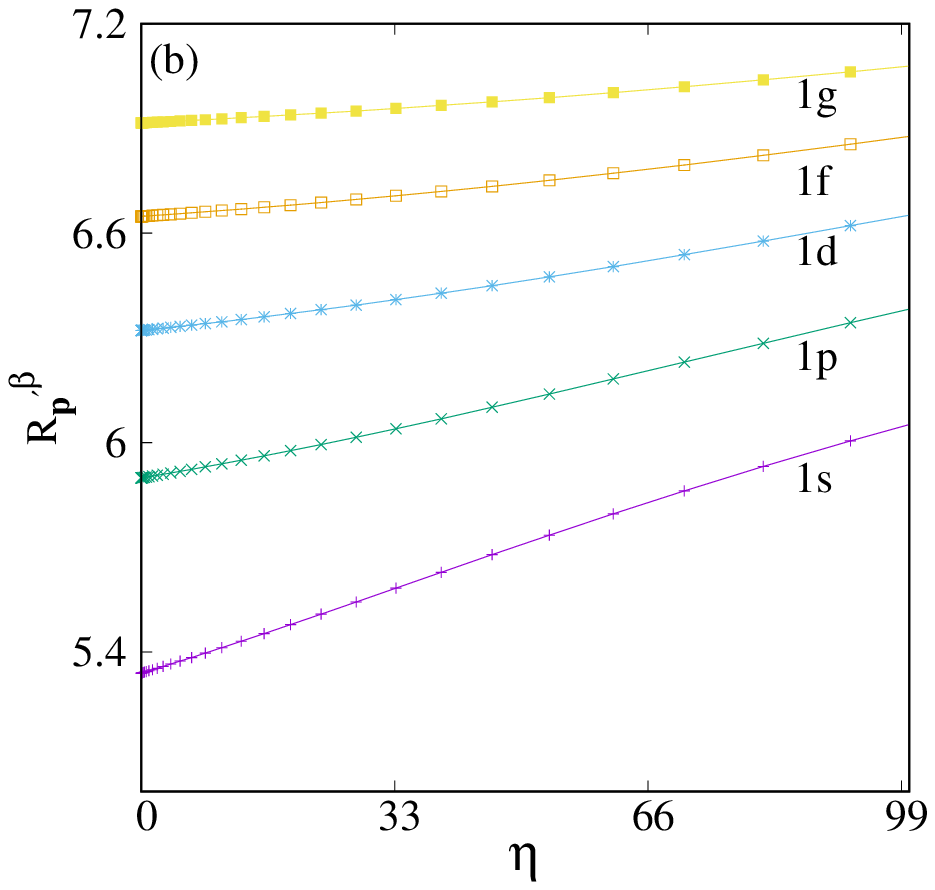}
\end{minipage}%
\hspace{0.02in}
\begin{minipage}[c]{0.32\textwidth}\centering
\includegraphics[scale=0.55]{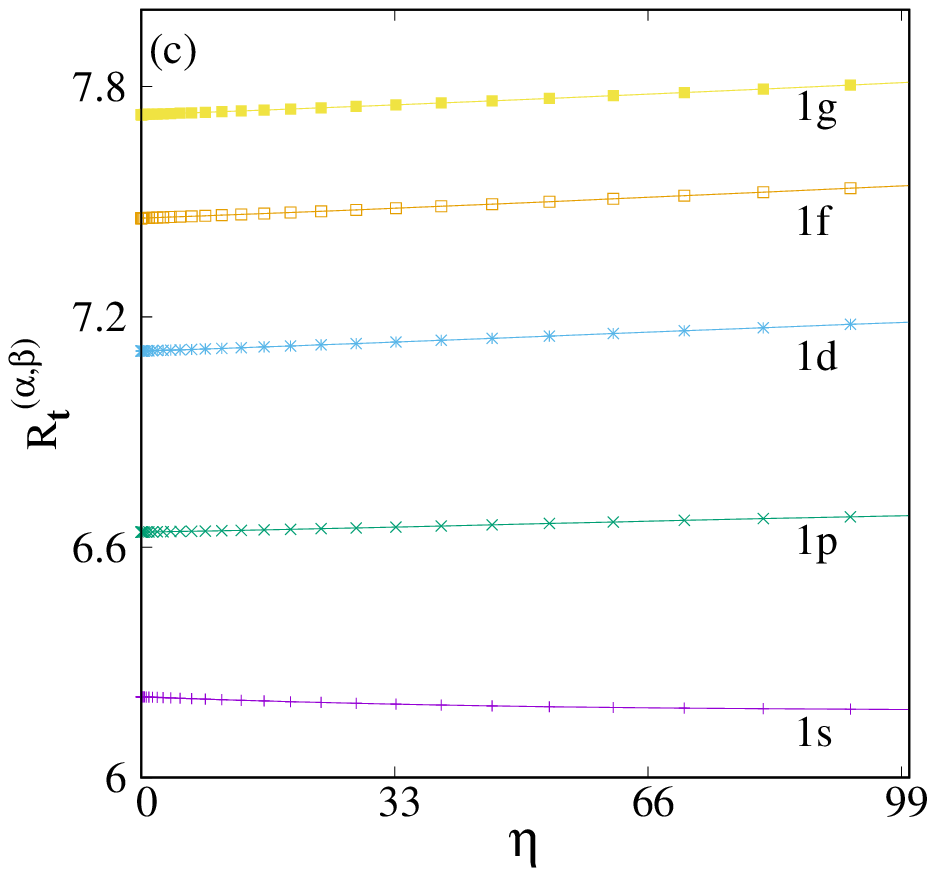}
\end{minipage}%
\caption{Plots of $R_{\rvec^{\prime}}^{\alpha}$, $R_{\pvec^{\prime}}^{\beta}$, $R_{t}^{(\alpha, \beta)}$ against $\eta$ for first five 
circular states of CHO in panels (a), (b), (c) respectively. See text for details.}
\end{figure}

Equation~(2) may be represented in the following form,
\begin{equation}
\begin{aligned}
\left[-\frac{1}{2} \ \frac{d^2}{dr^2} + \frac{l (l+1)} {2r^2} + \frac{1}{2}\omega r^{2} +V\Theta(r-r_{c}) \right] \psi_{n_r,l}(r)=
\mathcal{E}_{n,l}\ \psi_{n,l}(r), \\
\Theta(r-r_{c}) =0,~at~ r \leq |r_c|,   \ \ \ \ \ \Theta(r-r_{c}) =1, ~at~ r > |r_c|.
\end{aligned}
\end{equation}

Here, $\Theta(r-r_{c})$ is a Heaviside Theta function and $V$ is a constant, having very large value. The effect of localization and 
delocalization depends on $r_{c}$ and $\omega$. It has been observed that, the Hamiltonian can be generalized into a dimensionless form, 
so that one can correlate experimental observations with theoretical results \cite{zawadzki87,buttiker88,patil07}. Further, in 1D case, it is established that $\omega$
is proportional to the square root of the magnetic field parallel to the gradient of the confining potential \cite{zawadzki87}. Hence, 
it seems appropriate to study composite effect $r_c$ and $\omega$ with the aid of a single dimensionless parameter $\eta$. This will make our present 
study more interesting and appropriate from an experimental view point. It follows that,
\begin{equation}
\begin{aligned}
\mathcal{E}_{n_r,l}=\mathcal{E}_{n_r,l}\left(\frac{\hbar^{2}}{m}, \omega, r_c\right); \ \ \ 
\psi_{n_r,l}=\psi_{n_r,l}\left(\frac{\hbar^{2}}{m},\omega,r_c,r\right). 
\end{aligned}
\end{equation}

\begingroup           
\squeezetable
\begin{table}
\caption{$S_{\rvec^{\prime}},S_{\pvec^{\prime}}$, $S_{t}$ for $1s,~1p,~1d$ states in PISB and CHO (six selected $\eta$).
 See text for detail.}
\centering
\begin{ruledtabular}
\begin{tabular}{l|l|l|llllll}
state  & Property & PISB($\eta=0$) & $\eta=0.0001$  &  $\eta=0.0625$   &  $\eta=1.0$  & $\eta=5.0625$  &  $\eta=45.6976$  &  $\eta=104.8576$ \\
\hline
         & $S_{\rvec^{\prime}}$ & 0.675583 & 0.67558205 & 0.67493721 & 0.66522220 & 0.62260461 & 0.19387157 & $-$0.28934719 \\
1s       & $S_{\pvec^{\prime}}$ & 5.9416   & 5.941691   & 5.941941   & 5.945800   & 5.964491   & 6.266362   & 6.725853          \\
         & $S_{t}$     & 6.6172   & 6.617273   & 6.616878   & 6.611022   & 6.587096   & 6.460233   & 6.436505          \\
\hline 
         & $S_{\rvec^{\prime}}$ & 0.520372 & 0.52037321 & 0.52010134 & 0.51599338 & 0.49769459 & 0.28018517 & $-$0.06370302 \\
1p       & $S_{\pvec^{\prime}}$ & 6.5098   & 6.509889   & 6.509982   & 6.511416   & 6.518562   & 6.670296   &  7.018161   \\
         & $S_{t}$     & 7.0302   & 7.030263   & 7.030083   & 7.027410   & 7.016257   & 6.950482   &  6.954458   \\
\hline
         & $S_{\rvec^{\prime}}$ & 0.552449 & 0.55244843 & 0.55231997 & 0.55037642 & 0.54166146 & 0.42931507 & 0.20823953 \\
1d       & $S_{\pvec^{\prime}}$ & 7.0957   & 7.095786   & 7.095821   & 7.096367   & 7.099323   & 7.176126   & 7.415409 \\
         & $S_{t}$     & 7.6482   & 7.648235   & 7.648141   & 7.646743   & 7.640984   & 7.605441   & 7.623648
\end{tabular}
\end{ruledtabular}
\end{table}
\endgroup

After substitution of $r=r_{c} r^{\prime}$ into Eq.~(12), the modified dimensionless Schr\"odinger equation can be written as
\begin{equation}
\left[-\frac{1}{2} \ \frac{d^2}{dr^{\prime 2}} + \frac{l (l+1)} {2r^{\prime 2}} + \frac{1}{2} \eta r^{\prime 2} 
+V\theta(r^{\prime}-1) \right] \psi_{n_r,l}(r^{\prime})= \frac{mr_{c}^{2}}{\hbar^{2}}\mathcal{E}_{n_r,l} \ \psi_{n_r,l}(r^{\prime}),
\end{equation}

Where $r^{\prime}$ is a dimensionless variable and $\eta=\frac{m\omega r_{c}^{4}}{\hbar^{2}}$. At $\eta = 0$ this
represents the PISB Hamiltonian. The above conversion leads to, 
\begin{equation}
\begin{aligned}
\mathcal{E}_{n_r,l}\left(\frac{\hbar^{2}}{m}, \omega, r_c\right) & =\frac{\hbar^{2}}{mr_{c}^{2}}\mathcal{E}_{n_r,l}\left(1, \eta, 1\right), \\
\psi_{n_r,l}\left(\frac{\hbar^{2}}{m},\omega,r_c,r\right) & =\psi_{n_r,l}\left(1,\eta,1,r^{\prime}\right).
\end{aligned}
\end{equation}

\begin{figure}                         
\begin{minipage}[c]{0.32\textwidth}\centering
\includegraphics[scale=0.55]{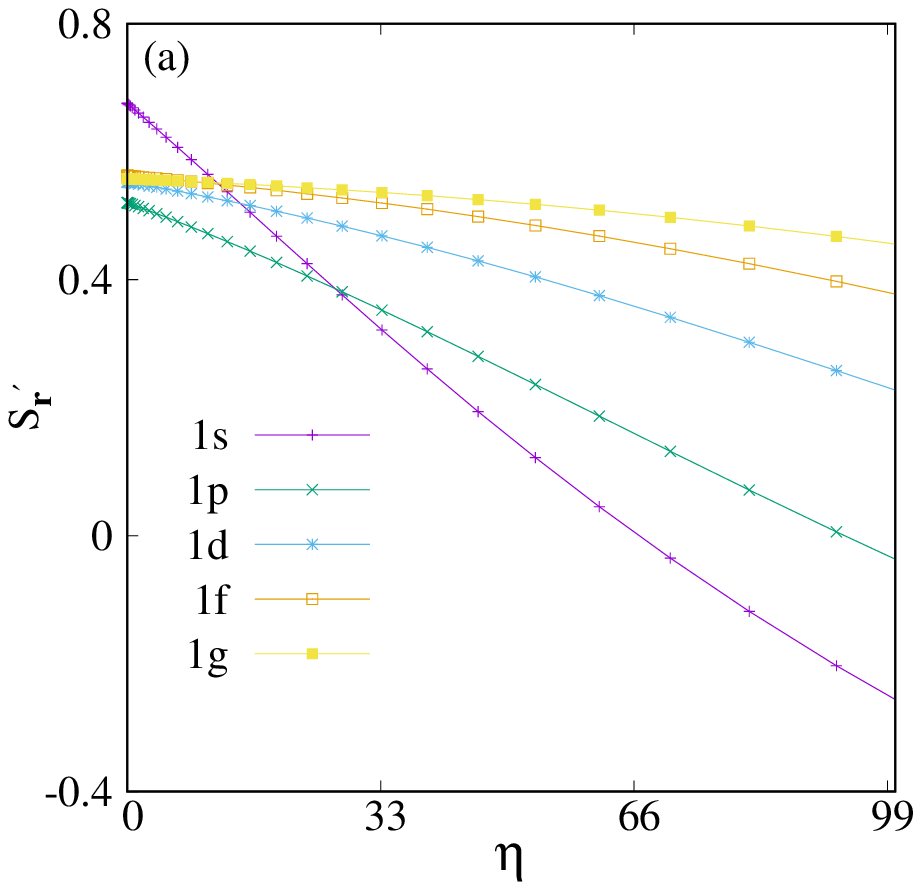}
\end{minipage}%
\hspace{0.02in}
\begin{minipage}[c]{0.32\textwidth}\centering
\includegraphics[scale=0.55]{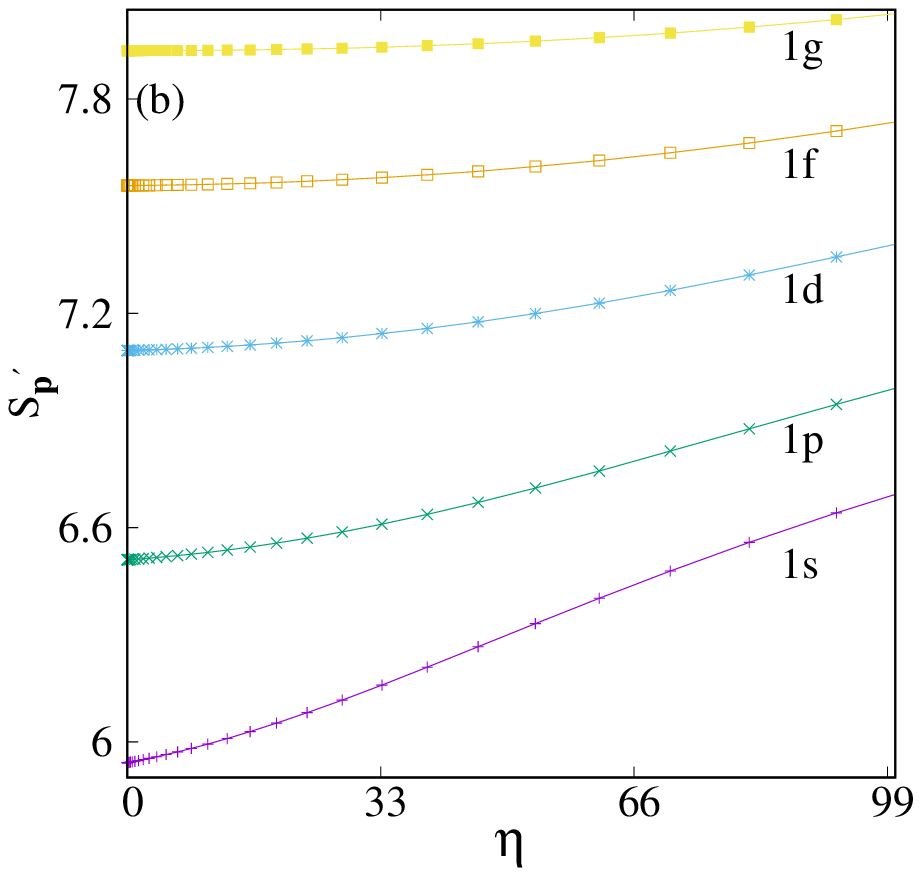}
\end{minipage}%
\hspace{0.02in}
\begin{minipage}[c]{0.32\textwidth}\centering
\includegraphics[scale=0.55]{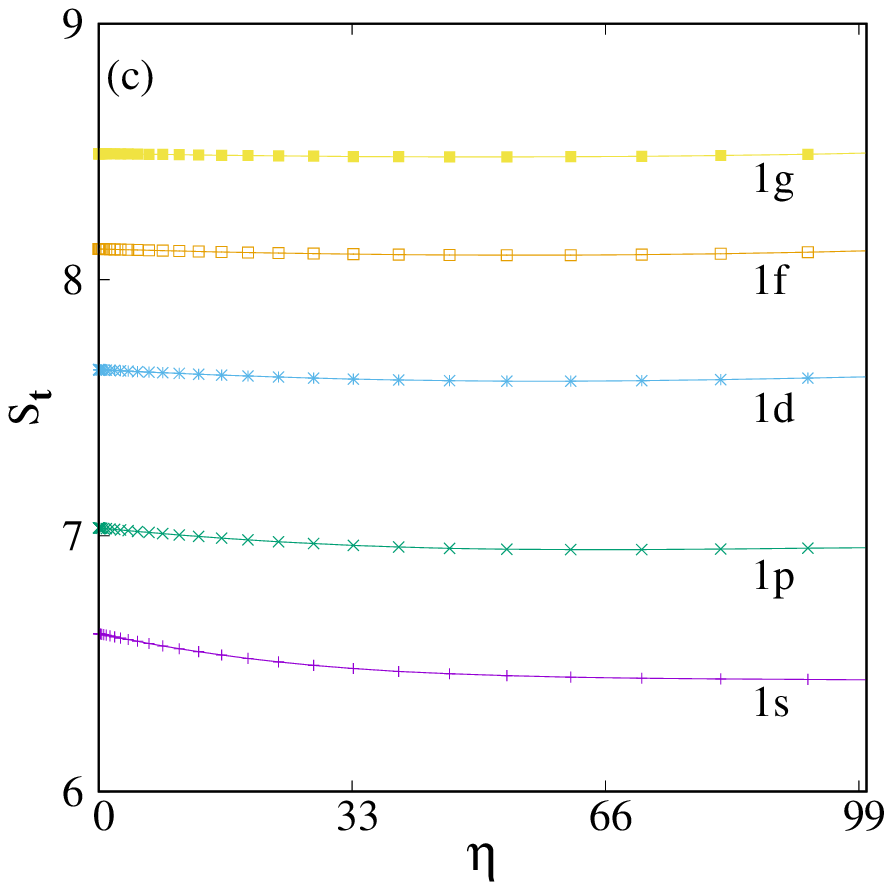}
\end{minipage}%
\caption{Plots of $S_{\rvec^{\prime}}$, $S_{\pvec^{\prime}}$, $S_{t}$ against $\eta$ for first five circular states of CHO in panels 
(a), (b), (c) respectively. See text for details.}
\end{figure}

Equation~(15) indicates that $\eta$ depends on the product of $\omega, m$ and quartic power of $r_c$. However, if we choose $m=\hbar=1$,
then the effective dependence remains on the product of $r_{c}^{4}$ and $\omega$. 

 Third column of Table~I at first portrays $R_{\rvec^{\prime}}^{\alpha},R_{\pvec^{\prime}}^{\beta}$ and 
$R_{t}^{(\alpha,\beta)}$
for $1s,~1p,~1d$ orbitals in PISB. Similarly, 4th-9th columns of this table imprints the same for $1s, 1p, 1d$ states in CHO at six selected 
$\eta$ values namely $0.0001, 0.0625, 1, 5.0625, 45.6976, 104.8576$. These results clearly indicate that, in CHO $R_{\rvec^{\prime}}^{\alpha}$
decreases and $R_{\pvec^{\prime}}^{\beta}$ increases with rise in $\eta$. In case of $1s$ state $R_{t}^{(\alpha,\beta)}$ lowers with $\eta$.
On the contrary,
for $1p, 1d$ states it progresses with elevation of $\eta$. Now, it is important to illustrate the behaviour R in CHO at $\eta \rightarrow 0$ region. 
A careful examination reveals that, in the neighbourhood of $\eta < 1$, CHO has R values comparable with PISB. This trend generally holds good for all other states 
as well. Hence, at low-$\eta$ region, CHO behaves like PISB. Now, panels (a), (b) and (c) of Fig.~1 delineate the variation of $R_{\rvec^{\prime}}^{\alpha}, 
R_{\pvec^{\prime}}^{\beta}$ and $R_{t}^{(\alpha,\beta)}$ respectively against $\eta$ for five lowest states of a CHO corresponding to $l=0$ to $4$. Panel (a) shows that, 
$R_{\rvec^{\prime}}^{\alpha}$ 
falls off with rise of $\eta$ implying greater localization at larger $\eta$. Interestingly at $\eta \rightarrow 0$ $R_{\rvec^{\prime}}^{\alpha}$ obeys 
the trend $R_{\rvec^{\prime}}^{\alpha}(1s)>R_{\rvec^{\prime}}^{\alpha}(1g)>R_{\rvec^{\prime}}^{\alpha}(1f)>R_{\rvec^{\prime}}^{\alpha}(1d)>
R_{\rvec^{\prime}}^{\alpha}(1p)$. But, at large $\eta$ region this trend modifies to $R_{\rvec^{\prime}}^{\alpha}(1g)>R_{\rvec^{\prime}}^{\alpha}(1f)
>R_{\rvec^{\prime}}^{\alpha}(1d)>R_{\rvec^{\prime}}^{\alpha}(1p)>R_{\rvec^{\prime}}^{\alpha}(1s)$. Panel (b) suggests that, 
$R_{\pvec^{\prime}}^{\beta}$ accelerates with growth of $\eta$. Finally, panel (c) depicts that, $R_{t}^{(\alpha,\beta)}$ for $1s$ state falls off with $\eta$
but for \emph{non-zero} $l$ states it enhances with increment of $\eta$. We also note that, at a fixed $n_{r}$ both $R_{\pvec^{\prime}}^{\beta}$ and $R_{t}^{(\alpha,\beta)}$ 
increase with increase in quantum number $l$.

\begingroup           
\squeezetable
\begin{table}
\caption{$E_{\rvec^{\prime}},E_{\pvec^{\prime}}$, $E_{t}$ for $1s,~1p,~1d$ states in PISB and CHO (six selected $\eta$).
 See text for detail.}
\centering
\begin{ruledtabular}
\begin{tabular}{l|l|l|llllll}
state  & Property & PISB($\eta=0$) & $\eta=0.0001$  &  $\eta=0.0625$   &  $\eta=1.0$  & $\eta=5.0625$  &  $\eta=45.6976$  &  $\eta=104.8576$ \\
\hline
         & $E_{\rvec^{\prime}}$ & 0.672078 & 0.6720791 & 0.67267502 & 0.68180978 & 0.72295672 & 1.23933182  & 2.09908616 \\
1s       & $E_{\pvec^{\prime}}$ & 0.003982 & 0.003982   & 0.003982   & 0.003960   & 0.003852   & 0.002818    & 0.001860   \\
         & $E_{t}$     & 0.002678 & 0.002679   & 0.002680   & 0.002700   & 0.002785   & 0.003492    & 0.003904   \\
\hline 
         & $E_{\rvec^{\prime}}$ & 0.803227 & 0.80322700 & 0.80351307 & 0.80784686 & 0.82740709 & 1.09024283 & 1.62179448  \\
1p       & $E_{\pvec^{\prime}}$ & 0.002277 & 0.002277   & 0.002277   & 0.002269   & 0.002236   & 0.001858   & 0.001354    \\
         & $E_{t}$     & 0.001829 & 0.001829   & 0.001829   & 0.001833   & 0.001850   & 0.002026   & 0.002197    \\
\hline
         & $E_{\rvec^{\prime}}$ & 0.851258 & 0.85125726 & 0.85139660 & 0.85350774 & 0.86304220 & 0.99491879  & 1.297877781 \\
1d       & $E_{\pvec^{\prime}}$ & 0.001378 & 0.001377   & 0.001377   & 0.001375   & 0.001363   & 0.001214    & 0.000967    \\
         & $E_{t}$     & 0.001173 & 0.001173   & 0.001173   & 0.001173   & 0.001176   & 0.001208    & 0.001255  
\end{tabular}
\end{ruledtabular}
\end{table}
\endgroup

Now we move on to S in Table~II, where $S_{\rvec^{\prime}}, S_{\pvec^{\prime}}$ and $S_{t}$ are probed for $1s, 1p, 1d$ states of PISB (3rd column) and CHO 
(at same particular set of $\eta$ as in Table~I). Like R, $S_{\rvec^{\prime}}$ progresses and $S_{\pvec^{\prime}}$ diminishes with growth in 
$\eta$. For $l=0$ states $S_{t}$ decreases with $\eta$, while, for $l \neq 0$ cases it mounts up. At $\eta \rightarrow 0$ region, like R, S in 
CHO also provides equivalent results to that of PISB. In order to gain further insight, Fig.~2 portrays $S_{\rvec^{\prime}}, S_{\pvec^{\prime}}$ and $S_{t}$ in 
left (a), middle (b) and right (c) panels, for lowest five $l(0-4)$ as a function of $\eta$. But unlike $R_{t}^{(\alpha,\beta)}$, $S_{t}$ for all these states 
lessen with increment in $\eta$. On the contrary, as observed in R, S's in $r$ space at $\eta \rightarrow 0$ obeys the same order, \emph{viz.}, 
$S_{\rvec^{\prime}}(1s)>S_{\rvec^{\prime}}(1g)
>S_{\rvec^{\prime}}(1f)>S_{\rvec^{\prime}}(1d)>S_{\rvec^{\prime}}(1p)$. As usual at $\eta \rightarrow \infty$ this 
trend modifies to $S_{\rvec^{\prime}}(1g)>S_{\rvec^{\prime}}(1f)>S_{\rvec^{\prime}}(1d)>S_{\rvec^{\prime}}(1p)>S_{\rvec^{\prime}}(1s)$. Here again analogous 
to $R_{\pvec^{\prime}}^{\beta} and R_{t}^{(\alpha,\beta)}$, both $S_{\pvec^{\prime}} and S_{t}$ advance with increase in $l$. 

\begin{figure}                         
\begin{minipage}[c]{0.32\textwidth}\centering
\includegraphics[scale=0.55]{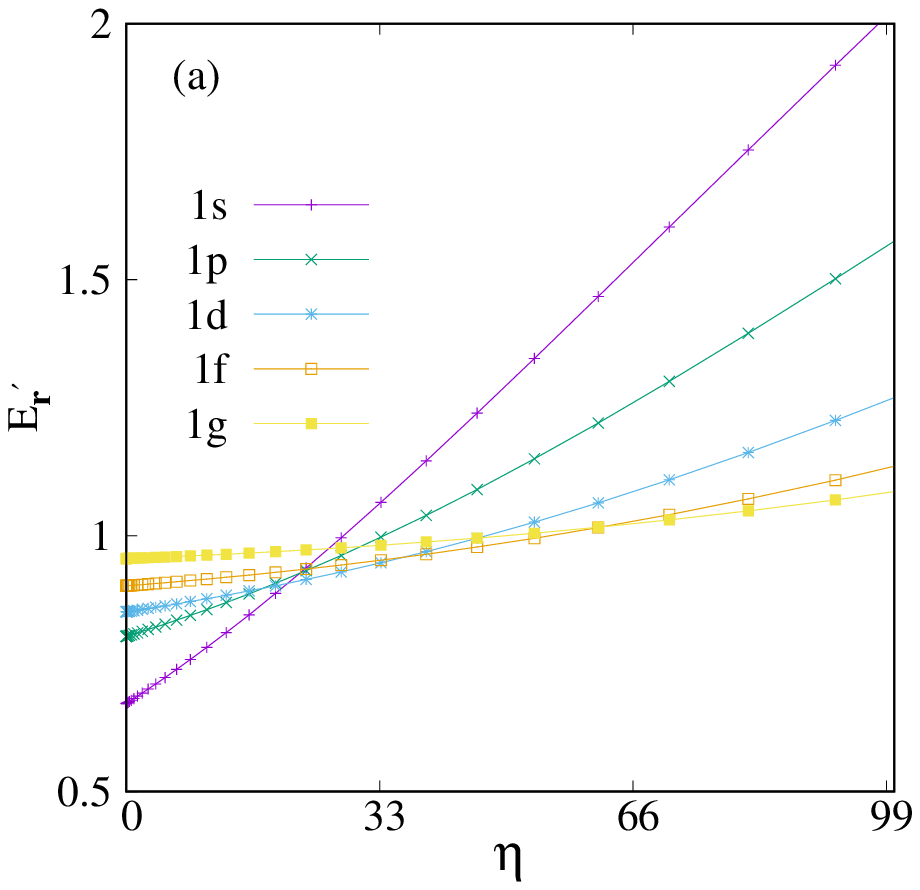}
\end{minipage}%
\hspace{0.02in}
\begin{minipage}[c]{0.32\textwidth}\centering
\includegraphics[scale=0.55]{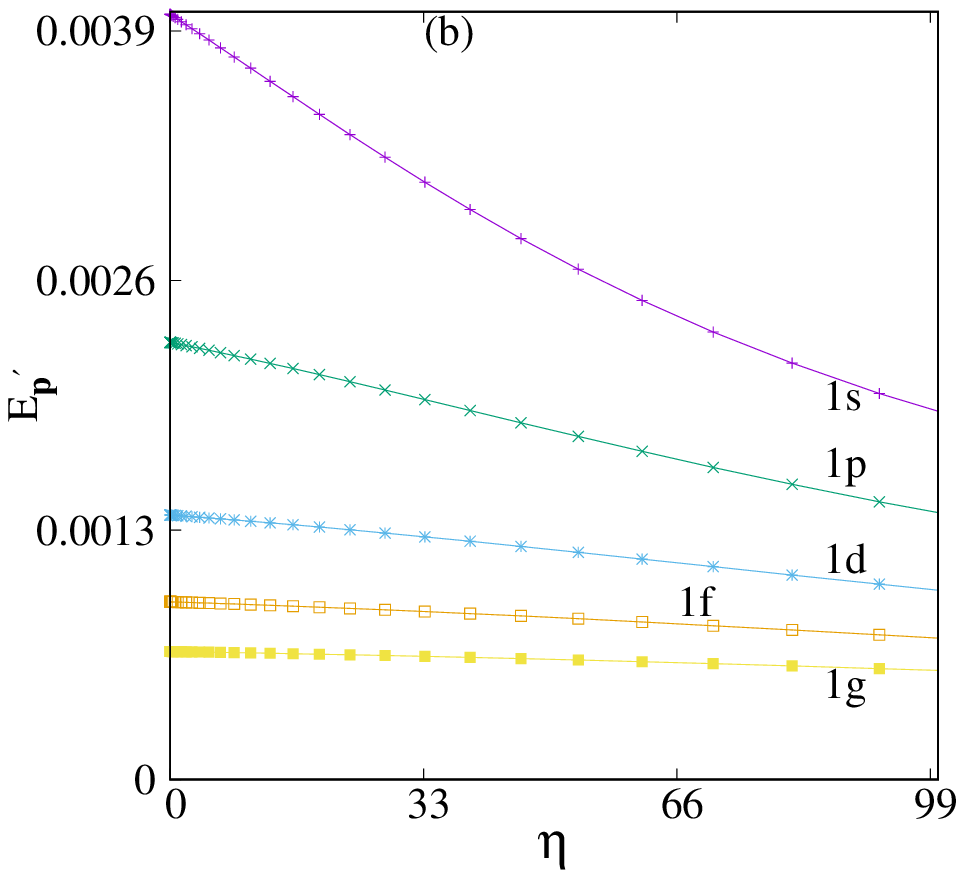}
\end{minipage}%
\hspace{0.02in}
\begin{minipage}[c]{0.32\textwidth}\centering
\includegraphics[scale=0.55]{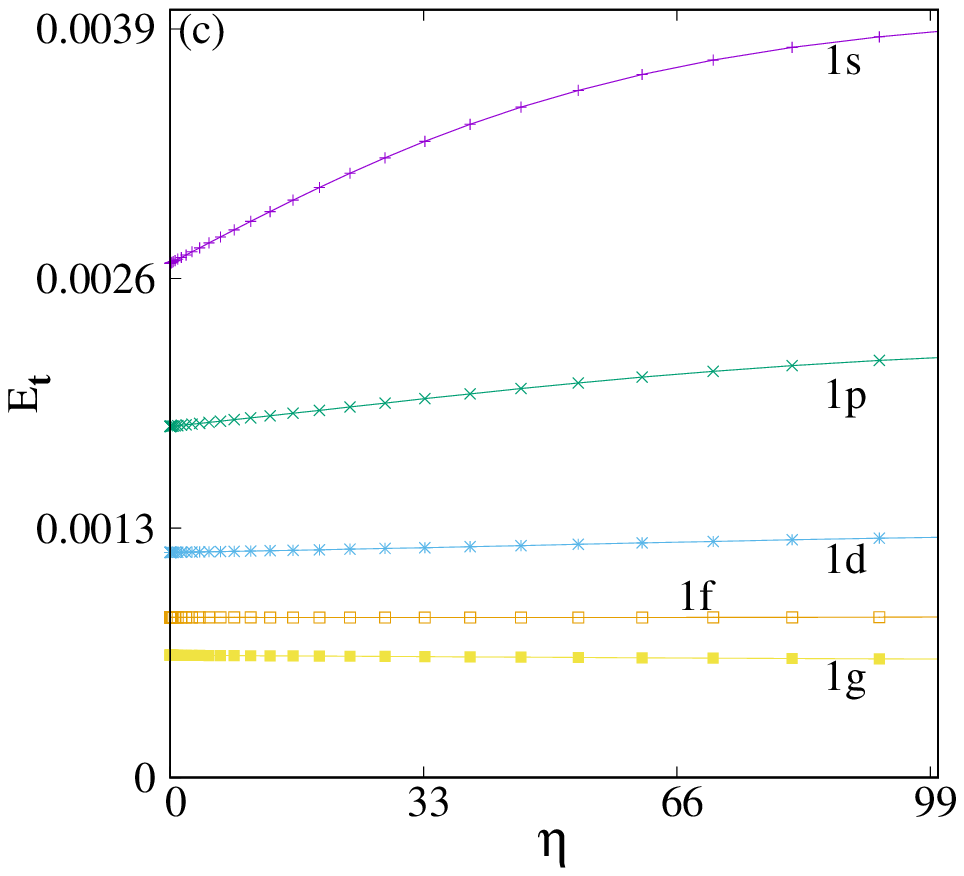}
\end{minipage}%
\caption{Plots of $E_{\rvec}$, $E_{\pvec}$, $E_{t}$ against $\eta$ for first five circular states of CHO in panels (a), (b), (c) respectively. 
See text for details.}
\end{figure}

Now we discuss $E$ in Table~III, by providing $E_{\rvec^{\prime}},E_{\pvec^{\prime}}$ and $E_{t}$ of $1s, 1p, 1d$ states of selected $\eta$ values used
in Table~I and II. Akin to R and S, E at $\eta \rightarrow 0$ delivers coequal result to that of PISB. But in other context, E shows complete reverse 
trends to what we have seen in R and S. $E_{\rvec^{\prime}}$, $E_{t}$ advance and $E_{\pvec^{\prime}}$ reduces with improvement in $\eta$. Above changes 
in $E_{\rvec^{\prime}},E_{\pvec^{\prime}}$ and $E_{t}$ are graphically displayed in Figure~3, in left (a), middle (b), right (c) panels for first
five circular states ($n_r=1$ and $l=0-4$). Here one can see that, $E_{\rvec^{\prime}},E_{t}$ decrease and $E_{\pvec^{\prime}}$ increases with progress of 
$\eta$. As $\eta$ approaches zero, $E_{\rvec^{\prime}}$ obeys the trend $E_{\rvec^{\prime}}(1g)>E_{\rvec^{\prime}}(1f)>E_{\rvec^{\prime}}(1d)>
E_{\rvec^{\prime}}(1p)>E_{\rvec^{\prime}}(1s)$ which gets reversed to $E_{\rvec^{\prime}}(1s)>E_{\rvec^{\prime}}(1p)>E_{\rvec^{\prime}}(1d)>
E_{\rvec^{\prime}}(1f)>E_{\rvec^{\prime}}(1g)$ at opposite $\eta$ limit.
Whereas, at a fixed $n_r$, both $E_{\pvec^{\prime}}$ and $E_{t}$ collapse with rise in $l$.   

\begingroup           
\squeezetable
\begin{table}
\caption{$R_{\rvec}^{\alpha},R_{\pvec}^{\beta}$ and $R^{\alpha,\beta}$ values for $1s,~2s,~1p,~2p,~1d,~2d$ orbitals in CHO at eight selected $r_c$ values. 
See text for detail.}
\centering
\begin{ruledtabular}
\begin{tabular}{llllllll}
$r_c$  &    $R_{\rvec}^{\alpha}$     & $R_{\pvec}^{\beta}$  &  $R_{t}^{\alpha,\beta}$  &  
$r_c$  &    $R_{\rvec}^{\alpha}$     & $R_{\pvec}^{\beta}$  &  $R_{t}^{\alpha,\beta}$  \\
\cline{1-4} \cline{5-8}
\multicolumn{4}{c}{$1s$}    &      \multicolumn{4}{c}{$2s$}    \\
\cline{1-4} \cline{5-8}
0.1      & $-$6.0366917844   & 12.247171978  & 6.2104801936 & 0.1   & $-$6.0653334752 & 14.2515610845 & 8.186227609   \\
0.2      & $-$3.9572613535   & 10.167740386  & 6.2104790325 & 0.2   & $-$3.9858896952 & 12.1721134683 & 8.186223773   \\
0.5      & $-$1.2088403559   & 7.41927227    & 6.21043191   & 0.5   & $-$1.2369267194 &  9.422993967  & 8.18606724    \\
1.0      & 0.86363060146   & 5.3460818801  & 6.2097124816 & 1.0     &  0.8438971150   &  7.339578635  & 8.18347575    \\
2.0      & 2.82653053607   & 3.3731259092  & 6.1996564453 & 2.0     &  2.9410466015   &  5.15805124   & 8.09909784    \\
5.0      & 3.63268067673   & 2.5410652239  & 6.1737459006 & 5.0     &  4.5764993107   &  2.6162033    & 7.1927026     \\
8.0      & 3.632690916310  & 2.5410540440  & 6.1737449603 & 8.0     &  4.5767695172   &  2.61482528   & 7.191594797   \\
$\infty$ & 3.6326909163101 & 2.5410540440  & 6.1737449603 &$\infty$ &  4.5767695172   &  2.614825285  & 7.1915948022  \\
\cline{1-4} \cline{5-8}
\multicolumn{4}{c}{$1p$}    &      \multicolumn{4}{c}{$2p$}    \\
\cline{1-4} \cline{5-8}
0.1    &  $-$6.1671363542   & 12.806502114 & 6.6393657598 & 0.1      & $-$6.29011971 & 14.178463580 & 7.88834387   \\
0.2    &  $-$4.0876997330   & 10.727065993 & 6.63936626   & 0.2      & $-$4.21067738 & 12.09902562  & 7.88834824  \\
0.5    &  $-$1.3390273792   & 7.978413948  & 6.639386568  & 0.5      & $-$1.46177311 & 9.35029899   & 7.88852588  \\
1.0    &   0.7373200936   & 5.902381495  & 6.639701588  & 1.0      & 0.618160513 & 7.273091288  & 7.891251801 \\
2.0    &   2.7629313961   & 3.882363728  & 6.6452951241 & 2.0      & 2.704922872 & 5.226357439  & 7.931280311 \\
5.0    &   3.8830108378   & 2.834907070  & 6.7179179078 & 5.0      & 4.575880956 & 2.9828085    & 7.558689456 \\
8.0    &   3.883056660633 & 2.8349473768 & 6.7180040374 & 8.0      & 4.57683522  & 2.982008644  & 7.558843864 \\
$\infty$ & 3.883056660633 & 2.8349473768 & 6.7180040374 &$\infty$  & 4.57683522  & 2.982008644  & 7.558843864 \\
\cline{1-4} \cline{5-8}
\multicolumn{4}{c}{$1d$}    &      \multicolumn{4}{c}{$2d$}    \\
\cline{1-4} \cline{5-8}
0.1    & $-$6.1181191683  & 13.2289542792 & 7.1108351109 & 0.1     & $-$6.2478841627       & 14.295117242 & 8.0472330793\\
0.2    & $-$4.0386800101  & 11.1495160962 & 7.1108360861 & 0.2     & $-$4.1684423307       & 12.21567885  & 8.047236519 \\
0.5    & $-$1.2899046230  & 8.400780312   & 7.110875689  & 0.5     & $-$1.4195583633       & 9.466934869  & 8.047376505 \\
1.0    &  0.7880364400  & 6.323450880   & 7.11148732   & 1.0     &  0.6600654534       & 7.389472850  & 8.049538303 \\
2.0    &  2.8408556597  & 4.2808923084  & 7.1217479681 & 2.0     &  2.7424775770       & 5.34296287   & 8.08544044  \\
5.0    &  4.2284239258  & 3.0468595     & 7.2752834258 & 5.0     &  4.8017869530       & 3.2330682    & 8.0348551   \\
8.0    &  4.22859084294 & 3.047026004   & 7.2756168469 & 8.0     &  4.804603250039670  & 3.2306644880 & 8.035267738 \\
$\infty$& 4.22859084294 & 3.047026004   & 7.2756168469 &$\infty$ &  4.804603250039670  & 3.2306644880 & 8.035267738 \\
\end{tabular}
\end{ruledtabular}
\end{table}
\endgroup

Upto now, we were concerned about the effect of change of $\eta$ in CHO. This investigation clearly reveals that, at $\eta \rightarrow 0$ CHO
behaves alike to PISB. But, since, $\eta \propto \omega r_{c}^{4}$, these results includes combined effect of both $\eta$ and $r_c$. In order to 
get a complete picture of confinement effect, these two factors need to be segregated. Now, we concentrate on analysing all these quantities
with respect to $r_c$. Later we also examine the behaviour of IE with change of $n_r$ at certain selected $r_c$ values namely $0.1, 2.5, 3, 5, 
\infty$. In both the cases we will keep $\omega$ fixed at one. Now, onwards we will use unprimed variables in IE suffixes. 

\begin{figure}                         
\begin{minipage}[c]{0.32\textwidth}\centering
\includegraphics[scale=0.55]{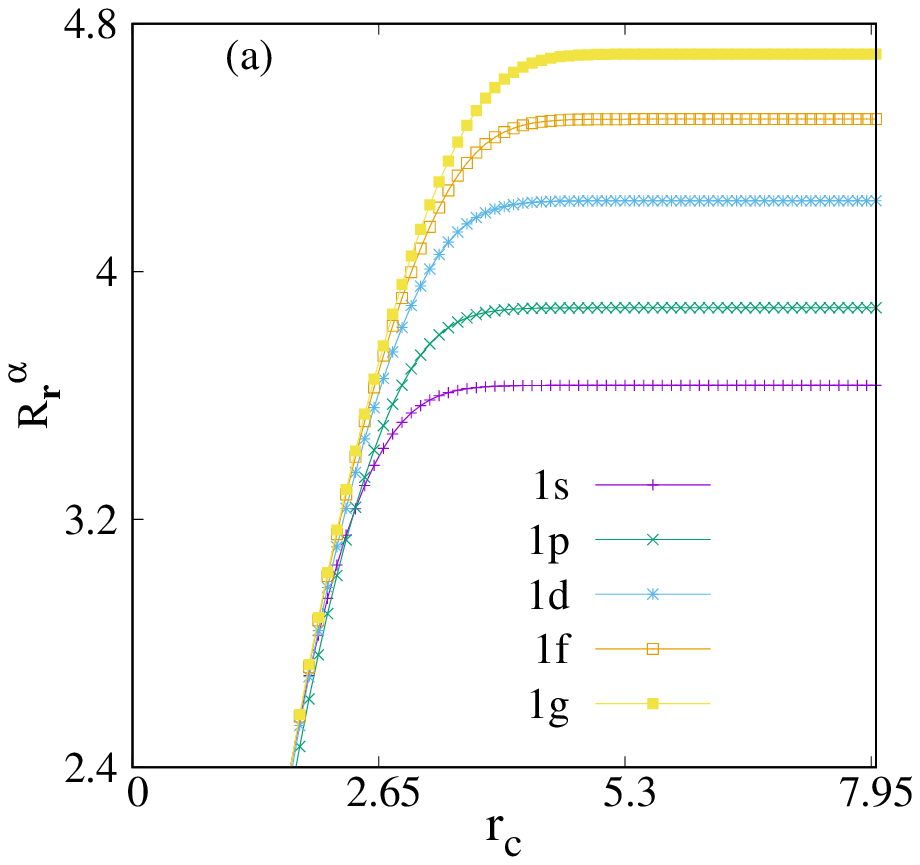}
\end{minipage}%
\hspace{0.02in}
\begin{minipage}[c]{0.32\textwidth}\centering
\includegraphics[scale=0.55]{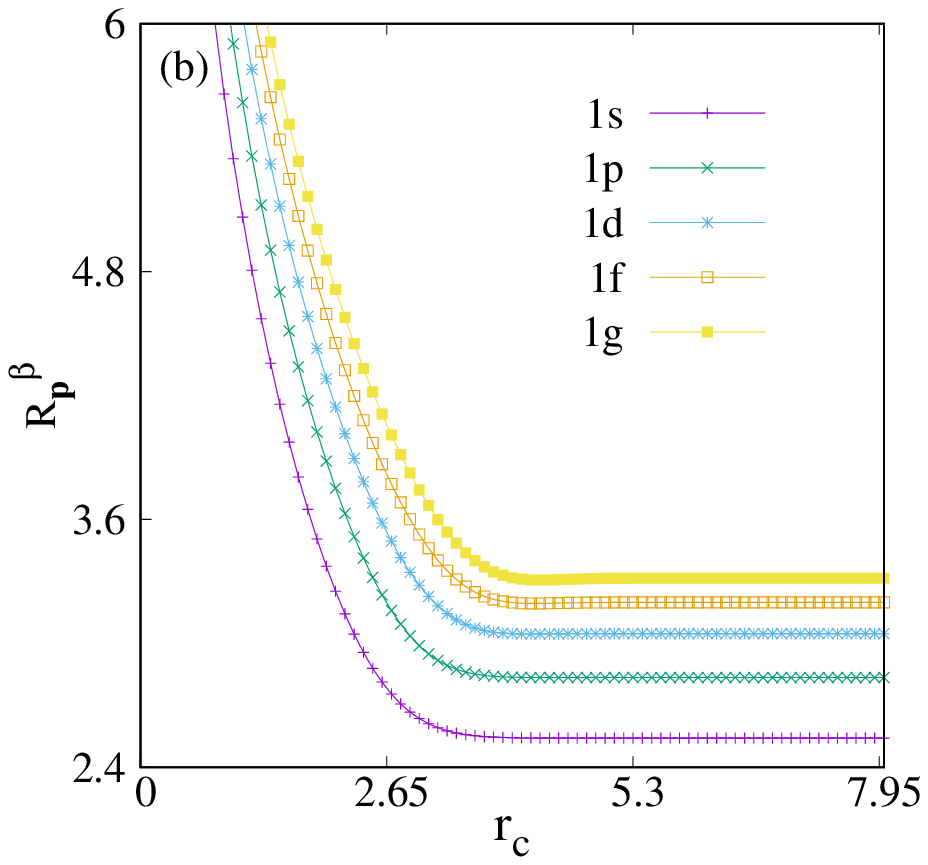}
\end{minipage}%
\hspace{0.02in}
\begin{minipage}[c]{0.32\textwidth}\centering
\includegraphics[scale=0.55]{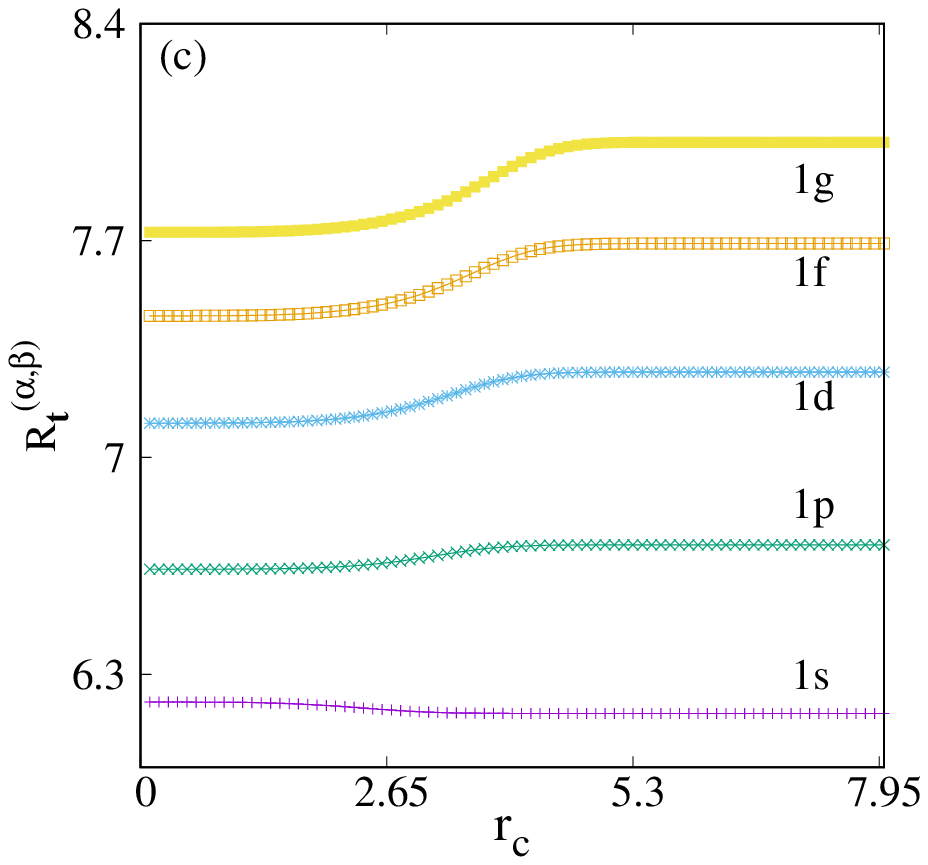}
\end{minipage}%
\caption{Plots of $R_{\rvec}^{\alpha}$, $R_{\pvec}^{\beta}$, $R_{t}^{(\alpha, \beta)}$ against $r_c$ for first five circular states of CHO  
in panels (a), (b), (c) respectively. See text for details.}
\end{figure}  

We will now study the variation of all these information measures with change of $r_c$. It is expected that, a progressively larger $r_c$
should lead to a delocalization of the system in such a fashion that, at $r_c \rightarrow \infty$ it should come out to IHO. Whereas, when 
$r_c \rightarrow 0$ impact of confinement is maximum. Here, calculation are pursued by choosing $r_c$ values starting from $0.1$ to $10$.
This, parametric increase in $r_c$ elicit the system from extremely confined environment to free situation.

To begin with, Table~IV impresses calculated $R_{\rvec}^{\alpha}$, $R_{\pvec}^{\beta}$ and $R_{t}^{(\alpha, \beta)}$ for first two $s,p$ and $d$
orbitals ($n_r=1,2$) of CHO at a selected set of eight $r_c$ values. In this and all following tables of CHO, IEs are furnished for these six 
states considering same set of $r_c$ values. $R_{\rvec}^{\alpha}$'s starting from particular negative values at very low $r_c$, continuously 
advance, finally merges to the respective IHO behaviour. In contrast, $R_{\pvec}^{\beta}$'s in for all these six states generally tend to diminish 
with $r_c$, again converging to IHO in the end. Consequently, the $R_{t}^{(\alpha, \beta)}$ for $1s$ and $2s$ states deplete with $r_c$ to reach 
the borderline values. However, for $l \neq 0$ states it enhances with $r_c$ to attain the limiting values. At very low $r_c$ values $n_r=1$ states
have higher $R_{\rvec}^{\alpha}$ values with respect to their $n_r=2$ counterparts. But, at moderate $r_c$ region this trend gets reverses. Moreover,
this crossover regions switch to higher $r_c$ values with rise of $l$ quantum number. This, observation infers that, the effect of confinement is 
more on higher $n_r$ states. There are no such crossover in $R_{\pvec}^{\beta}$ and $R_{t}^{(\alpha, \beta)}$ in any of these states. Unfortunately
no literature is available to make direct comparison with these computed values. Above observation is graphically depicted in Figure~4, where in 
segments (a)-(c), $R_{\rvec}^{\alpha}$, $R_{\pvec}^{\beta}$ and $R_{t}^{(\alpha, \beta)}$ of first five circular states with respected to $r_c$ are
portrayed. Panel (a) imprints that, for all of them, $R_{\rvec}^{\alpha}$'s quite steadily progress with $r_c$ and finally convene to IHO. Similarly,
from panel (b) it is clear that, $R_{\pvec}^{\beta}$ shows opposite pattern with $r_c$, before reaching IHO-limit. Panel (c) reveals that, for $l=0$
state $R_{t}^{(\alpha, \beta)}$ decreases with $r_c$. But, for $l \neq 0$ states reverse trend is observed. However, for all these five states 
$R_{t}^{(\alpha, \beta)}$'s finally converge to their respective IHO values.

\begin{figure}                                            
\begin{minipage}[c]{0.33\textwidth}\centering
\includegraphics[scale=0.5]{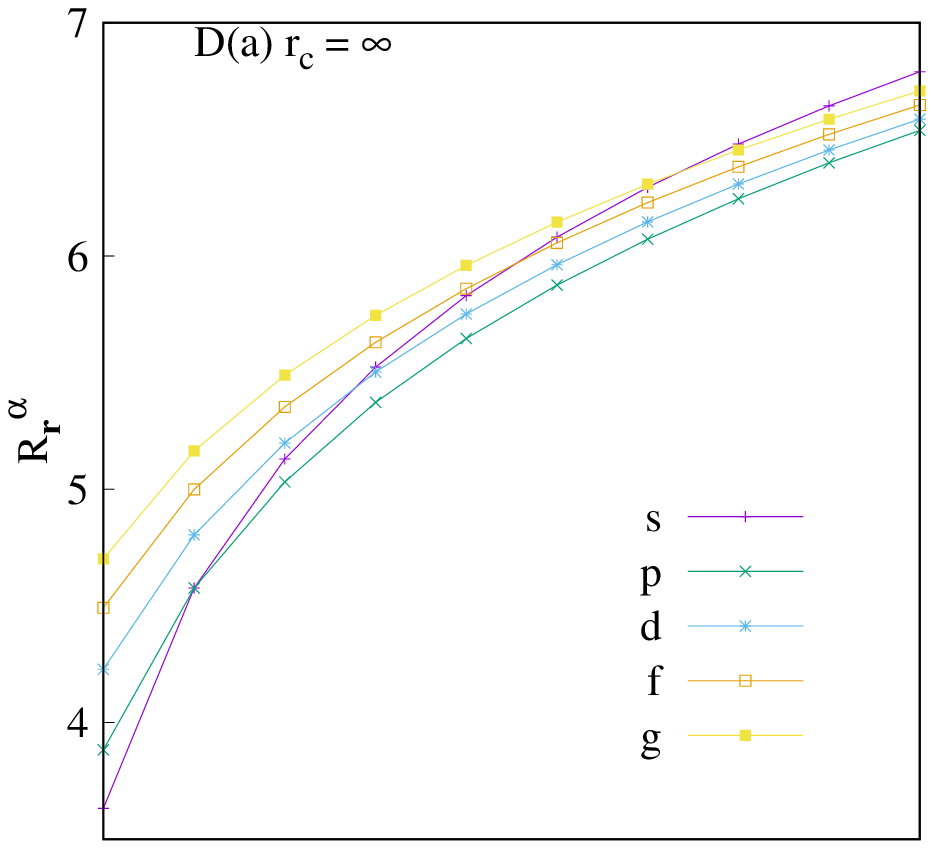}
\end{minipage}%
\begin{minipage}[c]{0.33\textwidth}\centering
\includegraphics[scale=0.48]{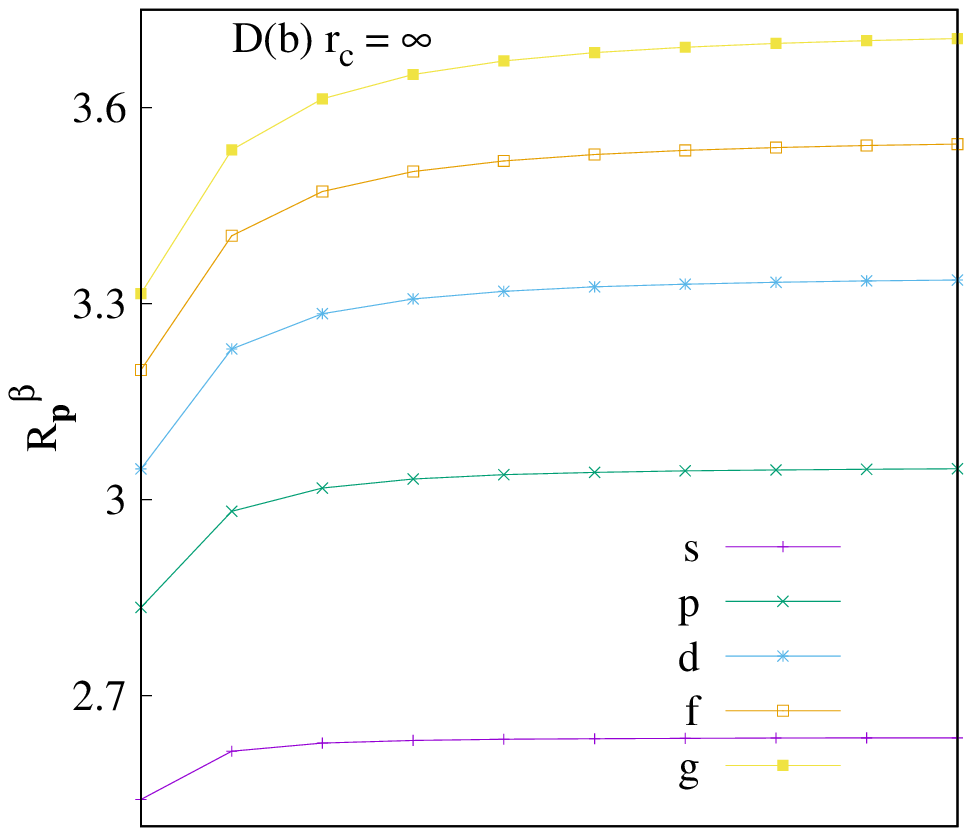}
\end{minipage}%
\begin{minipage}[c]{0.33\textwidth}\centering
\includegraphics[scale=0.48]{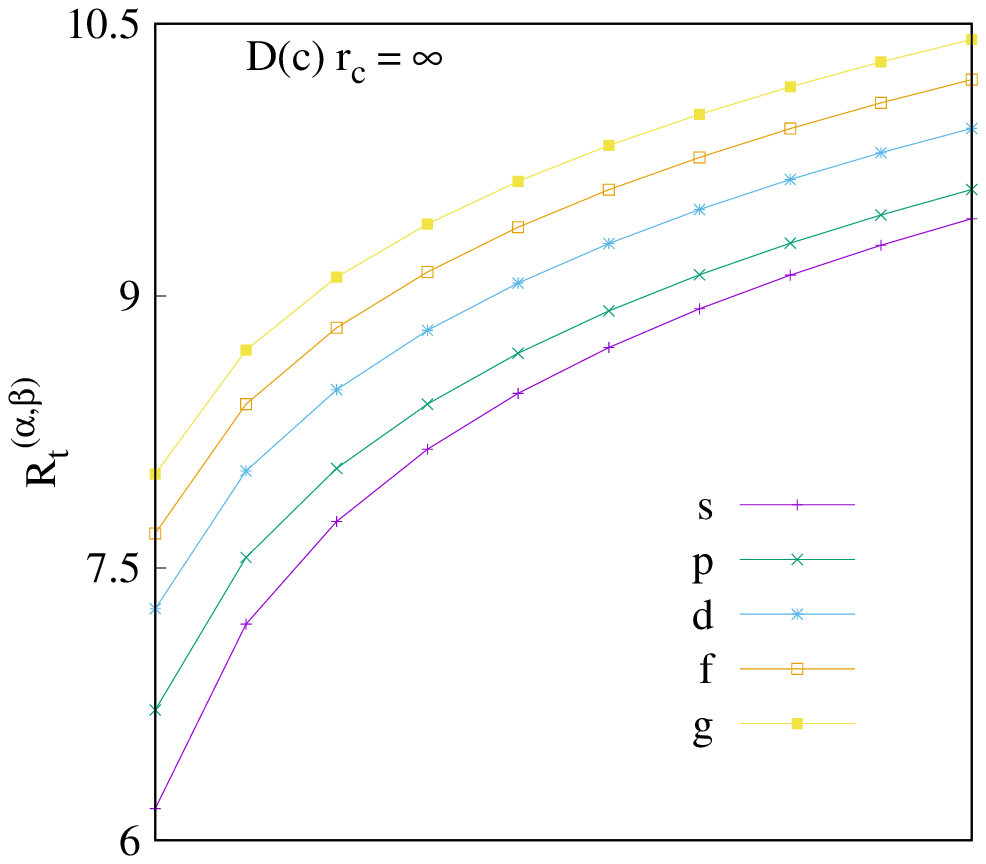}
\end{minipage}%
\hspace{0.2in}
\begin{minipage}[c]{0.33\textwidth}\centering
\includegraphics[scale=0.48]{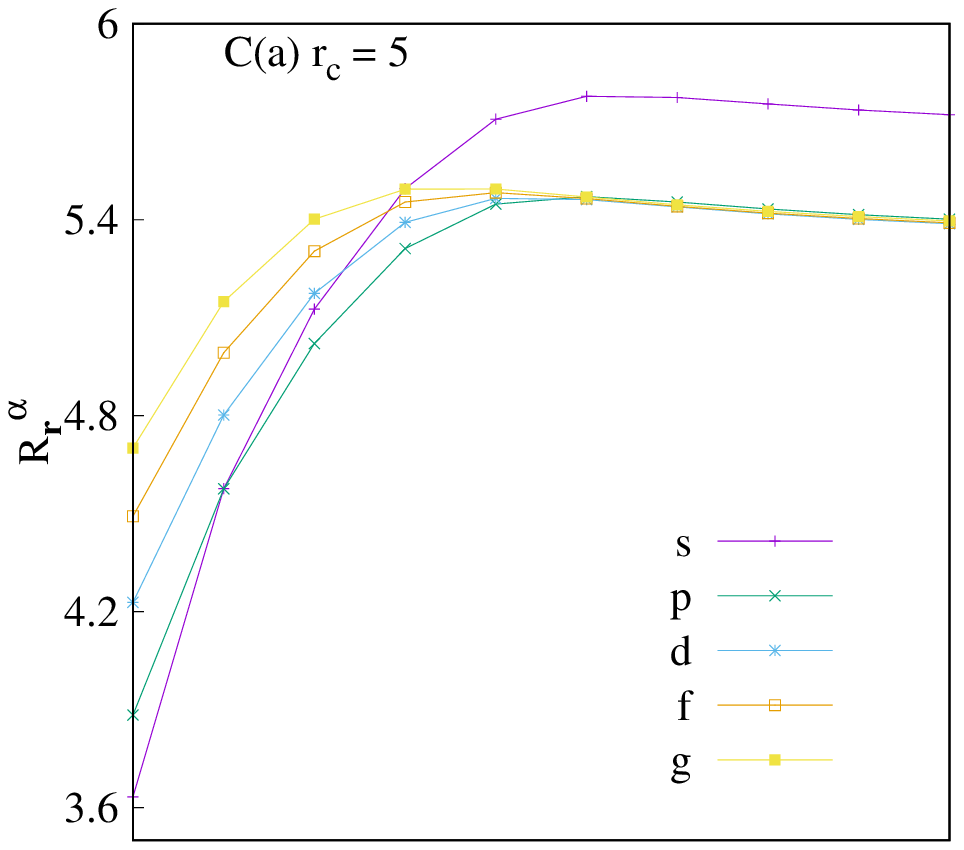}
\end{minipage}
\begin{minipage}[c]{0.33\textwidth}\centering
\includegraphics[scale=0.48]{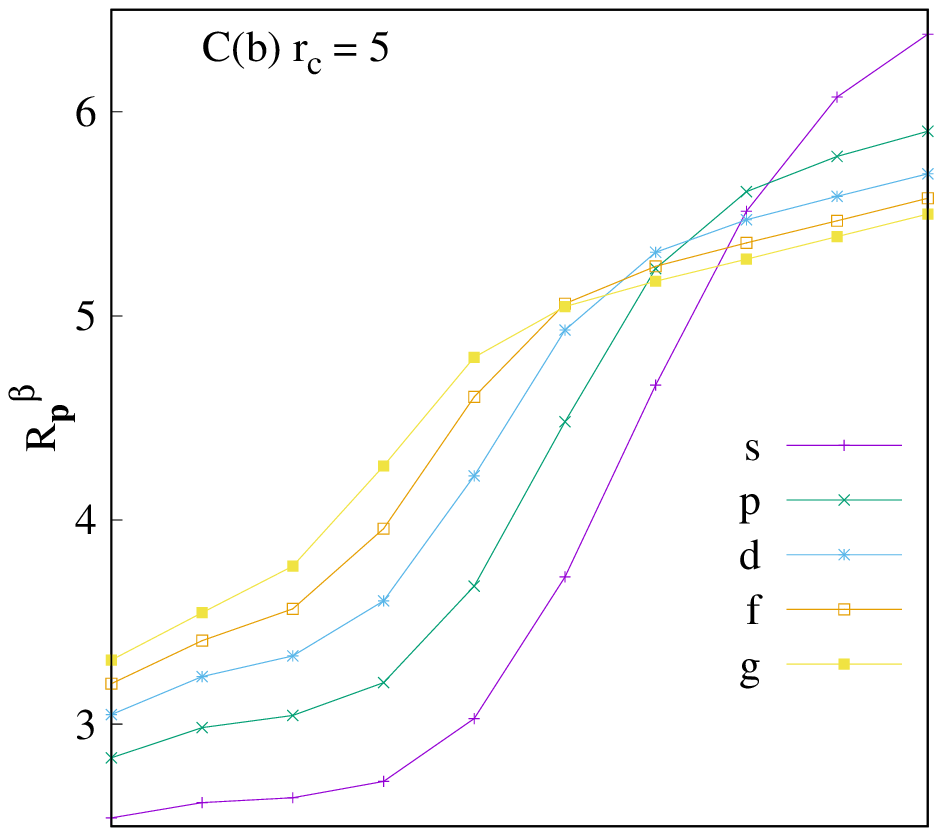}
\end{minipage}%
\begin{minipage}[c]{0.33\textwidth}\centering
\includegraphics[scale=0.48]{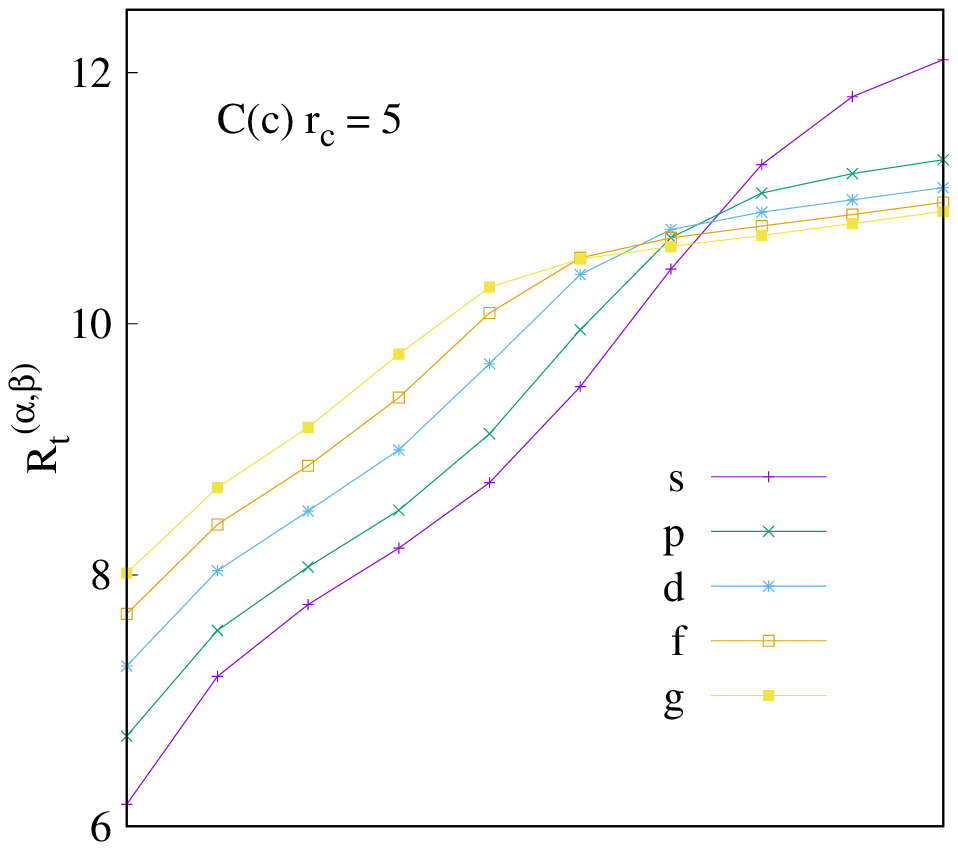}
\end{minipage}%
\hspace{0.2in}
\begin{minipage}[c]{0.33\textwidth}\centering
\includegraphics[scale=0.48]{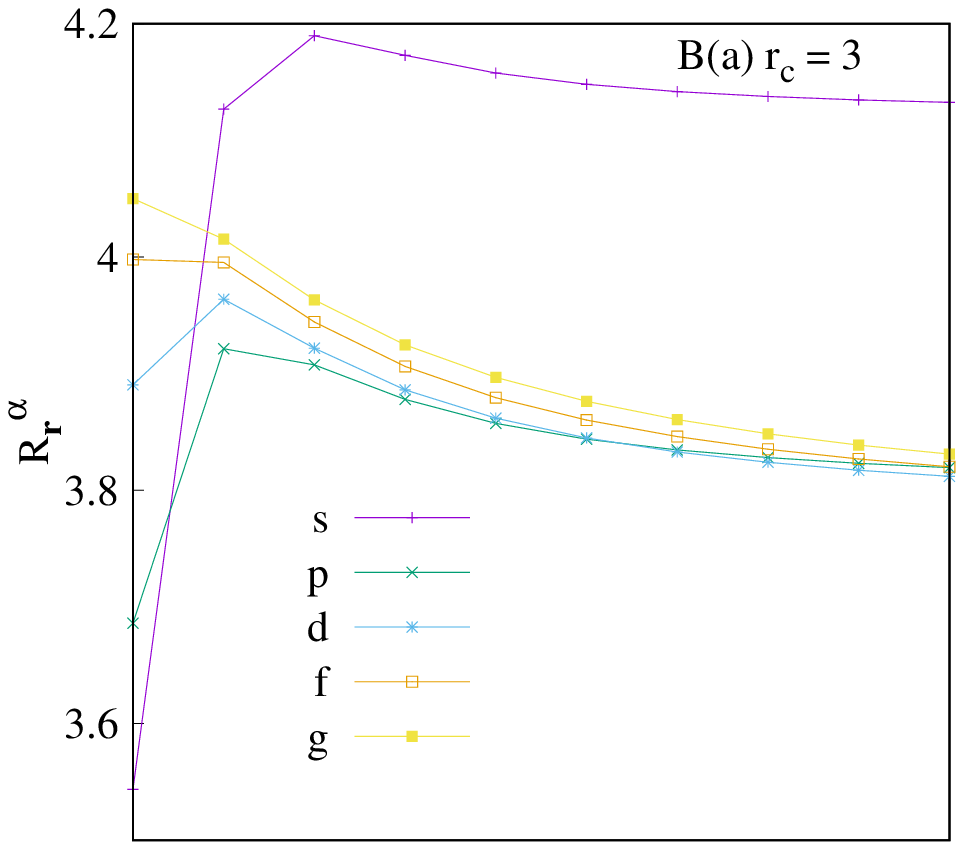}
\end{minipage}%
\begin{minipage}[c]{0.33\textwidth}\centering
\includegraphics[scale=0.48]{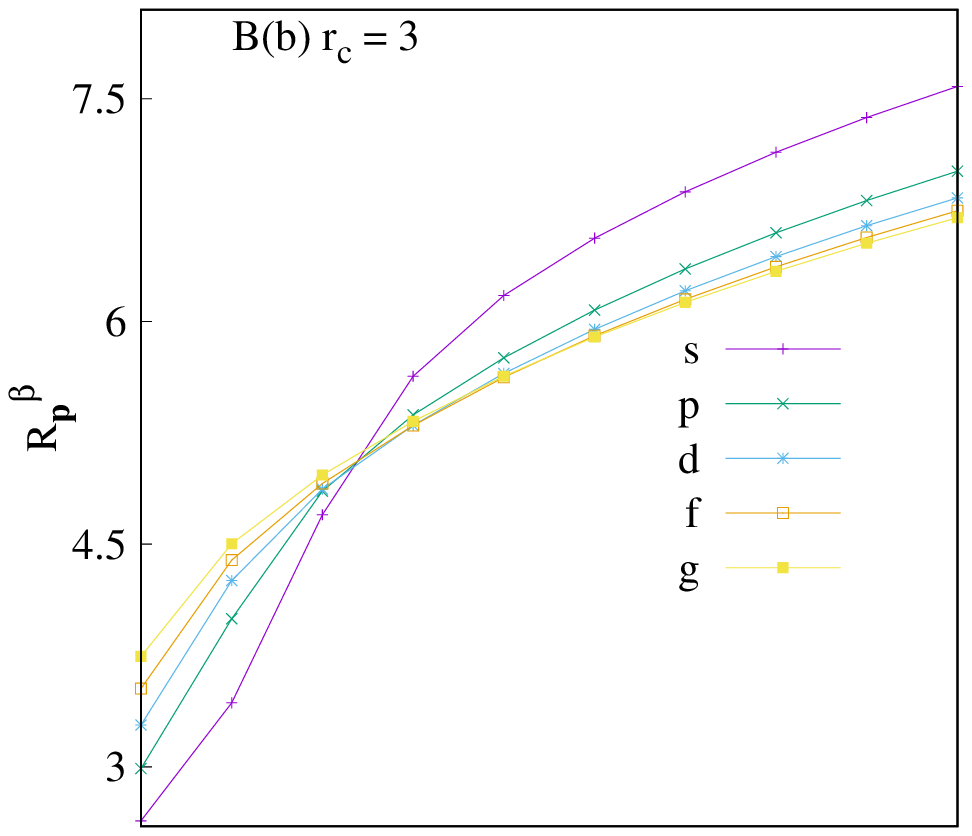}
\end{minipage}%
\begin{minipage}[c]{0.33\textwidth}\centering
\includegraphics[scale=0.48]{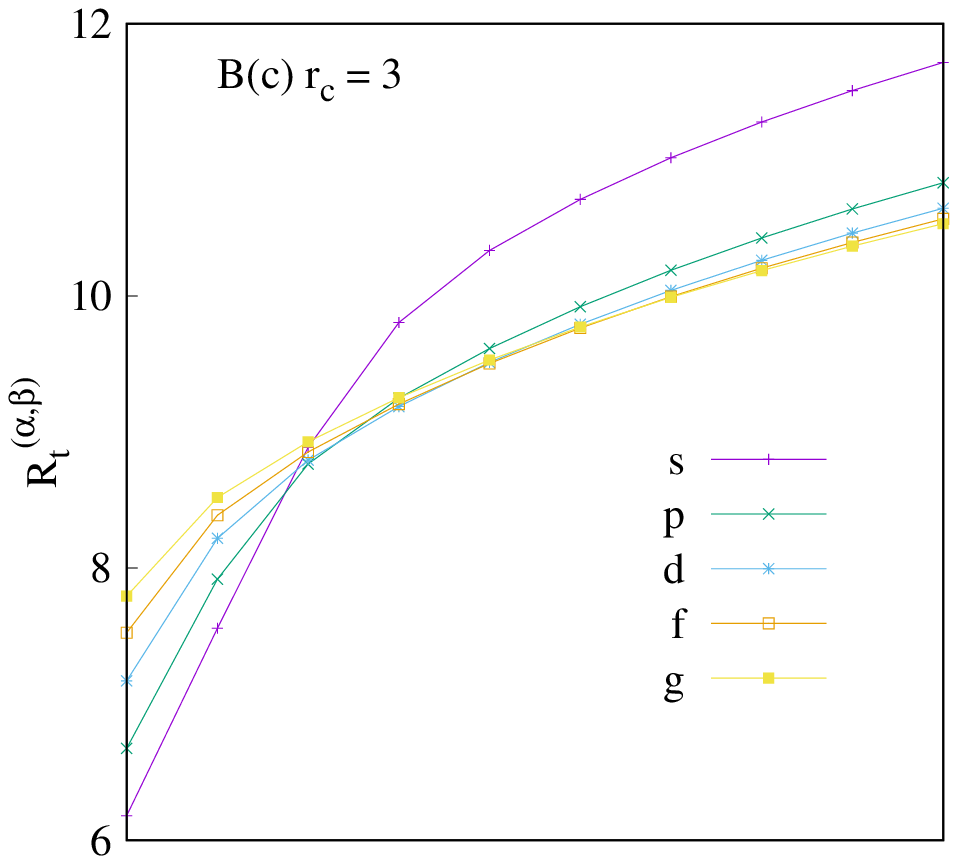}
\end{minipage}%
\hspace{0.2in}
\begin{minipage}[c]{0.33\textwidth}\centering
\includegraphics[scale=0.48]{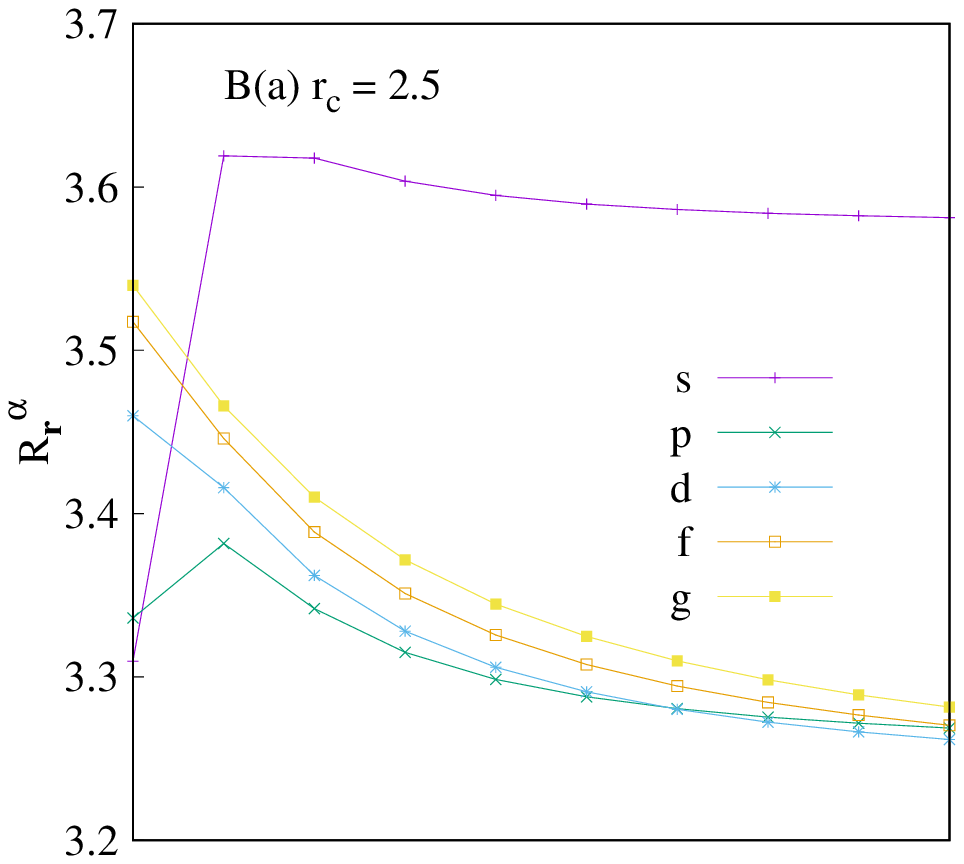}
\end{minipage}%
\begin{minipage}[c]{0.33\textwidth}\centering
\includegraphics[scale=0.48]{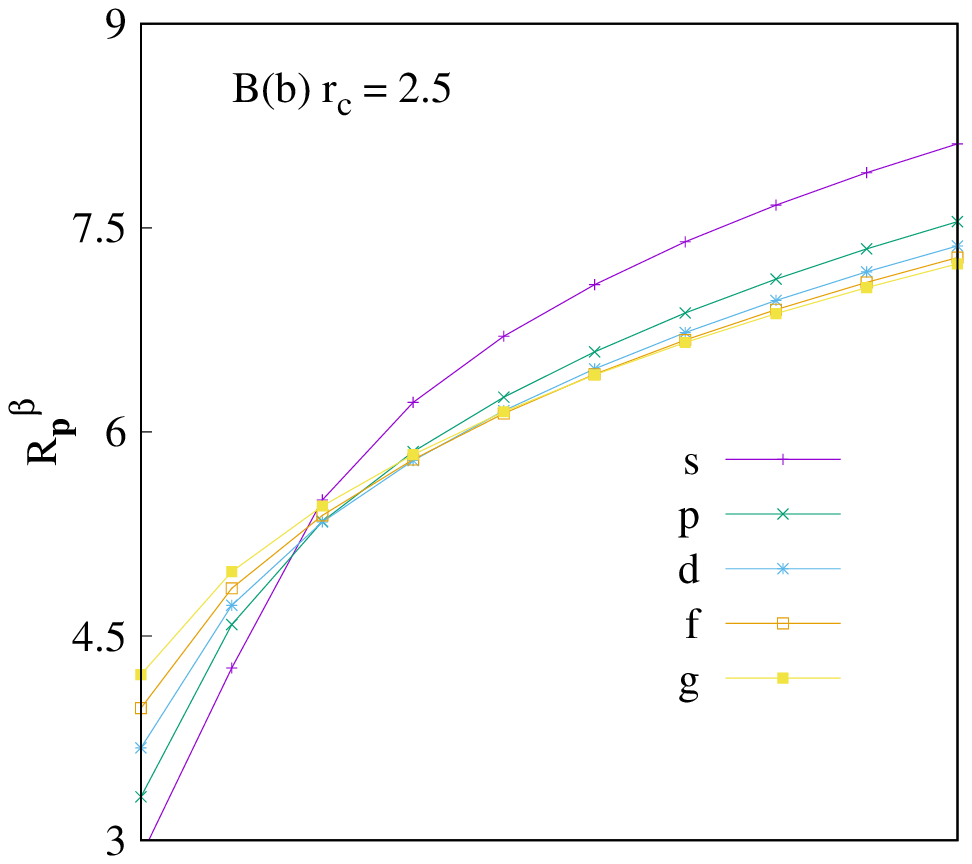}
\end{minipage}%
\begin{minipage}[c]{0.33\textwidth}\centering
\includegraphics[scale=0.48]{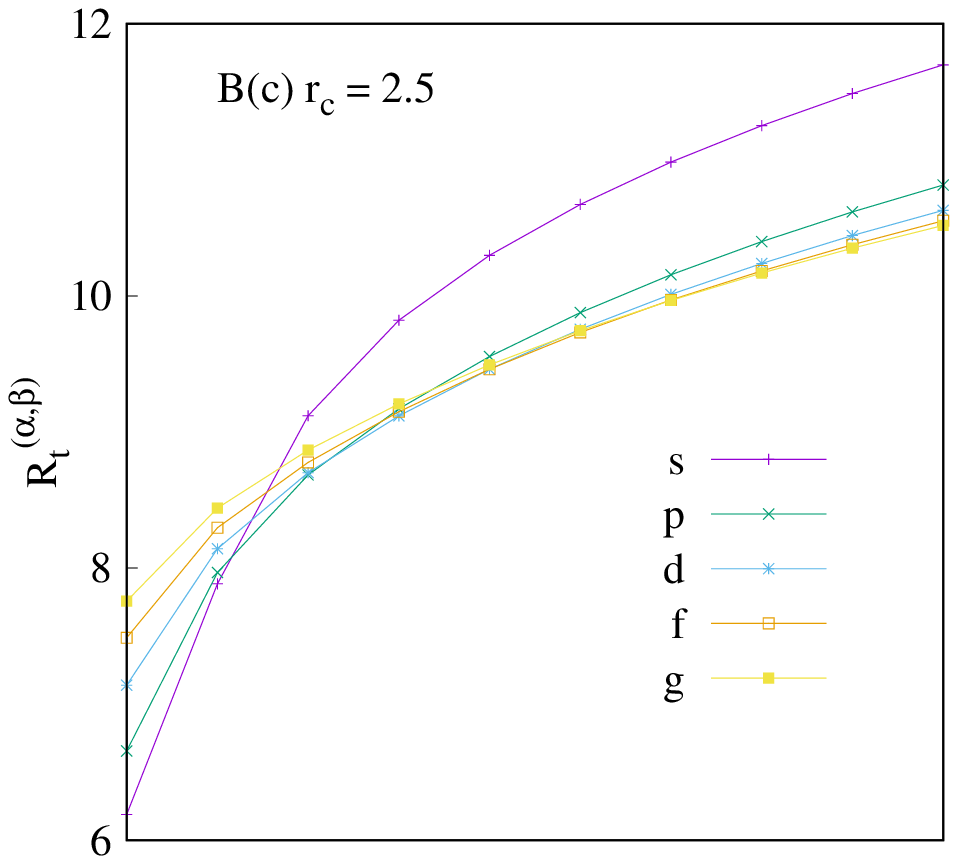}
\end{minipage}%
\hspace{0.2in}
\begin{minipage}[c]{0.33\textwidth}\centering
\includegraphics[scale=0.52]{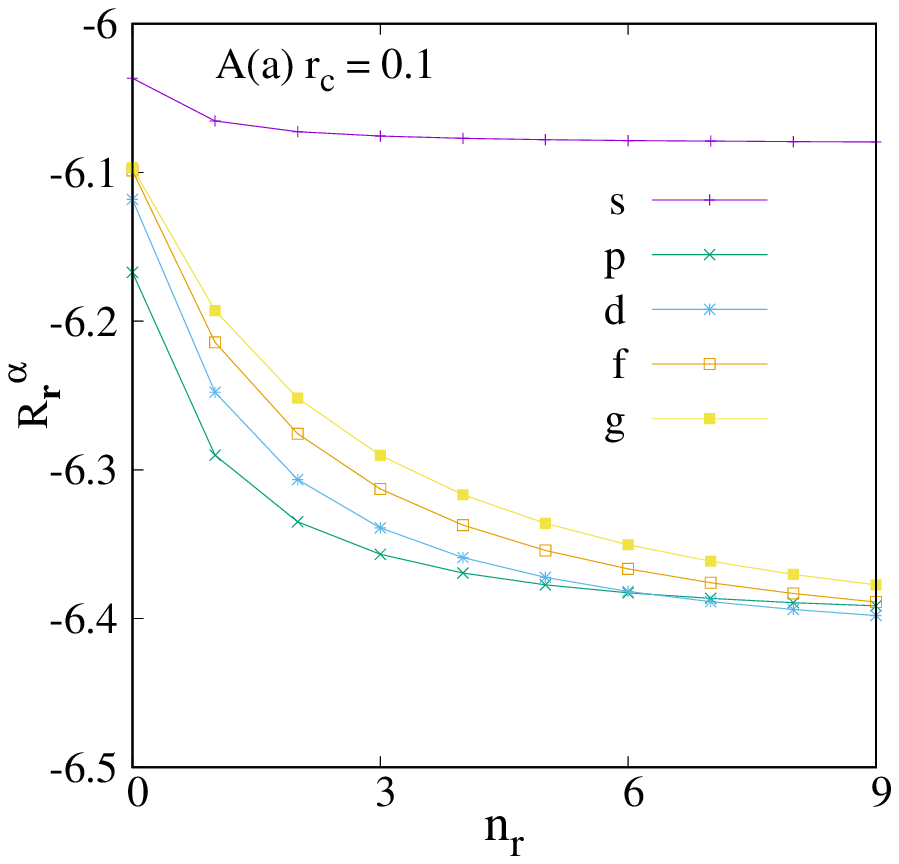}
\end{minipage}%
\begin{minipage}[c]{0.33\textwidth}\centering
\includegraphics[scale=0.52]{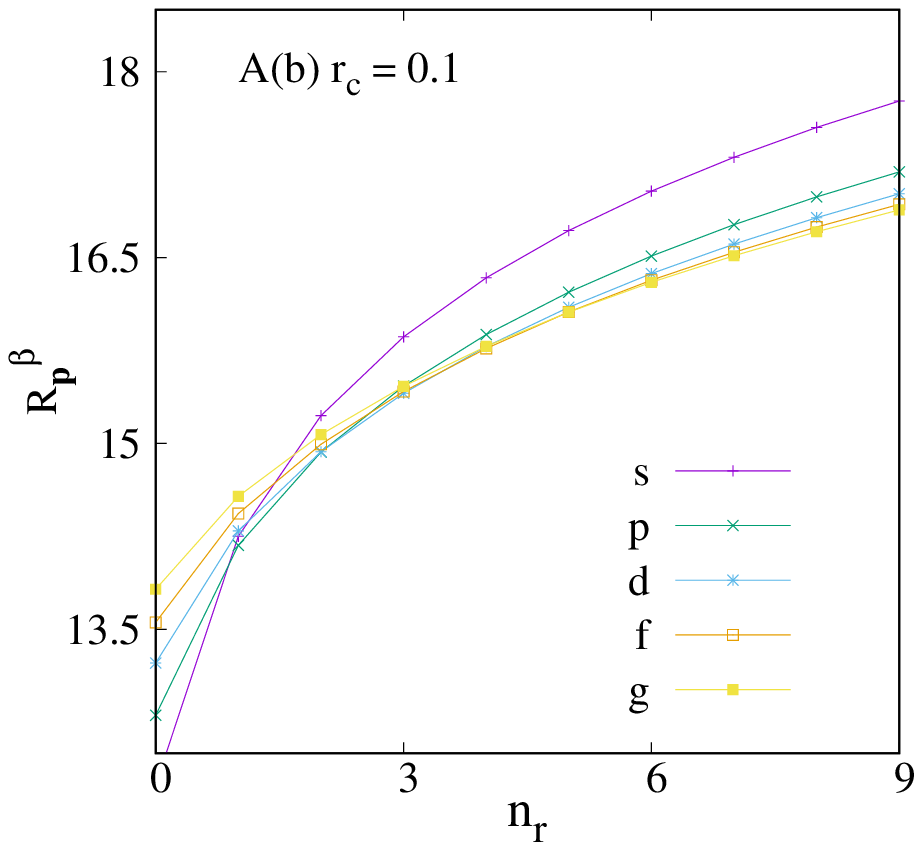}
\end{minipage}%
\begin{minipage}[c]{0.33\textwidth}\centering
\includegraphics[scale=0.52]{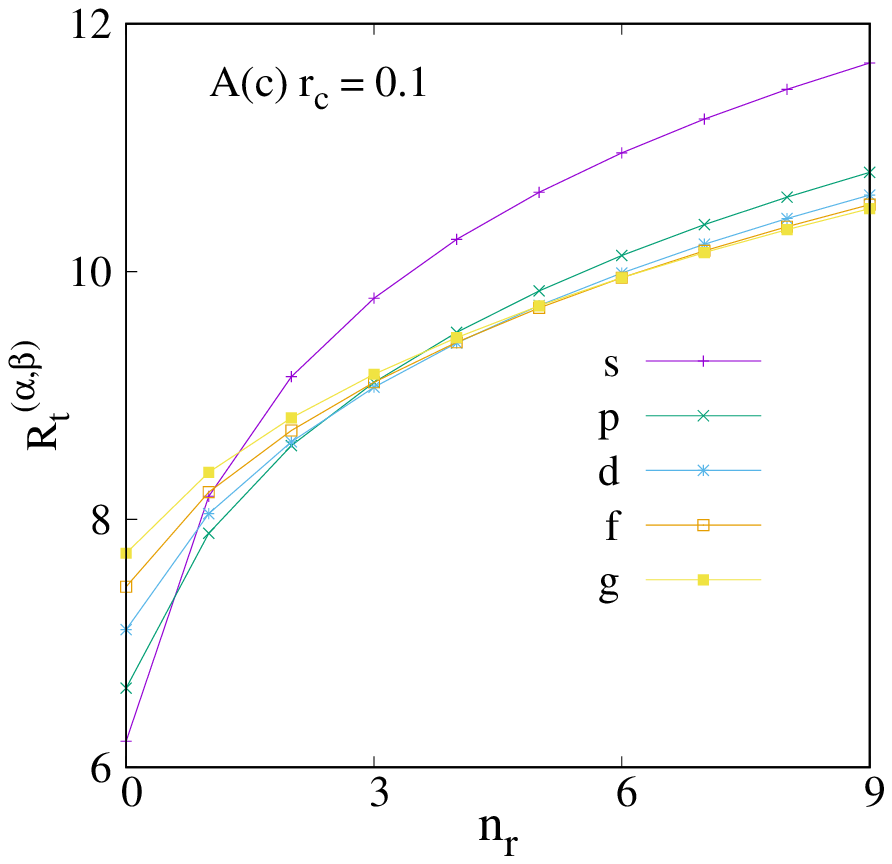}
\end{minipage}%
\caption{Plot of $R^{\alpha}_{\rvec}$ (a), $R^{\beta}_{\pvec}$ (a) and $R_{t}^{(\alpha, \beta)}$ (c) versus $n_{r}$ (at $\omega=1$) for $s,p,d,f,g$ 
states at five particular $r_{c}$'s of CHO, namely, $0.1,2.5,3,5,\infty$ in panels (A)-(E). $R_{t}^{(\alpha,\beta)}$'s 
for all these states obey the lower bound given in Eq.~(19). For more details, consult text.}
\end{figure}

To gain further knowledge, Figure~5 delineates $R^{\alpha}_{\rvec}$, $R^{\beta}_{\pvec}$ and $R_{t}^{(\alpha, \beta)}$, in left (a),
middle (b), right (c) panels, for lowest five node-less states as a function of $n_r$ (maximum of 9). Five different $r_c$'s are taken, that is, 
$0.1, 2.5, 3, 5, \infty$ in segments (A)-(E) from bottom to top. At $r_c=0.1$, for all $l$, $R^{\alpha}_{\rvec}$'s gradually falls off with $n_r$.
Albeit, it provides highest values for $l=0$ states. But, for \emph{non-zero} $l$ states, it grows up with rise in $l$ values. Hence, it can be 
concluded that, effect of confinement is maximum for $l=1$ states and minimum for $l=0$ states. However, higher $n_r$ states experience the 
confinement in greater extent. Both, $R^{\beta}_{\pvec}$ (a) and $R_{t}^{(\alpha, \beta)}$ show reverse trend. At, low $n_r$ values both these
quantities obey the trend $1g>1f>1d>1p>1s$. This, pattern gets inversed ($1s>1p>1d>1f>1g$) at higher $n_r$. These results clearly indicates that, At lower $r_c$ 
region quantum effect gets amplified as information content decreases, whereas, total information (uncertainty) increases with $n_r$. First
column (a), interesting show appearance of maximum in $R^{\alpha}_{\rvec}$ with regular advancement of $r_c$. Position of these maxima gets 
right shifted as $r_c$ intensifies. Apparently, there exists an interplay between two conjugate aspects: (i) radial confinement (localisation)
and (ii) accumulation of nodes with $n_r$ (delocalisation). As, $r_c$ progresses, delocalisation predominates for lower $n_r$ states. Hence., with 
continuous relaxation in confinement, states having higher $n_r$ value gets delocalised. At, $r_c \rightarrow \infty$, second effect prevails, 
system behaves as IHO. In second and third columns, one can sees that, both $R^{\beta}_{\pvec}$ and $R_{t}^{(\alpha, \beta)}$ always 
accelerate with $n_r$. At $r_c \rightarrow \infty$, these two quantities approaches to respective IHO-limits.

\begingroup           
\squeezetable
\begin{table}
\caption{ $S_{\rvec}, S_{\pvec}$ and $S_{t}$ values for $1s,~2s,~1p,~2p,~1d,~2d$ orbitals in CHO at eight selected $r_c$ values. See text for detail.}
\centering
\begin{ruledtabular}
\begin{tabular}{llllllll}
$r_c$  &    $S_{\rvec}$     & $S_{\pvec}$  &  $S_{t}$  &  
$r_c$  &    $S_{\rvec}$     & $S_{\pvec}$  &  $S_{t}$  \\
\cline{1-4} \cline{5-8}
\multicolumn{4}{c}{$1s$}    &      \multicolumn{4}{c}{$2s$}    \\
\cline{1-4} \cline{5-8}
0.1      & $-$6.232173222  & 12.8494      & 6.6172       & 0.1       & $-$6.4460987687 & 14.6389      &8.1928     \\
0.2      & $-$4.152747179  & 10.7700      & 6.6172       & 0.2       & $-$4.3666534417 & 12.5595      &8.1928     \\
0.5      & $-$1.404504328  & 8.0214       & 6.6168       & 0.5       & $-$1.6176276192 & 9.8106       &8.1929     \\
1.0      & 0.6652222004    & 5.9458       & 6.6110       & 1.0       &  0.4641636149 & 7.731        &8.195      \\
2.0      & 2.5846810393    & 3.9492       & 6.5338       & 2.0       &  2.5761673628 & 5.654        &8.230      \\
5.0      & 3.2170947394    & 3.21709491   & 6.4341896494 & 5.0       &  4.1507295460 & 4.1510       &8.3017     \\
8.0      & 3.2170948239    & 3.217094821  & 6.4341896449 & 8.0       &  4.1507455435 & 4.15074      &8.30148    \\
$\infty$ & 3.2170948239    & 3.2170948239 & 6.4341896478 &$\infty$   &  4.1507455435 & 4.1507455435 &8.301491087\\
\cline{1-4} \cline{5-8}
\multicolumn{4}{c}{$1p$}    &      \multicolumn{4}{c}{$2p$}    \\
\cline{1-4} \cline{5-8}
0.1    &   $-$6.38738206 & 13.417       & 7.029       & 0.1      & $-$6.651966568   & 14.7283        &  8.0763         \\
0.2    &   $-$4.30794705 & 11.338       & 7.030       & 0.2      & $-$4.572523919   & 12.6489        &  8.0763         \\
0.5    &   $-$1.55934019 & 8.5894       & 7.0300     & 0.5      & $-$1.823606736   & 9.9000         &  8.0763         \\
1.0    &   0.51599338    & 6.5114       & 7.0273      & 1.0      &  0.256528223     & 7.82132        &  8.07784        \\
2.0    &   2.5241140868  & 4.4663       & 6.9904      & 2.0      &  2.346915762     & 5.753          &  8.099          \\
5.0    &   3.4874566574  & 3.487448     & 6.974904    & 5.0      &  4.1477548396    & 4.1483         &  8.2960         \\
8.0    &   3.4874576660  & 3.487457668  & 6.974915334 & 8.0      &  4.14786196159   & 4.147863       &  8.295724       \\
$\infty$ & 3.4874576660  & 3.4874576660 & 6.974915332 &$\infty$  &  4.14786196159   & 4.14786196159  &  8.2957239232   \\
\cline{1-4} \cline{5-8}
\multicolumn{4}{c}{$1d$}    &      \multicolumn{4}{c}{$2d$}    \\
\cline{1-4} \cline{5-8}
0.1    &  $-$6.3553068427 & 14.0035      & 7.6481       & 0.1     & $-$6.5939939435     & 15.0676        & 8.4736       \\
0.2    &  $-$4.2758683878 & 11.9241      & 7.6482       & 0.2     & $-$4.5145520348     & 12.988         & 8.473        \\
0.5    &  $-$1.527121568  & 9.1753       & 7.6481       & 0.5     & $-$1.7656649279     & 10.2393        & 8.4736       \\
1.0    &  0.5503764295    & 7.0964       & 7.6467       & 1.0     &  0.3140078818     & 8.1605         & 8.4745       \\
2.0    &  2.5952812036    & 5.0319       & 7.6271       & 2.0     &  2.3974788669     & 6.091          & 8.488        \\
5.0    &  3.8426303929    & 3.84259239   & 7.68522278   & 5.0     &  4.3885945973     & 4.3909         & 8.7794       \\
8.0    &  3.8426381378    & 3.84263813   & 7.68527626   & 8.0     &  4.389113529281   & 4.38910        & 8.77821      \\
$\infty$& 3.8426381378    & 3.8426381378 & 7.6852762756 &$\infty$ &  4.389113529281   & 4.389113529281 & 8.7782270586 \\
\end{tabular}
\end{ruledtabular}
\end{table}
\endgroup

Now, we move to S in Table~V, where, $S_{\rvec}, S_{\pvec}$ and $S_{t}$ are presented for $1s,~2s~,1p~,2p~,1d~,2d$ states of CHO
at same set of $r_c$ as in Table~IV. Once again, no reference work exists for these, which could be compared. Like 
$R^{\alpha}_{\rvec}$, $S_{\rvec}$ also yield $(-)$ve values for all these six states at very low $r_c$ and then continuously 
progress, until reaching the borderline IHO values. However, like $R^{\beta}_{\pvec}$, $S_{\pvec}$ offers an opposite nature of 
$S_{\rvec}$, ($R^{\alpha}_{\rvec}$); from an initial $(+)$ve, consistently reduces to reach IHO. $S_{t}$'s for both $1s, 1p$ states 
decrease to reach IHO values. But, for $1d$ state it falls off, reaches a minimum and finally converges to IHO. 

\begin{figure}                         
\begin{minipage}[c]{0.32\textwidth}\centering
\includegraphics[scale=0.55]{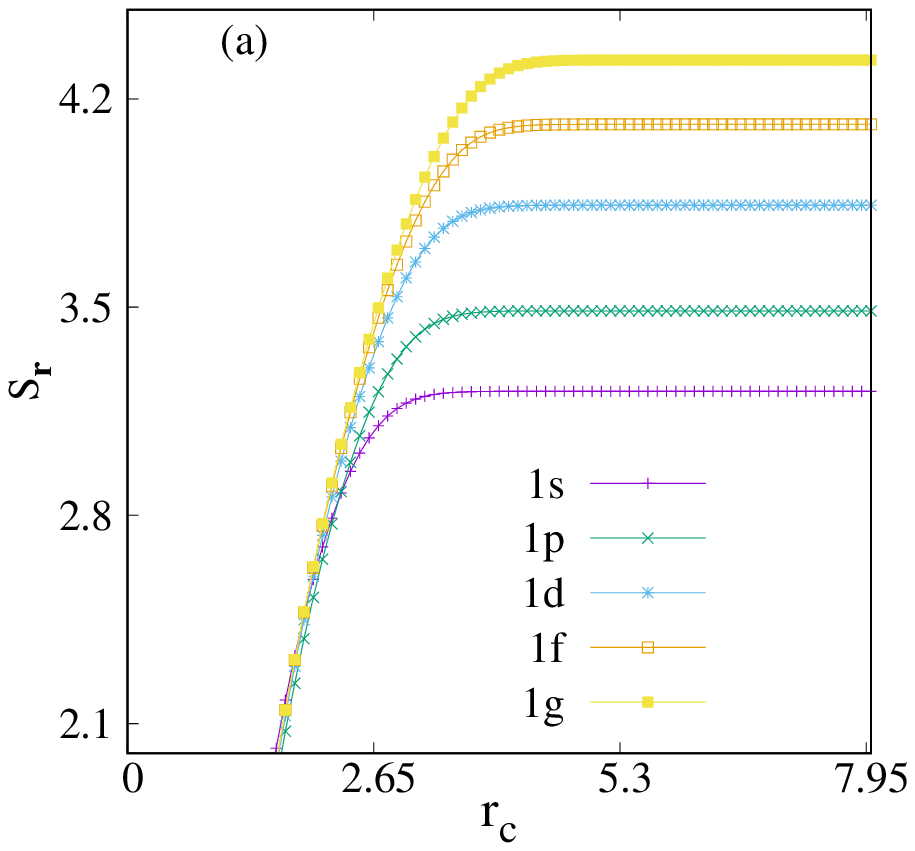}
\end{minipage}%
\hspace{0.02in}
\begin{minipage}[c]{0.32\textwidth}\centering
\includegraphics[scale=0.55]{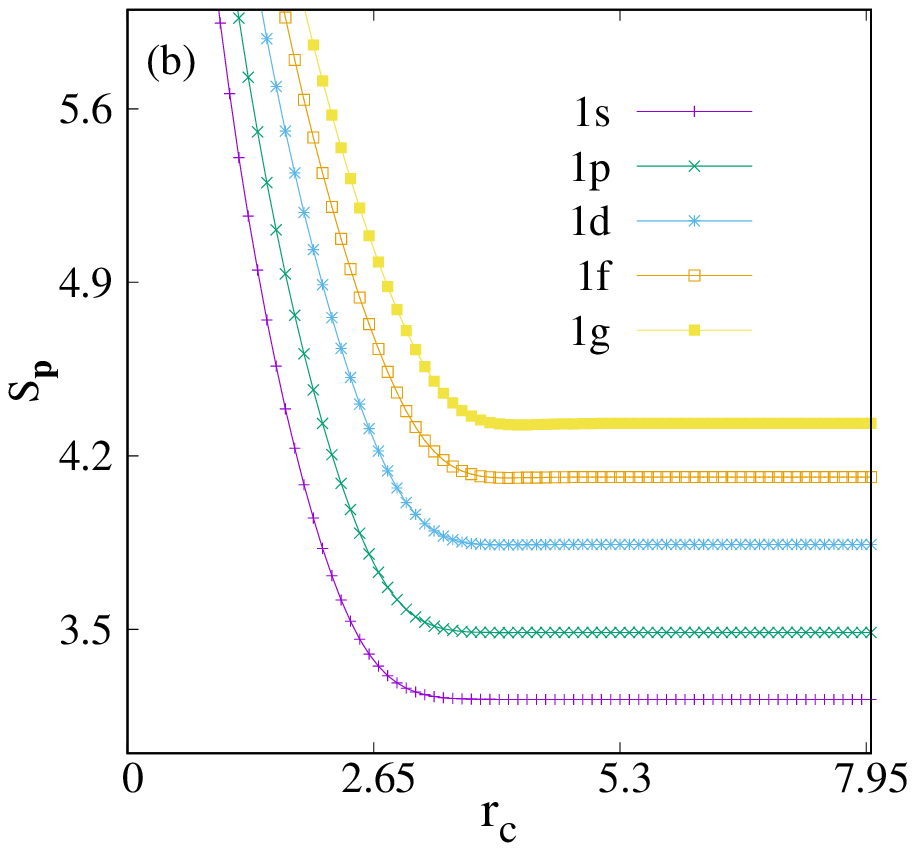}
\end{minipage}%
\hspace{0.02in}
\begin{minipage}[c]{0.32\textwidth}\centering
\includegraphics[scale=0.55]{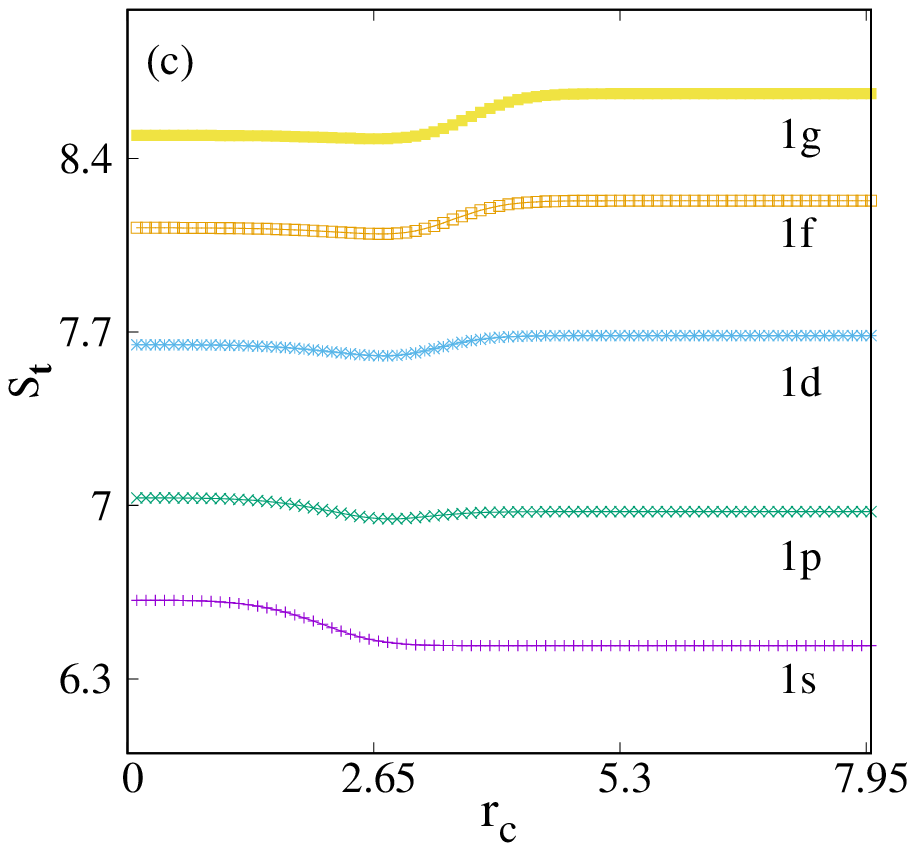}
\end{minipage}%
\caption{Plots of $S_{\rvec}$, $S_{\pvec}$, $S_{t}$ against $r_c$ for first five circular states of CHO in panels (a), (b), (c) respectively. 
See text for details.}
\end{figure}

Next, Figure~6 indicates behavioral patterns of $S_{\rvec}$, $S_{\pvec}$, $S_{t}$ with $r_c$ in segments (a)-(c), for same five states Figure~6. 
It is important to point out that, panels (a),(b),(c) of both Figures 4 and 6 deliver similar style. For all these states $S_{\rvec}$'s mount up 
with $r_c$ and finally convene to corresponding $r$-space IHO, while $S_{\pvec}$'s decrement before attaining that. Panel (c) shows that, 
for $1s,1p$ states $S_{t}$'s decrease with $r_c$ and finally merge to IHO, while for $1d,1f,1g$ states, there appears a minimum before reaching 
the limiting IHO values.     

\begin{figure}                                            
\begin{minipage}[c]{0.33\textwidth}\centering
\includegraphics[scale=0.48]{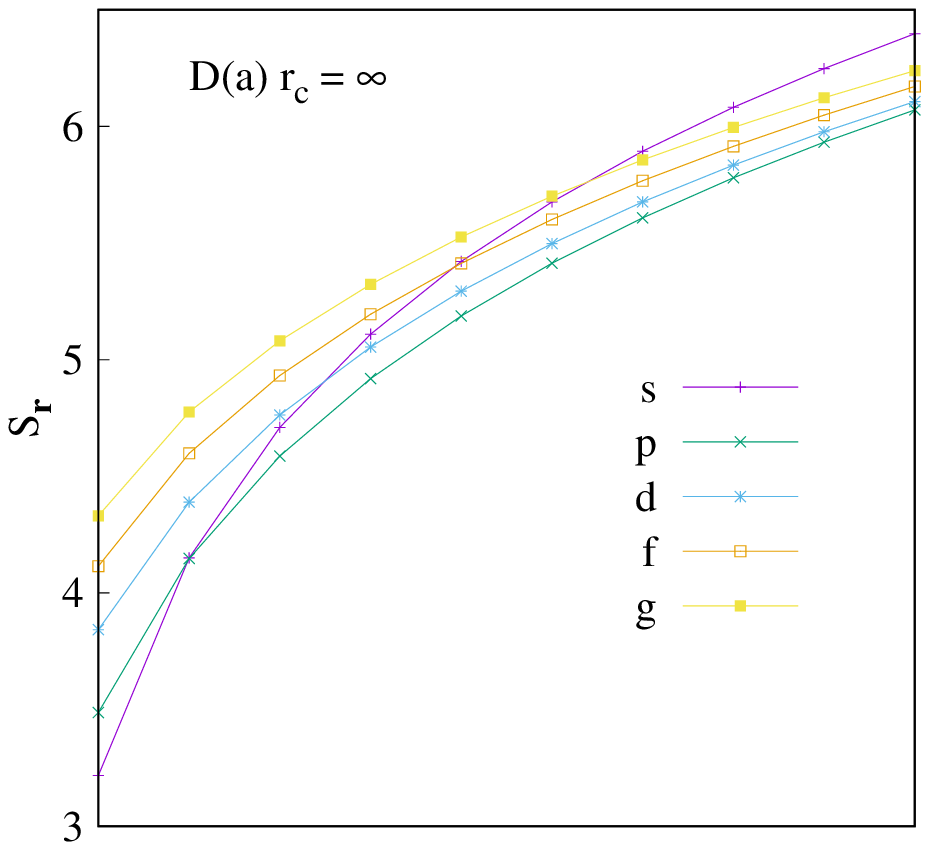}
\end{minipage}%
\begin{minipage}[c]{0.33\textwidth}\centering
\includegraphics[scale=0.48]{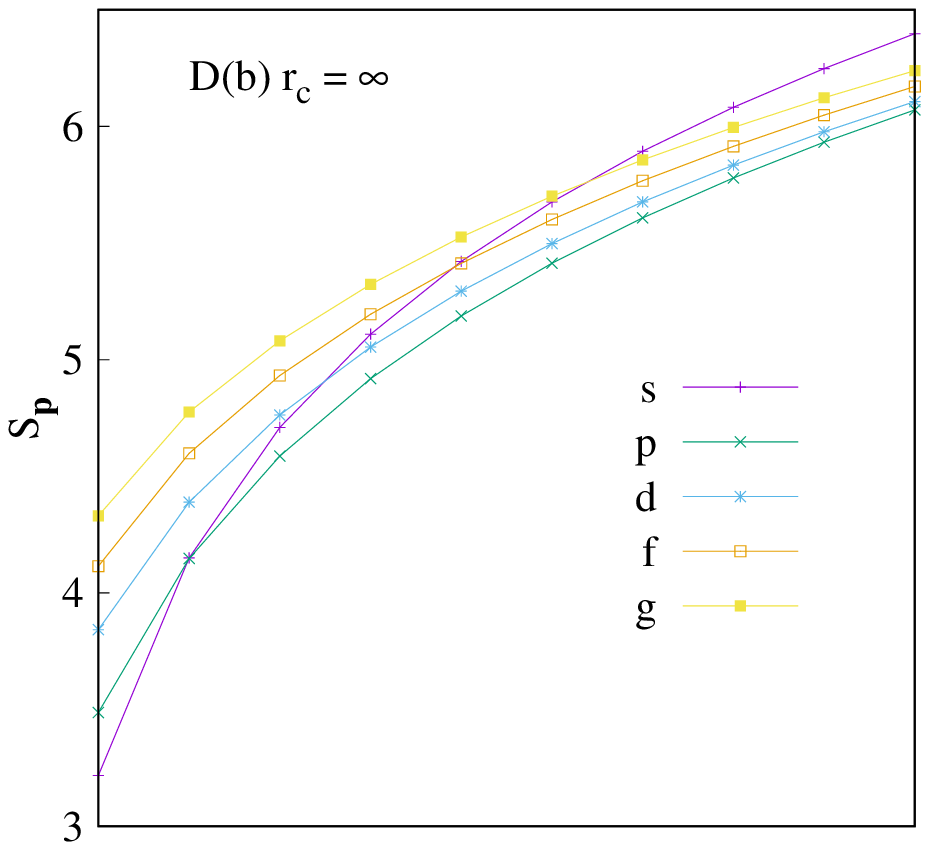}
\end{minipage}%
\begin{minipage}[c]{0.33\textwidth}\centering
\includegraphics[scale=0.48]{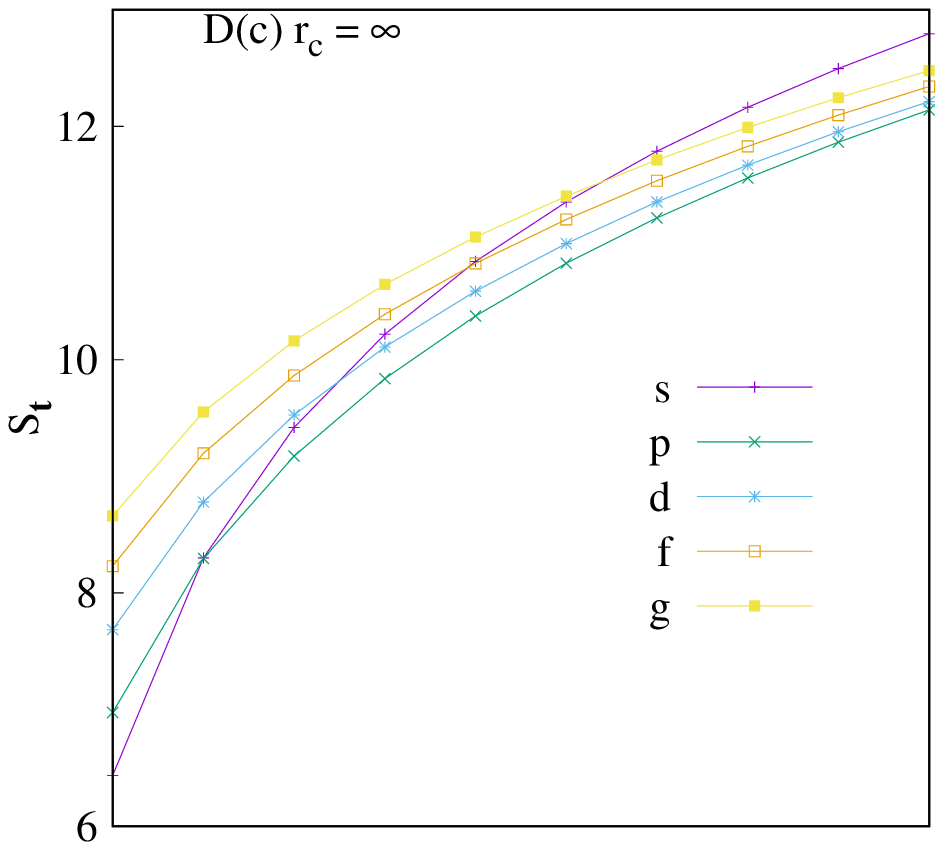}
\end{minipage}%
\hspace{0.2in}
\begin{minipage}[c]{0.33\textwidth}\centering
\includegraphics[scale=0.48]{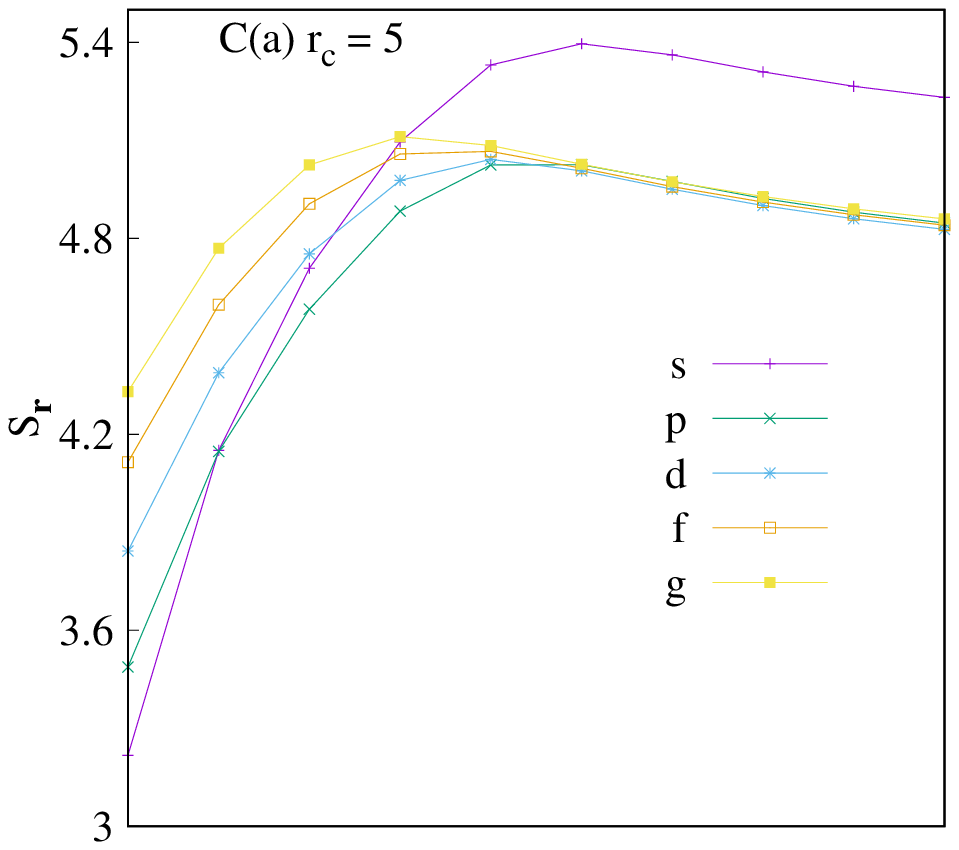}
\end{minipage}
\begin{minipage}[c]{0.33\textwidth}\centering
\includegraphics[scale=0.48]{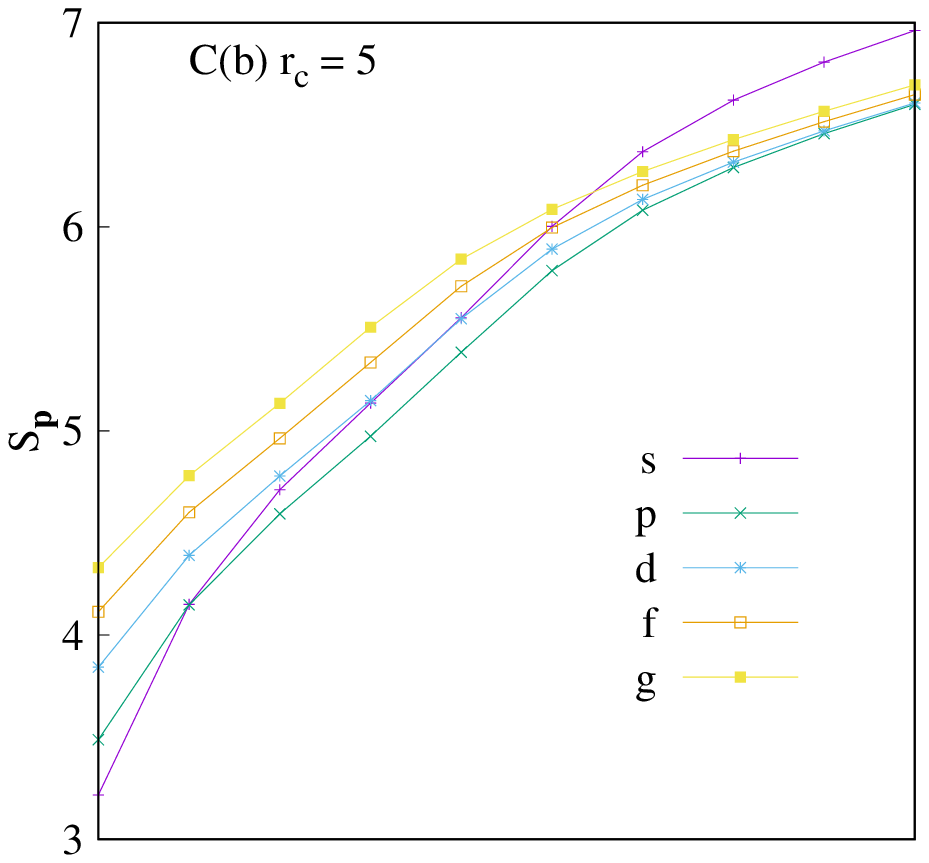}
\end{minipage}%
\begin{minipage}[c]{0.33\textwidth}\centering
\includegraphics[scale=0.48]{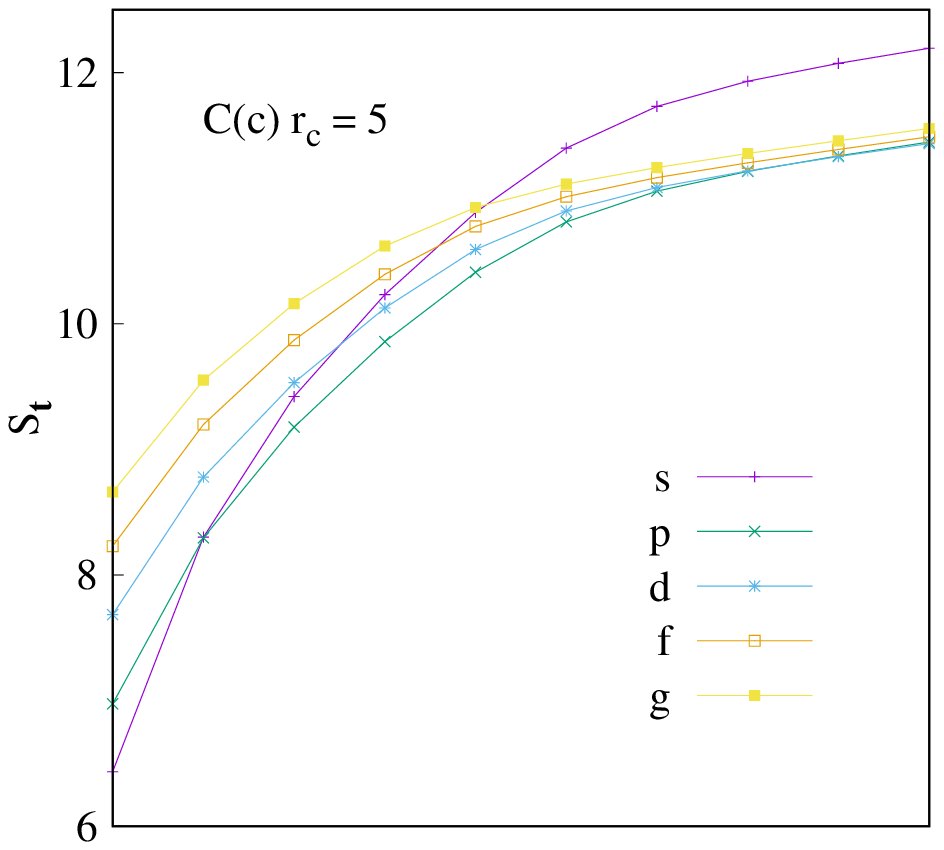}
\end{minipage}%
\hspace{0.2in}
\begin{minipage}[c]{0.33\textwidth}\centering
\includegraphics[scale=0.48]{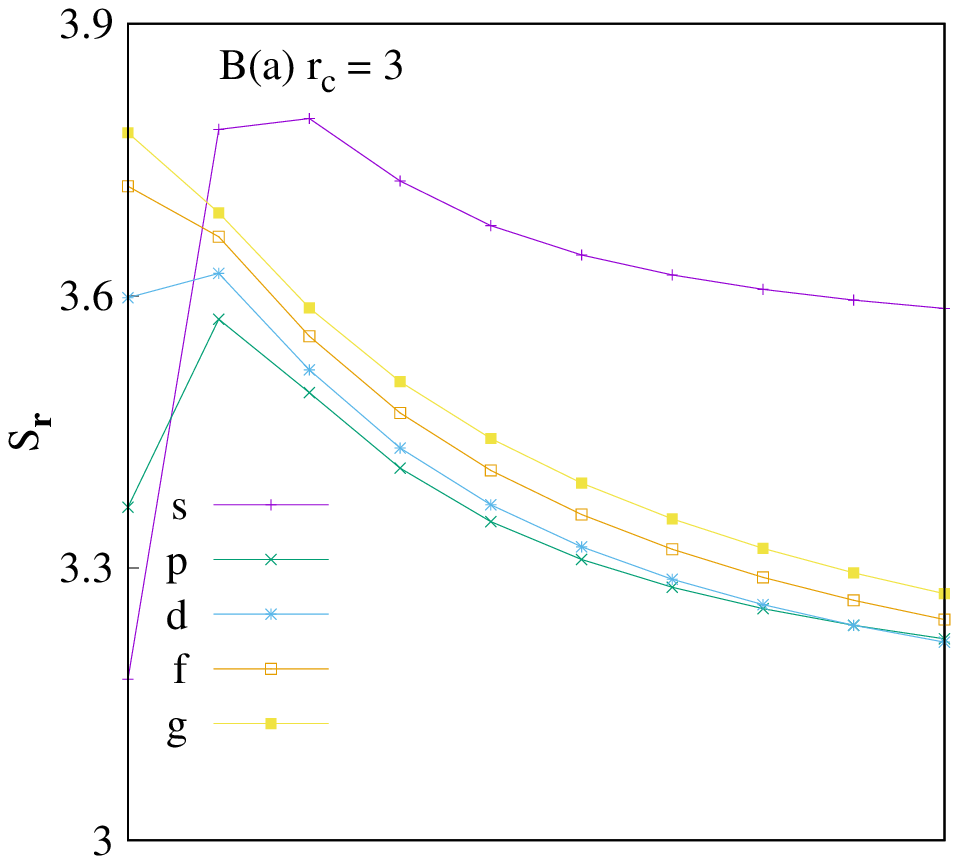}
\end{minipage}%
\begin{minipage}[c]{0.33\textwidth}\centering
\includegraphics[scale=0.48]{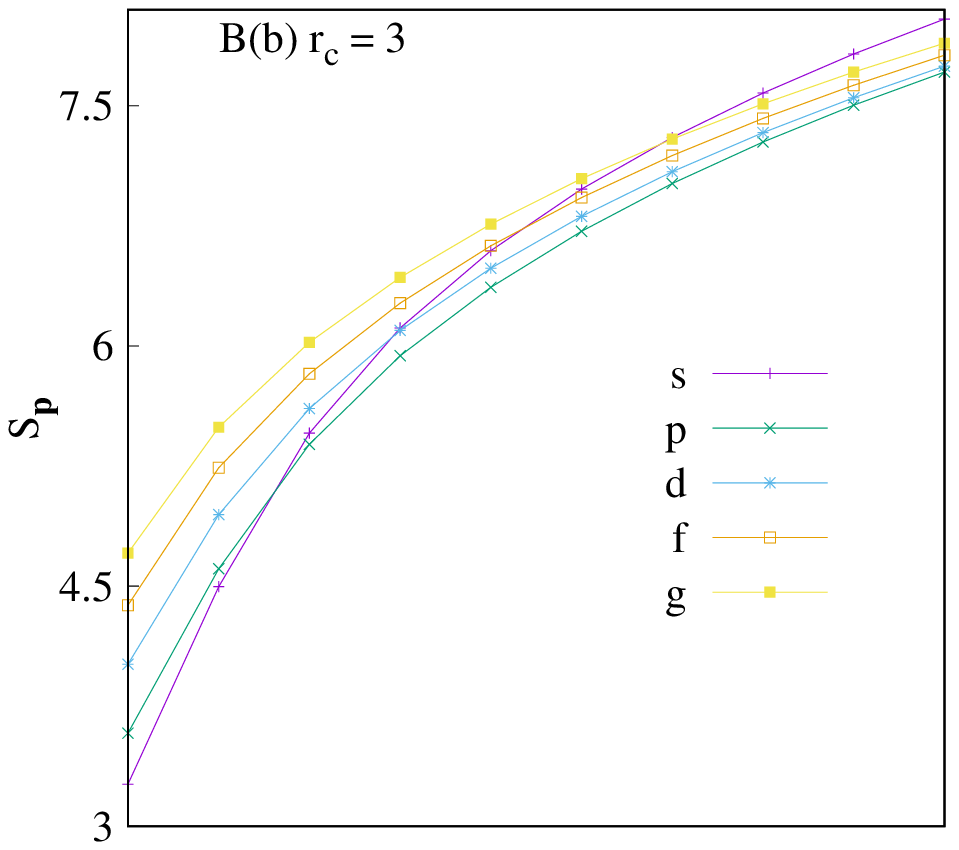}
\end{minipage}%
\begin{minipage}[c]{0.33\textwidth}\centering
\includegraphics[scale=0.48]{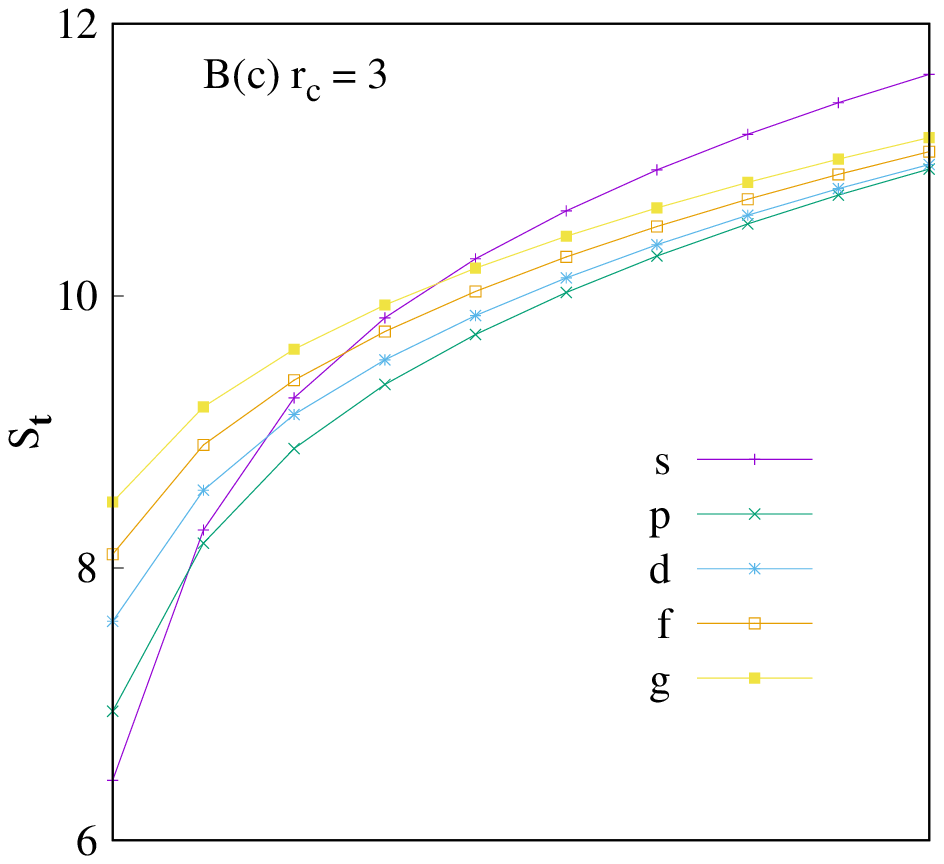}
\end{minipage}%
\hspace{0.2in}
\begin{minipage}[c]{0.33\textwidth}\centering
\includegraphics[scale=0.48]{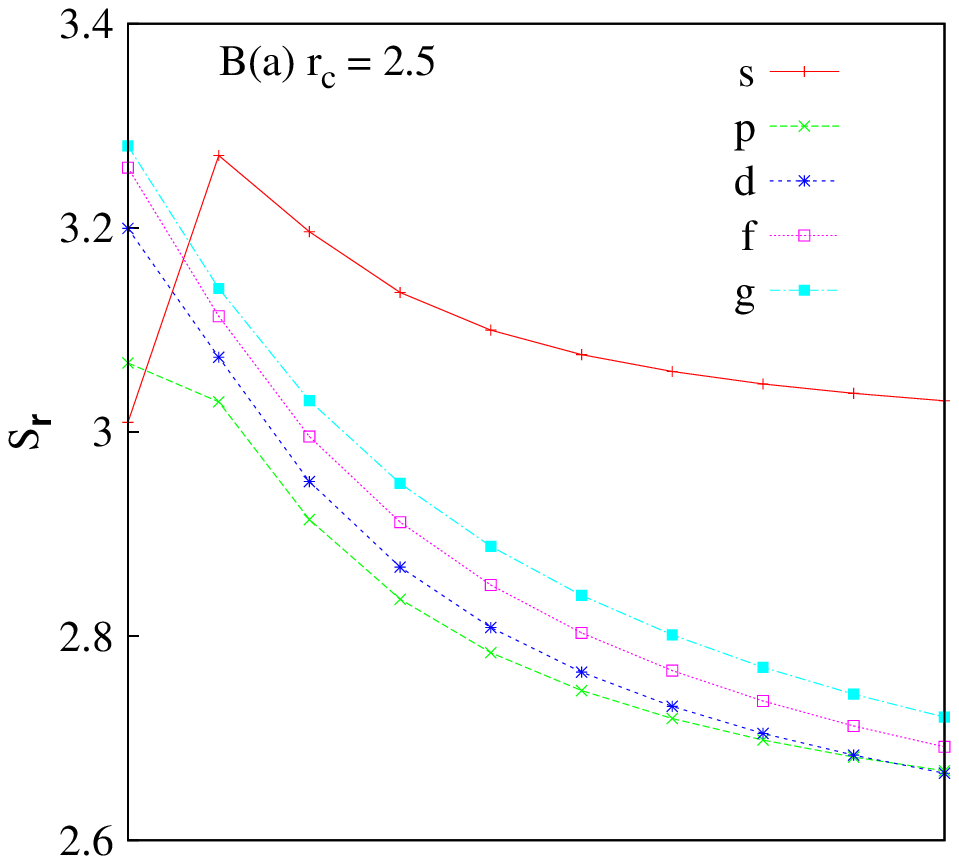}
\end{minipage}%
\begin{minipage}[c]{0.33\textwidth}\centering
\includegraphics[scale=0.48]{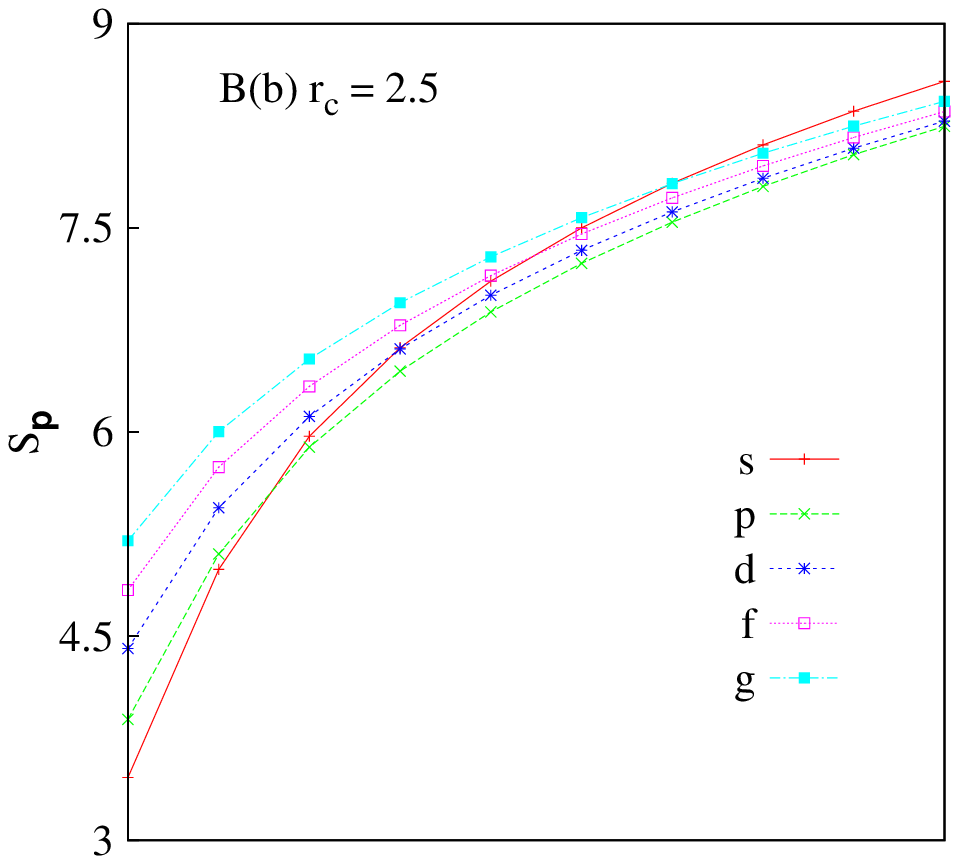}
\end{minipage}%
\begin{minipage}[c]{0.33\textwidth}\centering
\includegraphics[scale=0.48]{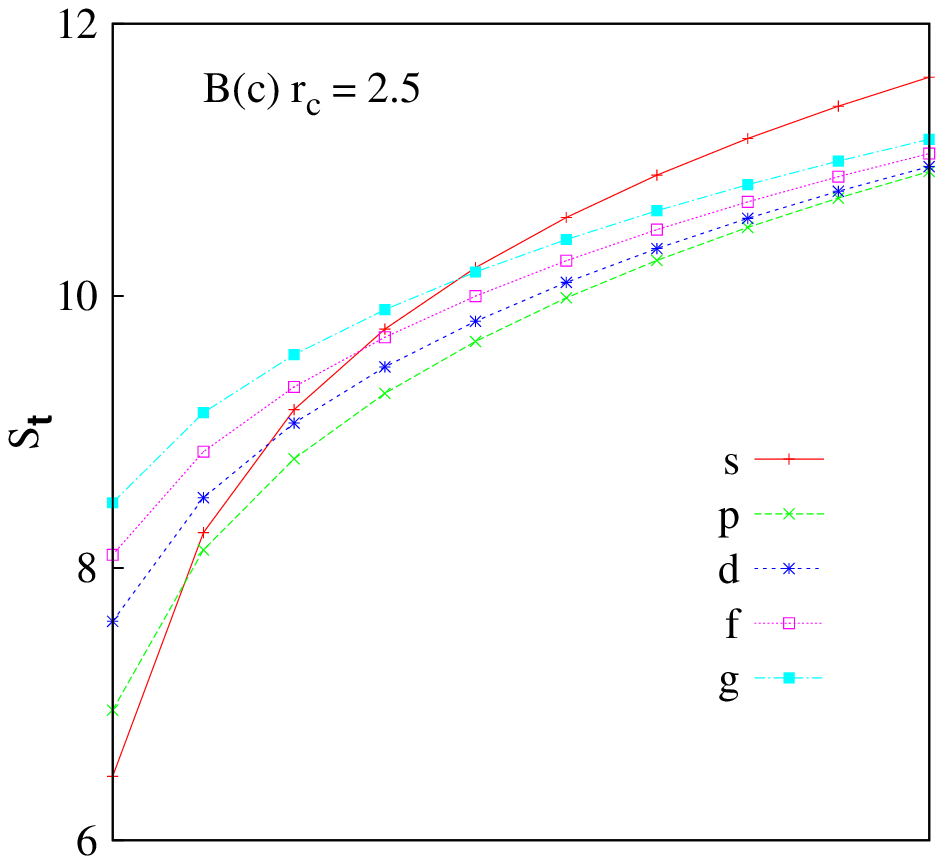}
\end{minipage}%
\hspace{0.2in}
\begin{minipage}[c]{0.33\textwidth}\centering
\includegraphics[scale=0.52]{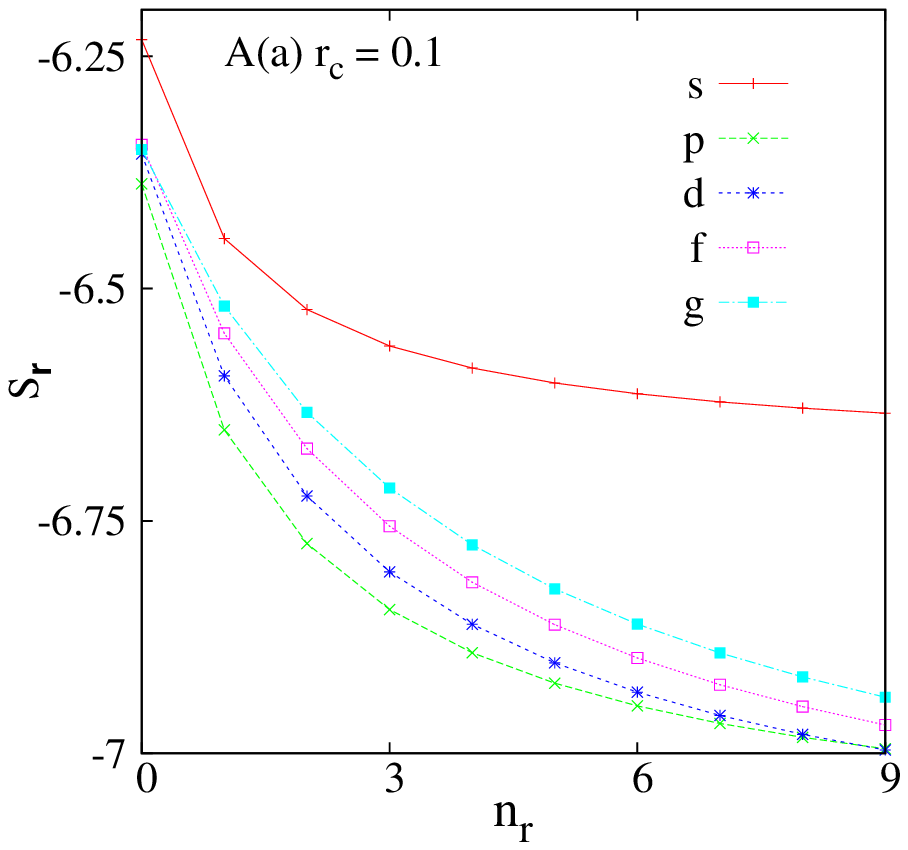}
\end{minipage}%
\begin{minipage}[c]{0.33\textwidth}\centering
\includegraphics[scale=0.52]{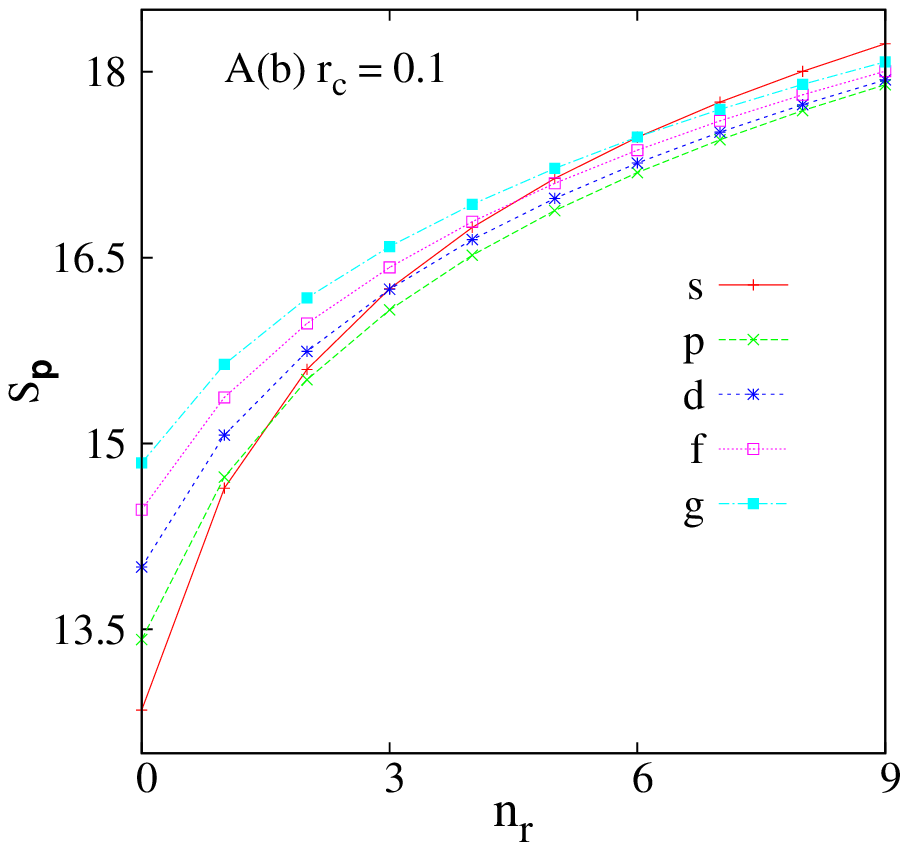}
\end{minipage}%
\begin{minipage}[c]{0.33\textwidth}\centering
\includegraphics[scale=0.52]{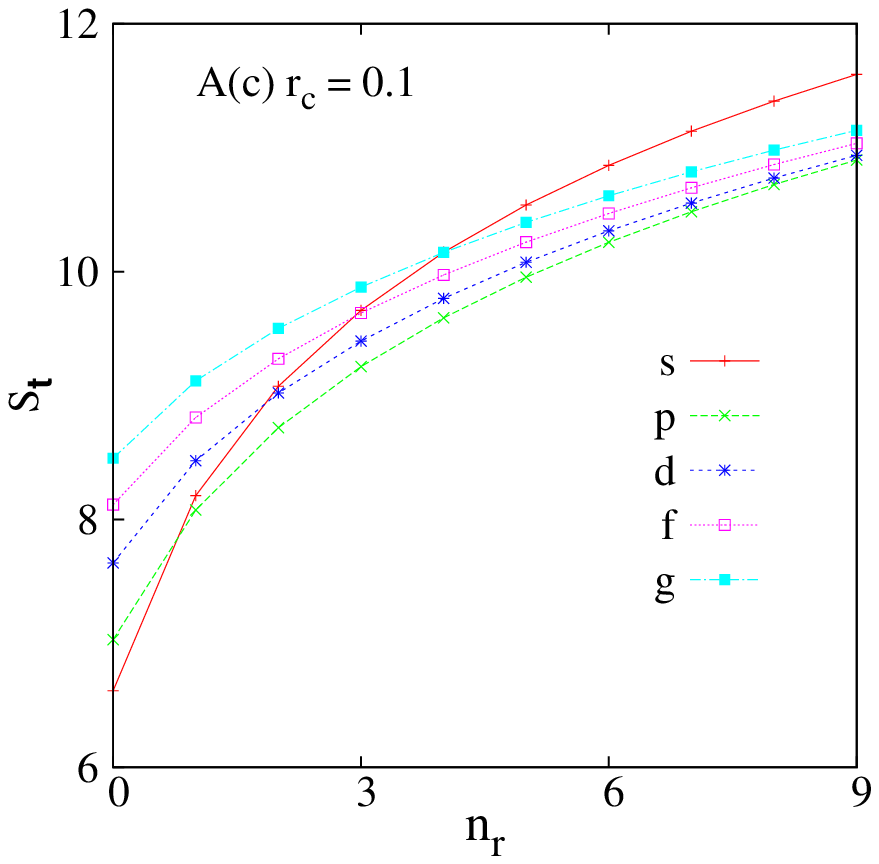}
\end{minipage}%
\caption{Plot of $S_{\rvec}$ (a), $S_{\pvec}$ (a) and $S_{t}$ (c) versus $n_{r}$  (at $\omega=1$) for $s,p,d,f,g$ 
states at five particular $r_{c}$'s of CHO, namely, $0.1,2.5,3,5,\infty$ in panels (A)-(E). $S_{t}$'s 
for all these states obey the lower bound given in Eq.~(19). For more details, consult text.}
\end{figure}
                    
In Figure~7, $S_{\rvec}$ (a), $S_{\pvec}$ (b), $S_{t}$ (c) of $l=0-4$ states are plotted against $n_r$ at same five $r_c$ of Figure~5, in panels (A)-(E)
from bottom to top. Again, the graphs in Figure~7 imprint analogous shape and propensity to that of Figure~5. Thus in coherence with $R^{\alpha}_{\rvec}$
at $r_c=0.1$, for five $l$, $S_{\rvec}$ gets lowered in A(a), while $S_{\pvec}$'s and $S_{t}$'s improve with $n_r$ in A(b) and A(c), respectively. This
reinforces our previous epilogue (as in R in Figure~4) that, at very low $r_c$, effect of confinement is more prevalent in high-lying states, signifying 
a intensification of quantum nature in such circumstances. As usual, like $R^{\alpha}_{\rvec}$ here also, the first column ((a)) of Figure~7 render 
the appearance of maximum in $S_{\rvec}$ plots with gradual growth of $r_c$. Their position gets shifted to right as $r_c$ improves. This observation 
indicates that, at $r_c \rightarrow \infty$ system behaves like IHO.  

\begingroup           
\squeezetable
\begin{table}
\caption{$E_{\rvec}, E_{\pvec}$ and $E_{t}$ values for $1s,~2s,~1p,~2p,~1d,~2d$ orbitals in CHO at eight selected $r_c$ values. See text for detail.}
\centering
\begin{ruledtabular}
\begin{tabular}{llllllll}
$r_c$  &    $E_{\rvec}$     & $E_{\pvec}$  &  $E_{t}$  &  
$r_c$  &    $E_{\rvec}$     & $E_{\pvec}$  &  $E_{t}$  \\
\cline{1-4} \cline{5-8}
\multicolumn{4}{c}{$1s$}    &      \multicolumn{4}{c}{$2s$}    \\
\cline{1-4} \cline{5-8}
0.1      & 672.0719164	& 0.0000039863	& 0.0026791    & 0.1       & 1453.1909702895	  &    0.00000057	  &      0.00082825 \\
0.2      & 84.01080088	& 0.0000318904	& 0.0026791    & 0.2       & 181.6485572148	  &    0.000004559     	  &      0.000828257 \\
0.5      & 5.3814002356	& 0.0004980894	& 0.002680418  & 0.5       & 11.6246913489	  &    0.000071246	  &      0.000828213 \\
1.0      & 0.6818097823	& 0.0039601229	& 0.0027000505 & 1.0       & 1.4515093698	  &    0.0005701466	  &      0.0008275732 \\
2.0      & 0.1056762183	& 0.0284605182	& 0.0030075999 & 2.0       & 0.1784022679	  &    0.004660181	  &      0.000831386 \\
5.0      & 0.0634936361	& 0.0634934018	& 0.004031427  & 5.0       & 0.0406758398	  &    0.0406482969	  &      0.0016534036 \\
8.0      & 0.0634936347	& 0.0634936349	& 0.0040314417 & 8.0       & 0.040670749	  &    0.0406756097	  &      0.0016543075 \\
$\infty$ & 0.0634936347	& 0.0634936347	& 0.0040314416 &$\infty$   & 0.0406756097	  &    0.0406756097	  &      0.0016545052 \\
\cline{1-4} \cline{5-8}
\multicolumn{4}{c}{$1p$}    &      \multicolumn{4}{c}{$2p$}    \\
\cline{1-4} \cline{5-8}
0.1      & 803.22700816   &	0.0000022775	& 0.0018293897 & 0.1      & 1454.974575234	& 0.0000005896	& 0.000857877   \\
0.2      & 100.40423515   &	0.0000182203	& 0.0018294011 & 0.2      & 181.8717931823	& 0.000004717	& 0.0008578847   \\
0.5      & 6.4281046025   &	0.0002846331	& 0.0018296512 & 0.5      & 11.6397197874	& 0.0000736934	& 0.0008577713   \\
1.0      & 0.8078468658   &	0.0022696156	& 0.0018335018 & 1.0      & 1.454810759	        & 0.0005883981	& 0.0008560079   \\
2.0      & 0.1107979053   &	0.0171292823	& 0.0018978886 & 2.0      & 0.1813047648	& 0.004569953	& 0.0008285543   \\
5.0      & 0.0476202385   &	0.0476216277	& 0.0022677533 & 5.0      & 0.0324128865	& 0.0323758305	& 0.0010493941   \\
8.0      & 0.047620224    &	0.0476202241	& 0.0022676857 & 8.0      & 0.032411515	        & 0.0324115148	& 0.0010505063   \\
$\infty$ & 0.047620224    &	0.047620224	& 0.0022676857 &$\infty$  & 0.032411515	        & 0.032411515	& 0.0010505063   \\
\cline{1-4} \cline{5-8}
\multicolumn{4}{c}{$1d$}    &      \multicolumn{4}{c}{$2d$}    \\
\cline{1-4} \cline{5-8}
0.1    &  851.25726418	& 0.000001378	& 0.0011730159 & 0.1     & 1368.86082394   & 0.0000004746	& 0.0006496614\\
0.2    &  106.40757650	& 0.0000110238	& 0.0011730187 & 0.2     & 171.107627763   & 0.0000037968	& 0.0006496761\\
0.5    &  6.8111728555	& 0.0001722254	& 0.0011730568 & 0.5     & 10.9509523492   & 0.0000593197	& 0.0006496078\\
1.0    &  0.8535077417	& 0.0013750845	& 0.0011736453 & 1.0     & 1.3689869817	   & 0.0004737266	& 0.0006485256 \\
2.0    &  0.1114898571	& 0.0106198478	& 0.0011840053 & 2.0     & 0.1712007721	   & 0.0036813808	& 0.0006302552 \\
5.0    &  0.0357152613	& 0.0357193674	& 0.0012757265 & 5.0     & 0.0249812774	   & 0.0249259554	& 0.0006226822 \\
8.0    &  0.0357151695	& 0.0357151695	& 0.0012755733 & 8.0     & 0.0249755634	   & 0.0249755633	& 0.0006237788 \\
$\infty$& 0.0357151695	& 0.0357151695	& 0.0012755733 &$\infty$ & 0.0249755634	   & 0.0249755634	& 0.0006237788 \\
\end{tabular}
\end{ruledtabular}
\end{table}

At this stage we move on to explore the last measure of this study, that is, E in Table VI. A cross-section of $E_{\rvec}, E_{\pvec}$ 
and $E_{t}$ for $1s,2s,1p,2p,1d,2d$ states of CHO (same set of $r_c$ values used for previous measures) is offered. One notices that, 
$E_{\rvec}$ decreases (as opposed to $R^{\alpha}_{\rvec}, S_{\rvec}$) while $E_{\pvec}$ accelerates (as opposed to $R^{\beta}_{\pvec}, S_{\pvec}$)
with progress in $r_c$. However, behaviour of $E_{t}$ with $r_c$ varies from state to state.
For $1s,1p,1d$ states it advances with $r_c$. But, for $2s$ and $2p$ states it passes through a minimum and in case of $2d$ state it always 
falls off with increase in boundary.       

\begin{figure}                         
\begin{minipage}[c]{0.32\textwidth}\centering
\includegraphics[scale=0.55]{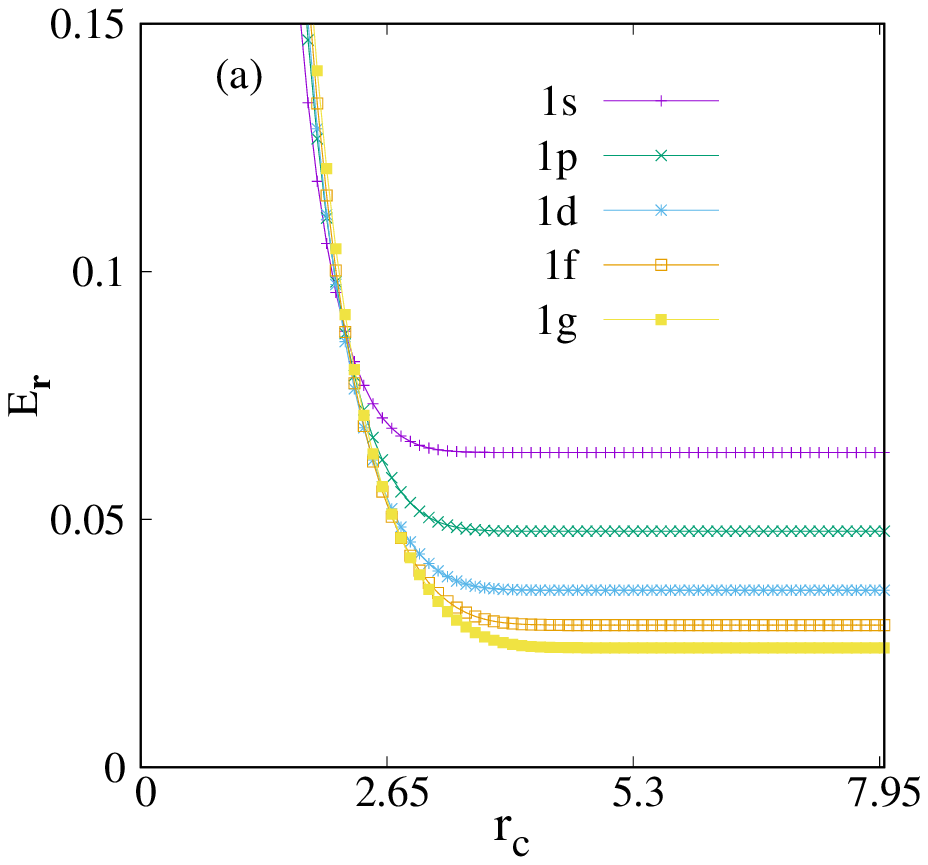}
\end{minipage}%
\hspace{0.02in}
\begin{minipage}[c]{0.32\textwidth}\centering
\includegraphics[scale=0.55]{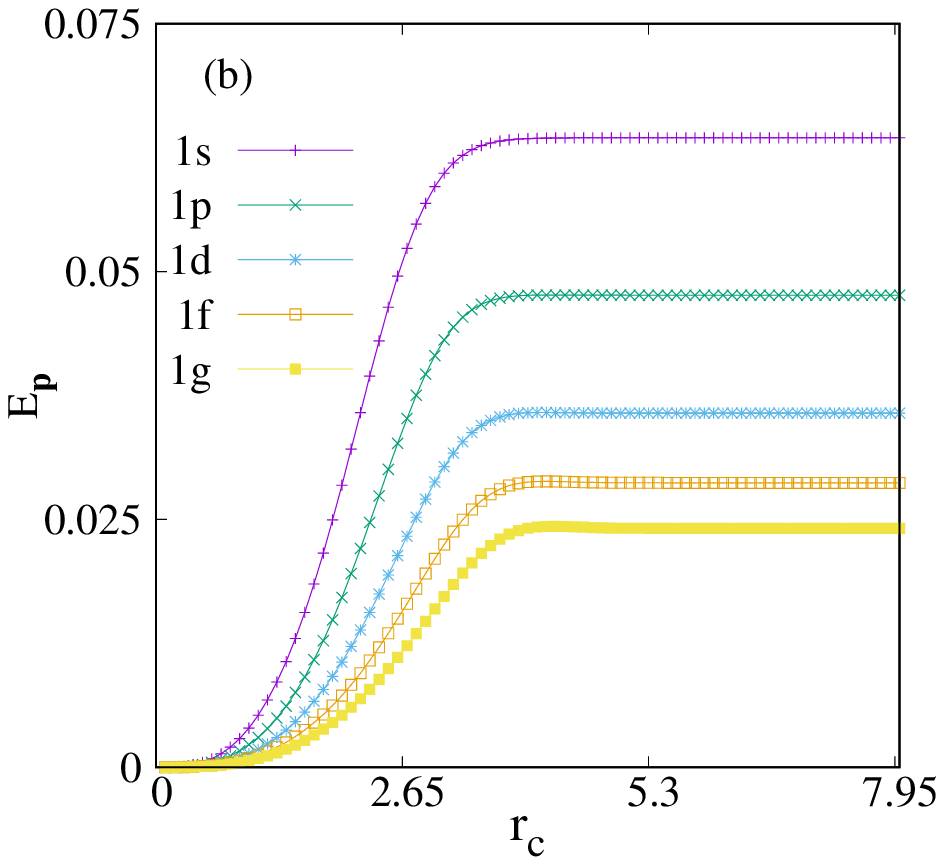}
\end{minipage}%
\hspace{0.02in}
\begin{minipage}[c]{0.32\textwidth}\centering
\includegraphics[scale=0.55]{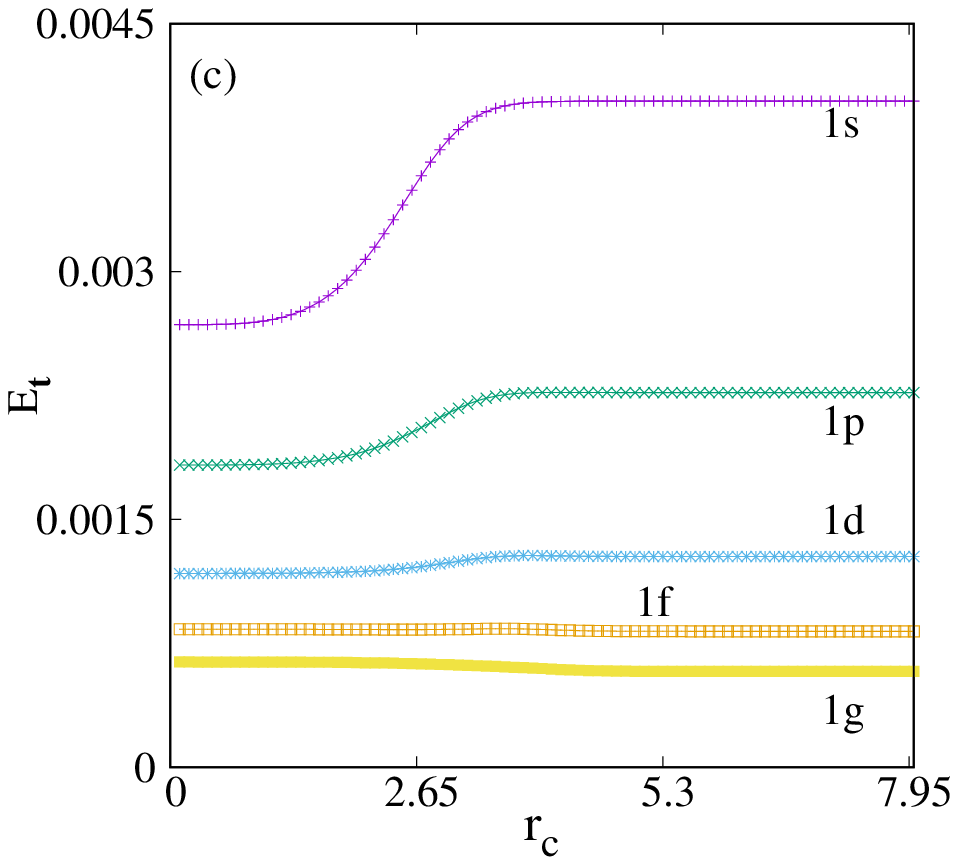}
\end{minipage}%
\caption{Plots of $E_{\rvec}$, $E_{\pvec}$, $E_{t}$ against $r_c$ for first five circular states of CHO in panels (a), (b), (c) respectively. 
See text for details.}
\end{figure}

These changes in $E_{\rvec}, E_{\pvec}$ and $E_{t}$ with $r_c$ are graphically displayed in Figure~8, in left (a), middle (b) and right (c) panels
for first five circular states. One notices that, 
$E_{t}$ for $1s,1p,1d$ states increases with $r_c$,  while for $1f,1g$ states it decreases. Interestingly, at large $r_c$, both  $E_{\rvec}$
and $E_{\pvec}$ decrease with increase in $l$. 

In Figure~9,  $E_{\rvec}, E_{\pvec}$ and $E_{t}$ are portrayed (in columns (a),(b),(c)) for $l=0-4$ states as functions of $n_r$ at five different
$r_c$ values (in segments A-E). At the lowest $r_c$ considered, $E_{\rvec}$ progresses with $n_r$. However, the first column (a) suggests that, a minimum 
appears in $E_{\rvec}$ graphs as $r_c$ is extended. Also the positions of these minima gets right shifted with increment in $r_c$. On the contrary, 
for all concern $r_c$ values, both $E_{\pvec}$, $E_{t}$ diminish with $n_r$.    

\begin{figure}                                            
\begin{minipage}[c]{0.33\textwidth}\centering
\includegraphics[scale=0.48]{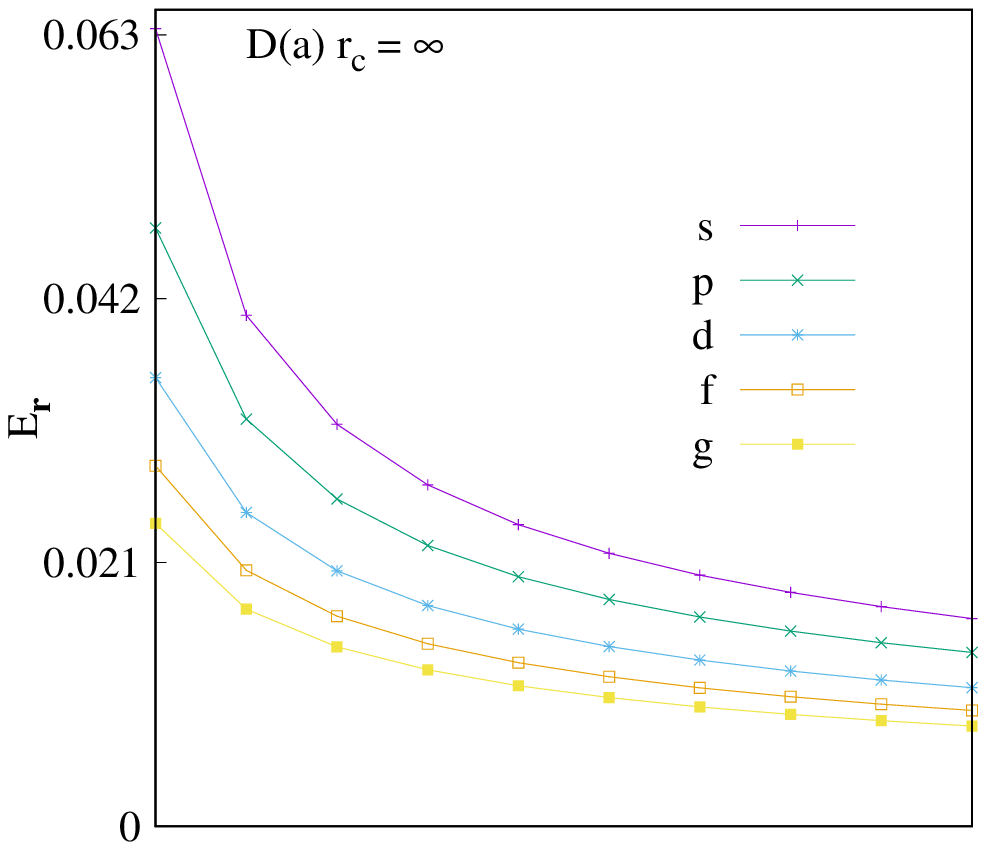}
\end{minipage}%
\begin{minipage}[c]{0.33\textwidth}\centering
\includegraphics[scale=0.48]{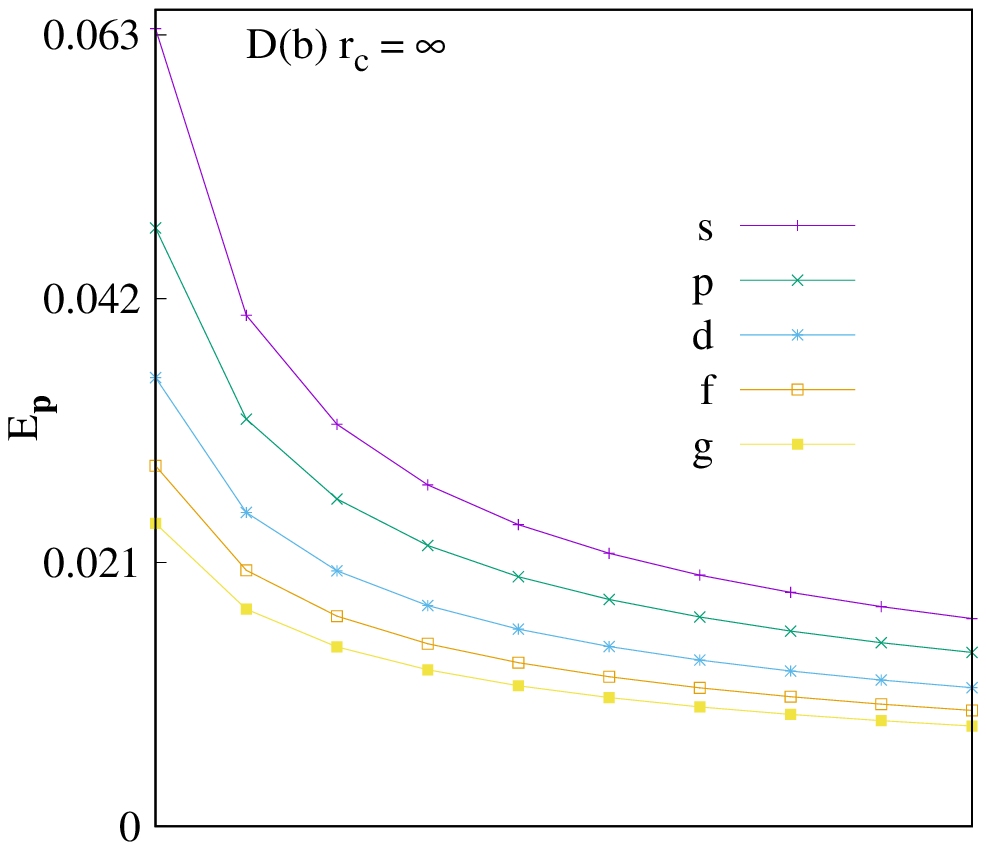}
\end{minipage}%
\begin{minipage}[c]{0.33\textwidth}\centering
\includegraphics[scale=0.48]{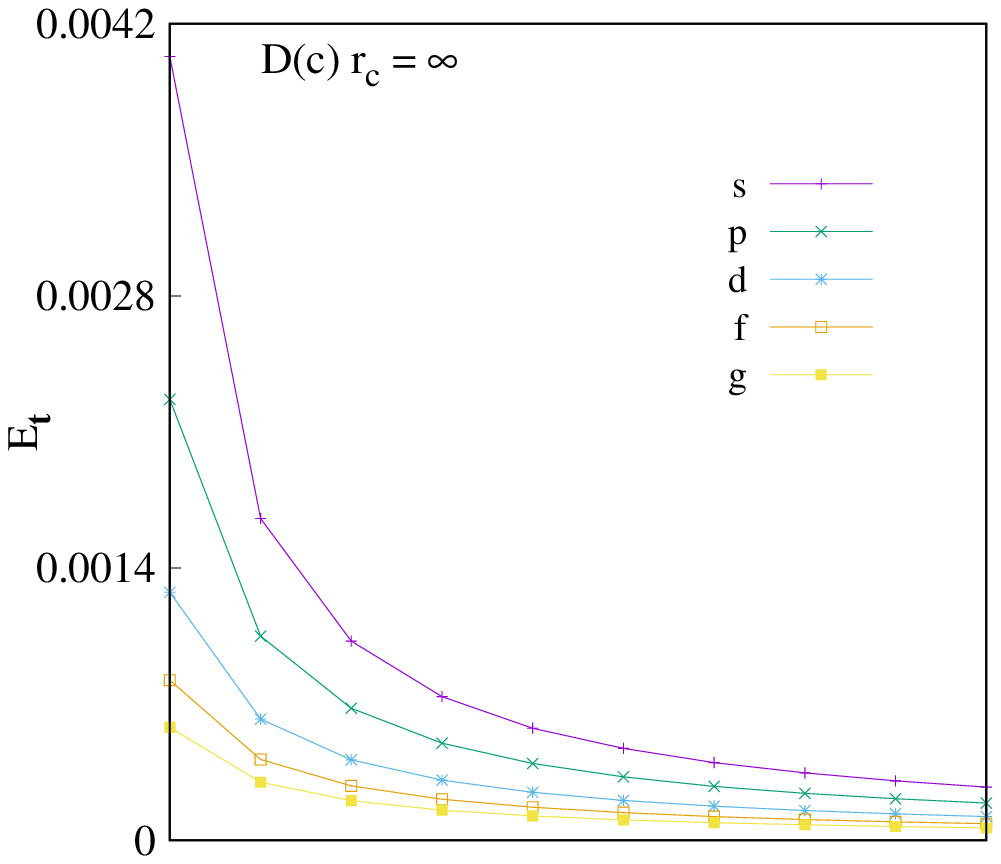}
\end{minipage}%
\hspace{0.2in}
\begin{minipage}[c]{0.33\textwidth}\centering
\includegraphics[scale=0.48]{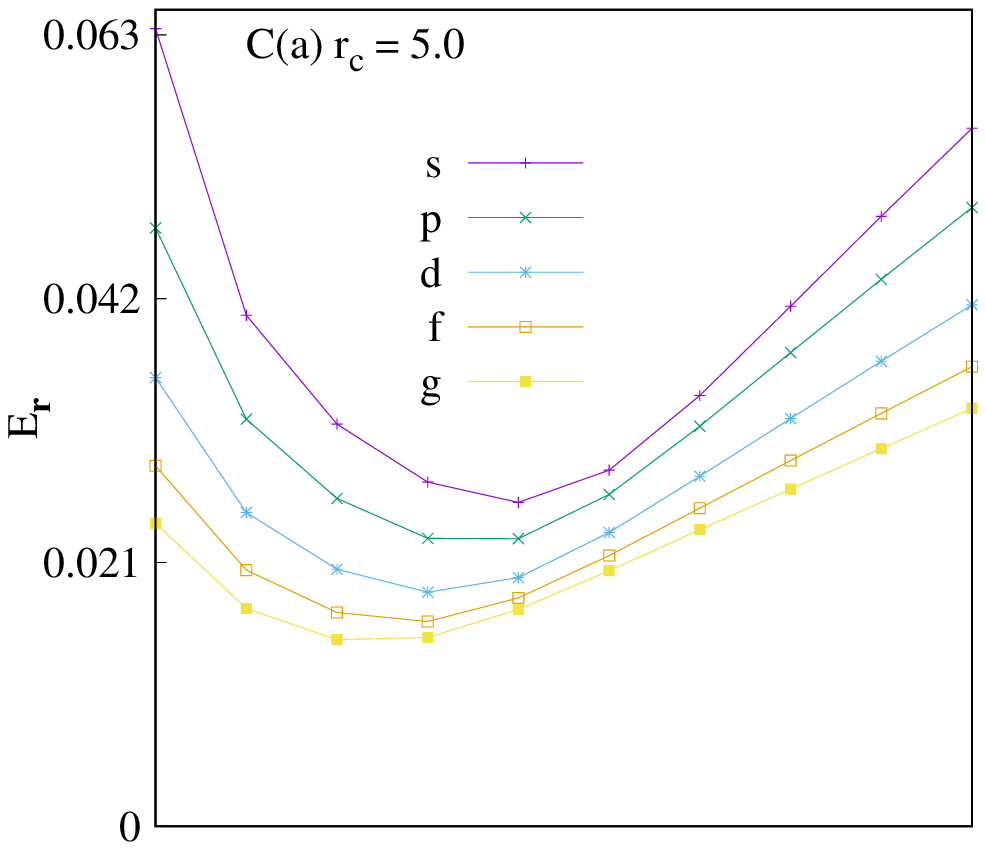}
\end{minipage}
\begin{minipage}[c]{0.33\textwidth}\centering
\includegraphics[scale=0.48]{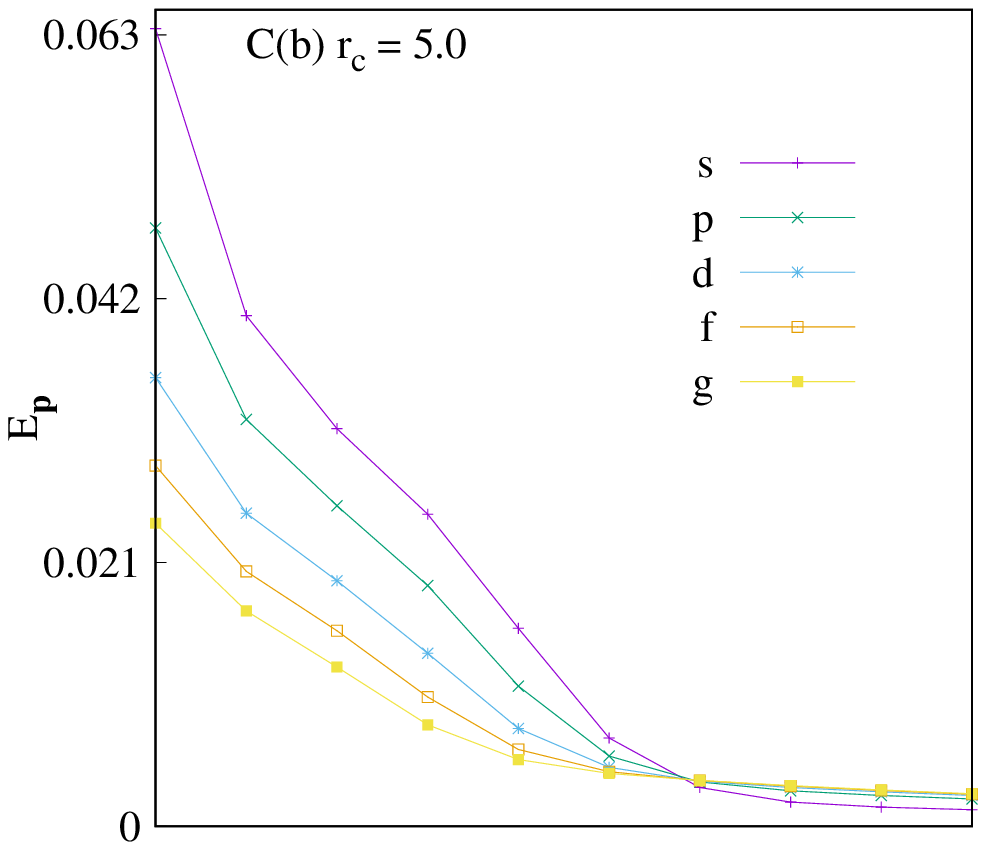}
\end{minipage}%
\begin{minipage}[c]{0.33\textwidth}\centering
\includegraphics[scale=0.48]{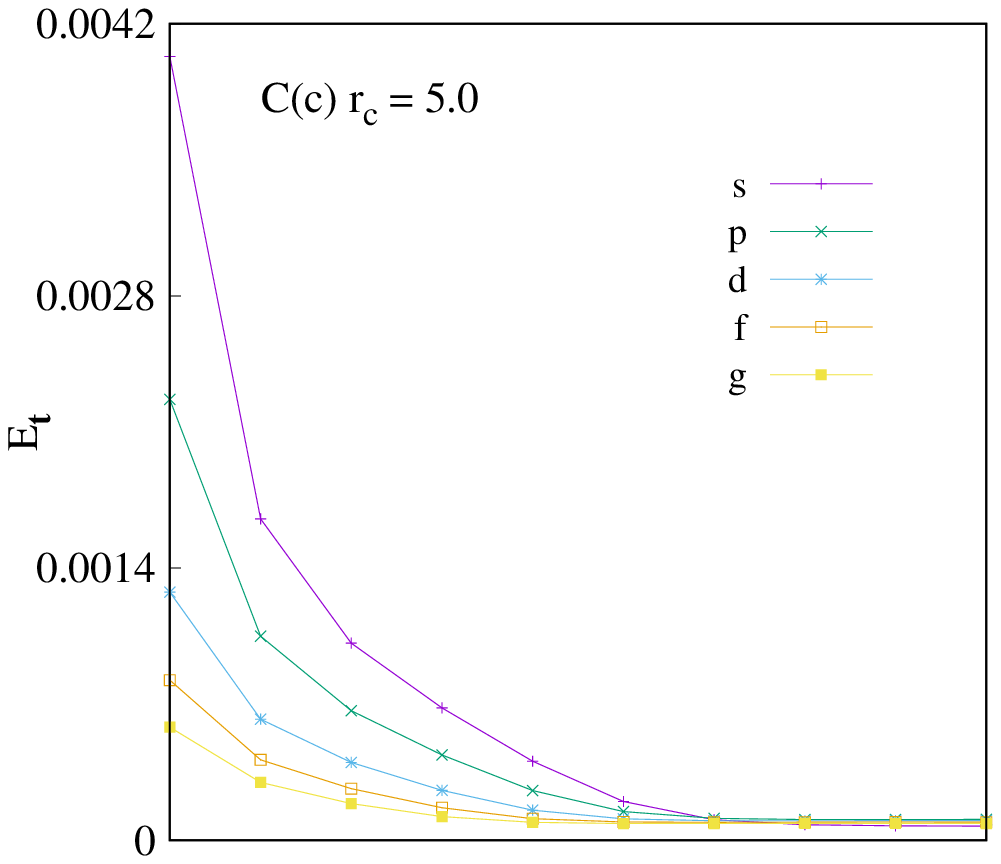}
\end{minipage}%
\hspace{0.2in}
\begin{minipage}[c]{0.33\textwidth}\centering
\includegraphics[scale=0.48]{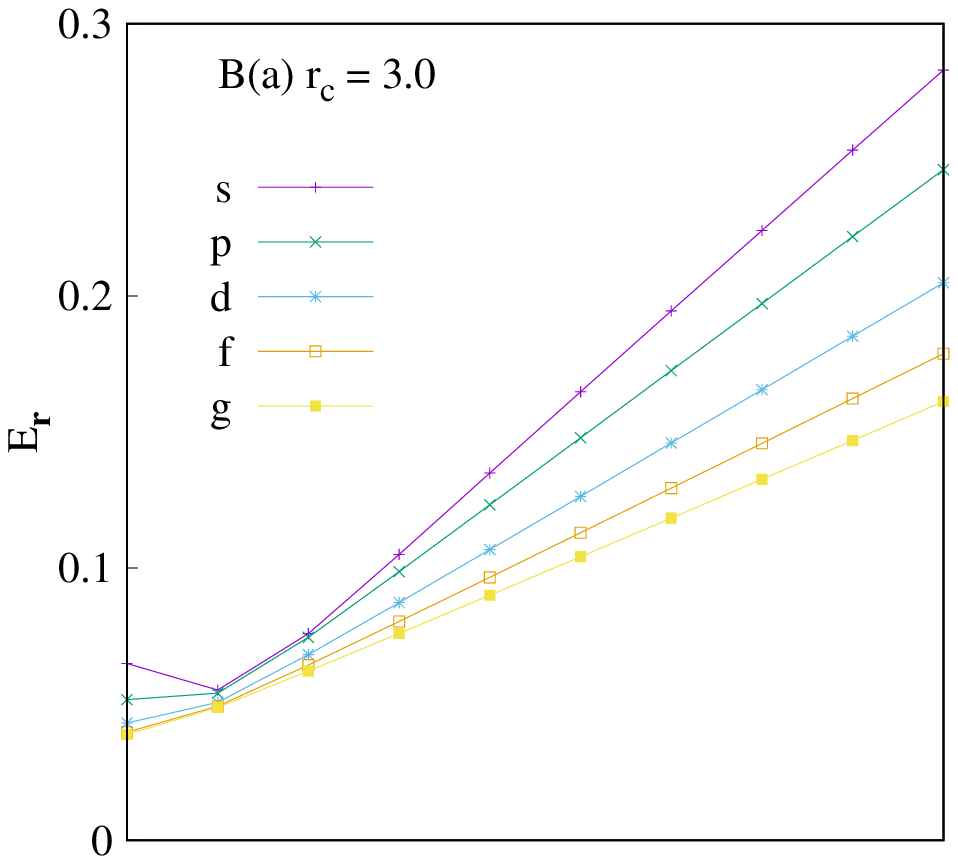}
\end{minipage}%
\begin{minipage}[c]{0.33\textwidth}\centering
\includegraphics[scale=0.48]{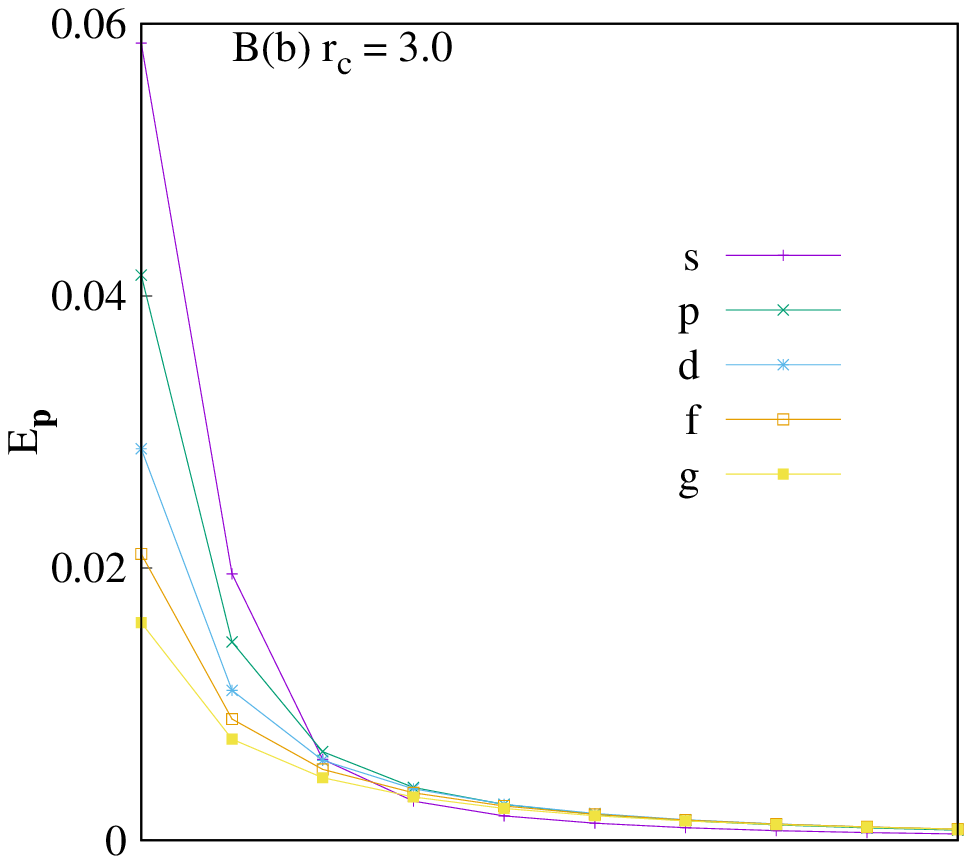}
\end{minipage}%
\begin{minipage}[c]{0.33\textwidth}\centering
\includegraphics[scale=0.48]{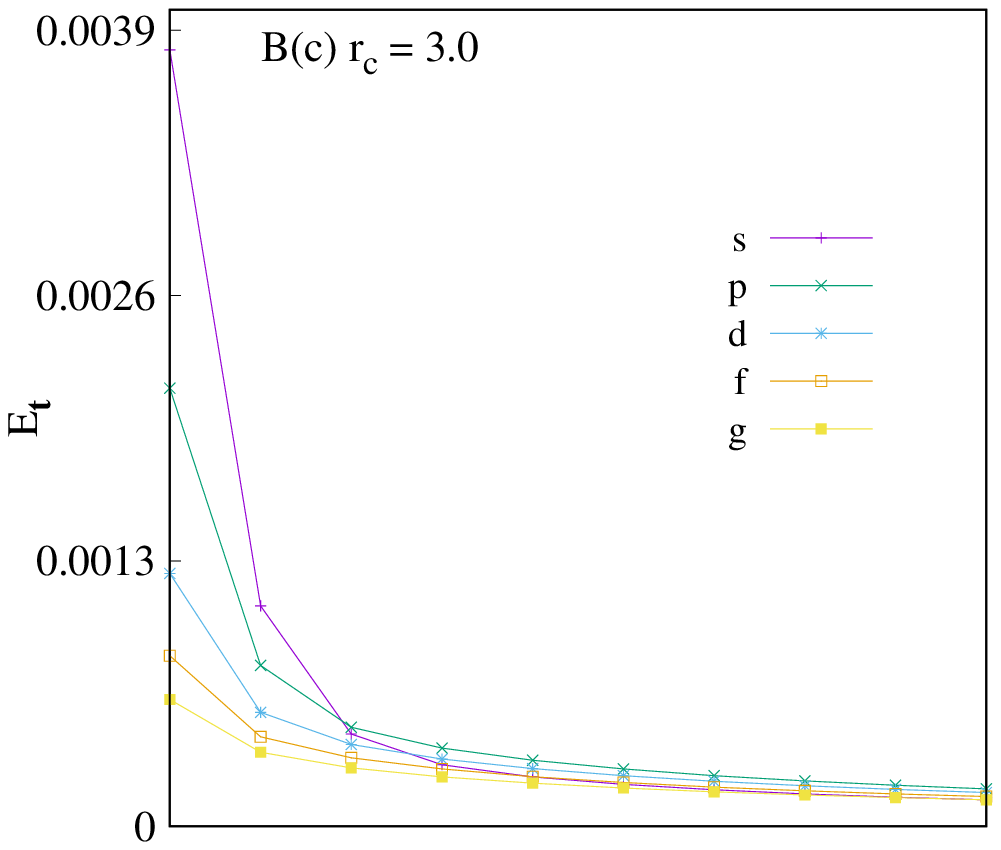}
\end{minipage}%
\hspace{0.2in}
\begin{minipage}[c]{0.33\textwidth}\centering
\includegraphics[scale=0.48]{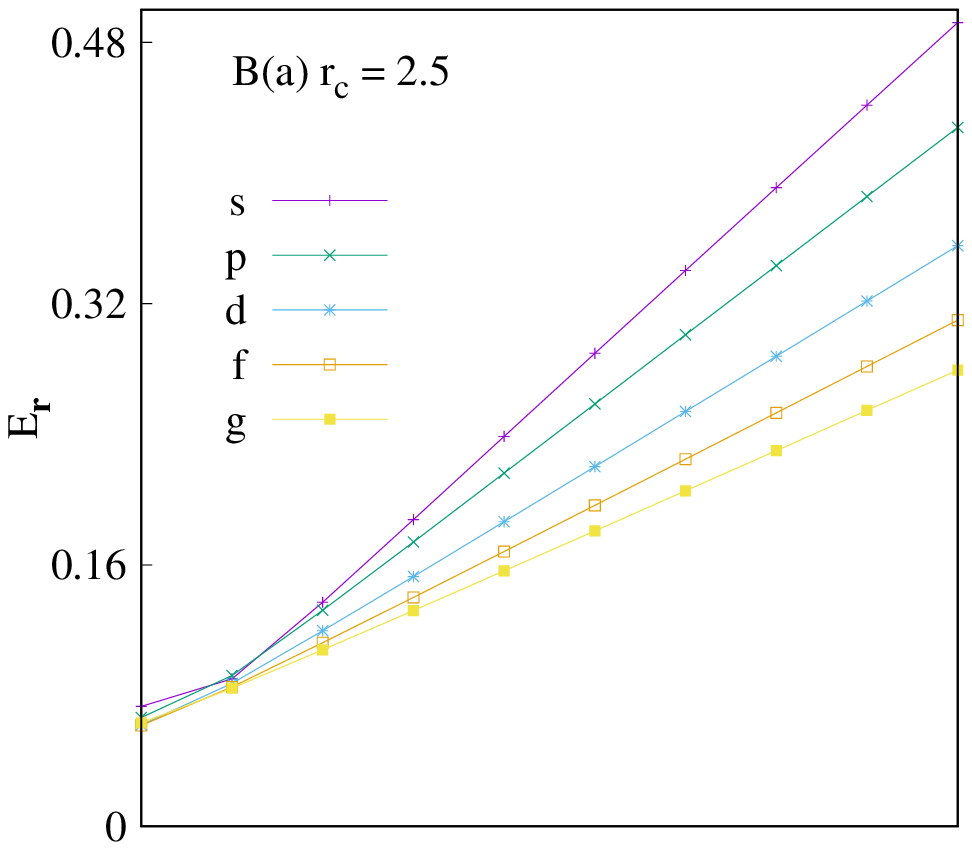}
\end{minipage}%
\begin{minipage}[c]{0.33\textwidth}\centering
\includegraphics[scale=0.48]{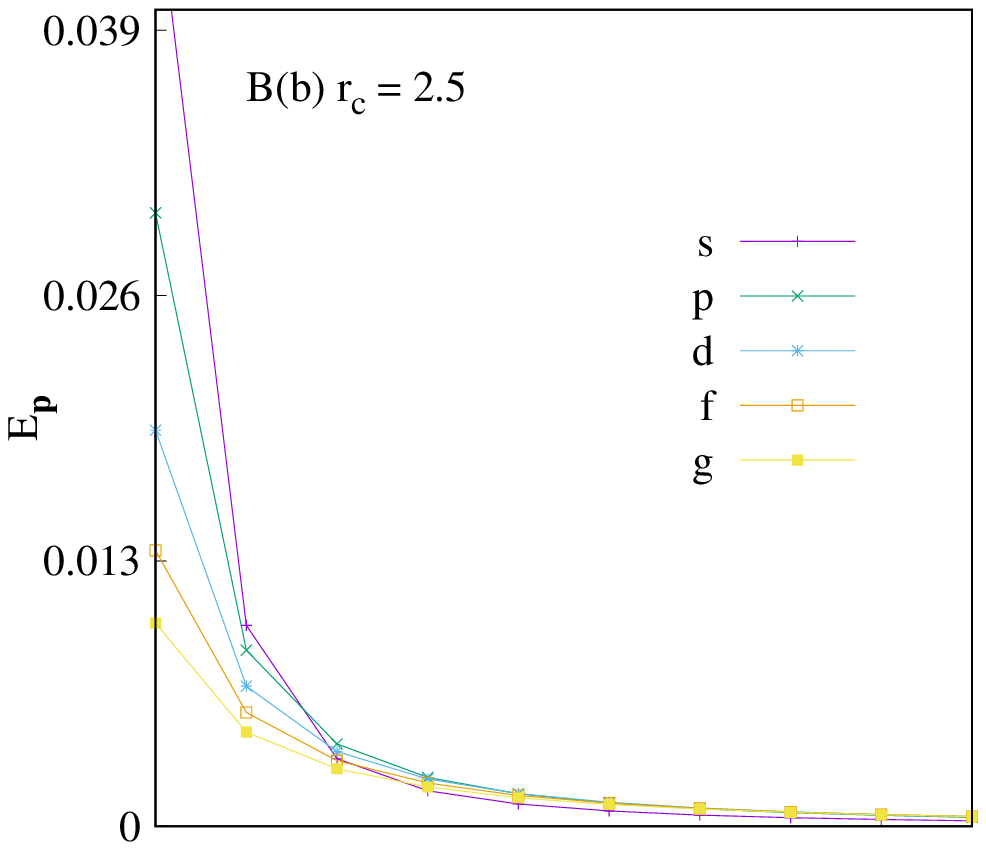}
\end{minipage}%
\begin{minipage}[c]{0.33\textwidth}\centering
\includegraphics[scale=0.48]{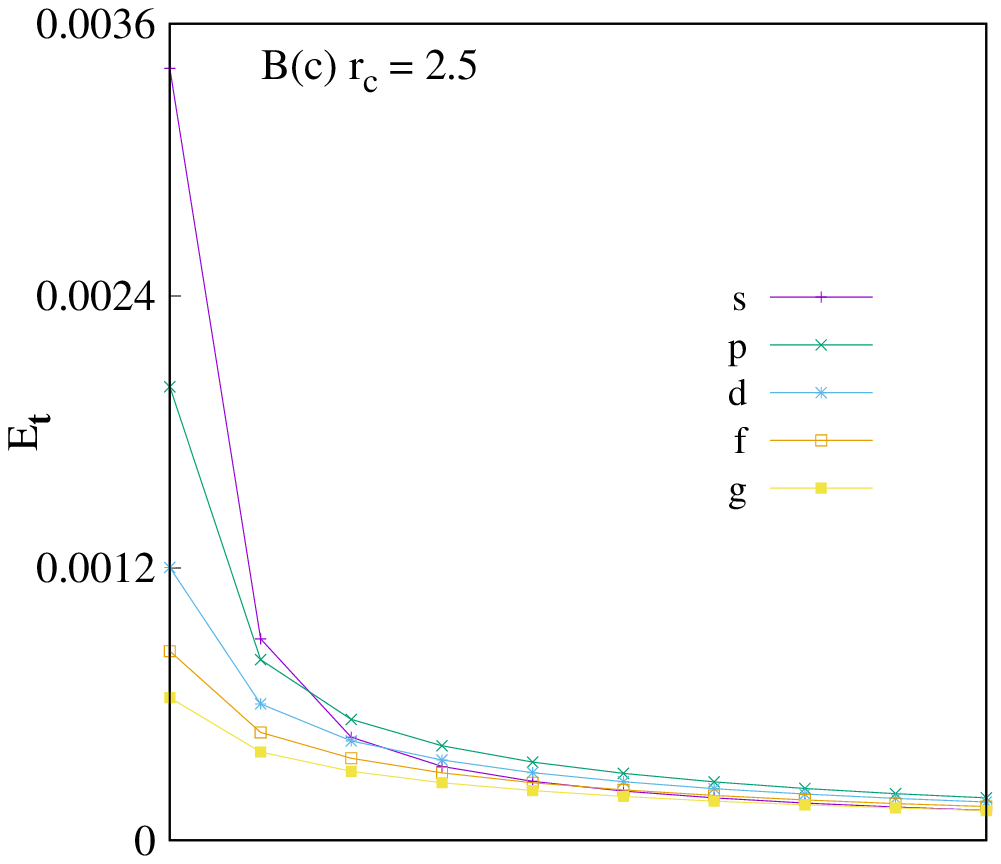}
\end{minipage}%
\hspace{0.2in}
\begin{minipage}[c]{0.33\textwidth}\centering
\includegraphics[scale=0.52]{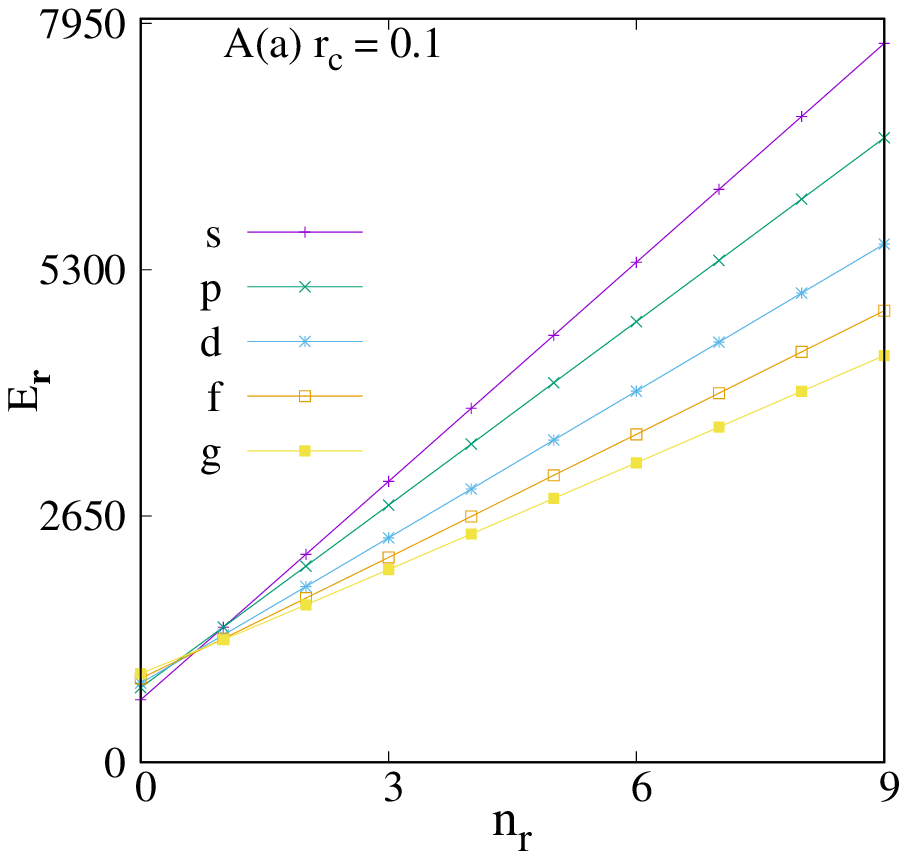}
\end{minipage}%
\begin{minipage}[c]{0.33\textwidth}\centering
\includegraphics[scale=0.52]{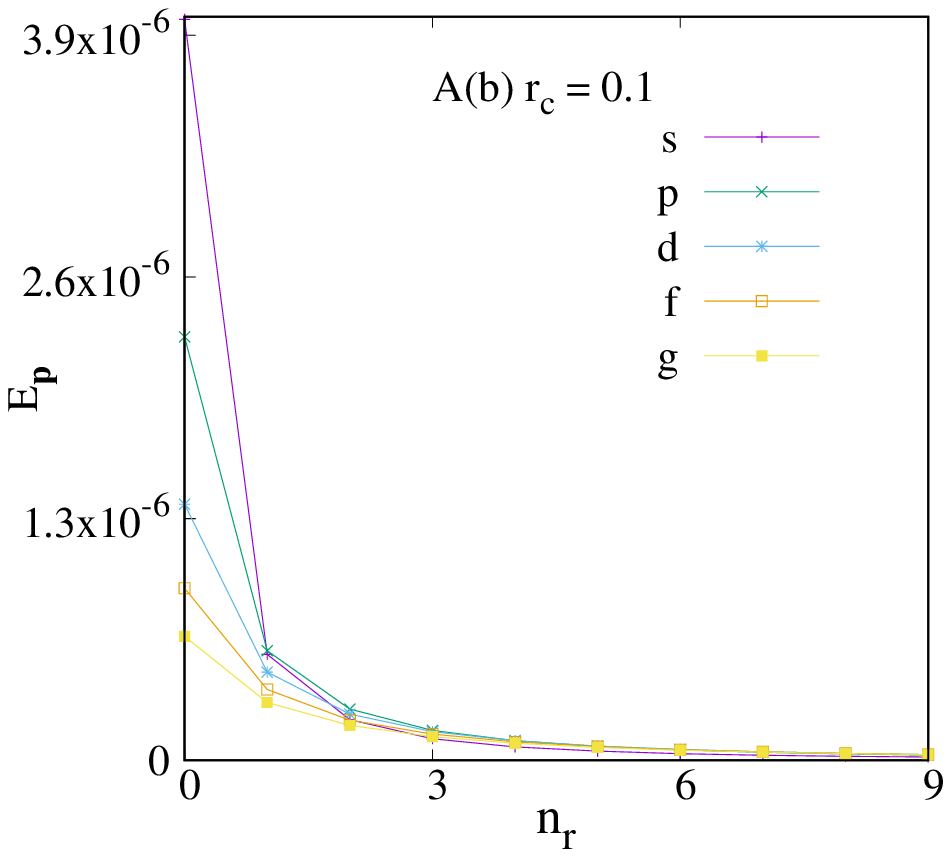}
\end{minipage}%
\begin{minipage}[c]{0.33\textwidth}\centering
\includegraphics[scale=0.52]{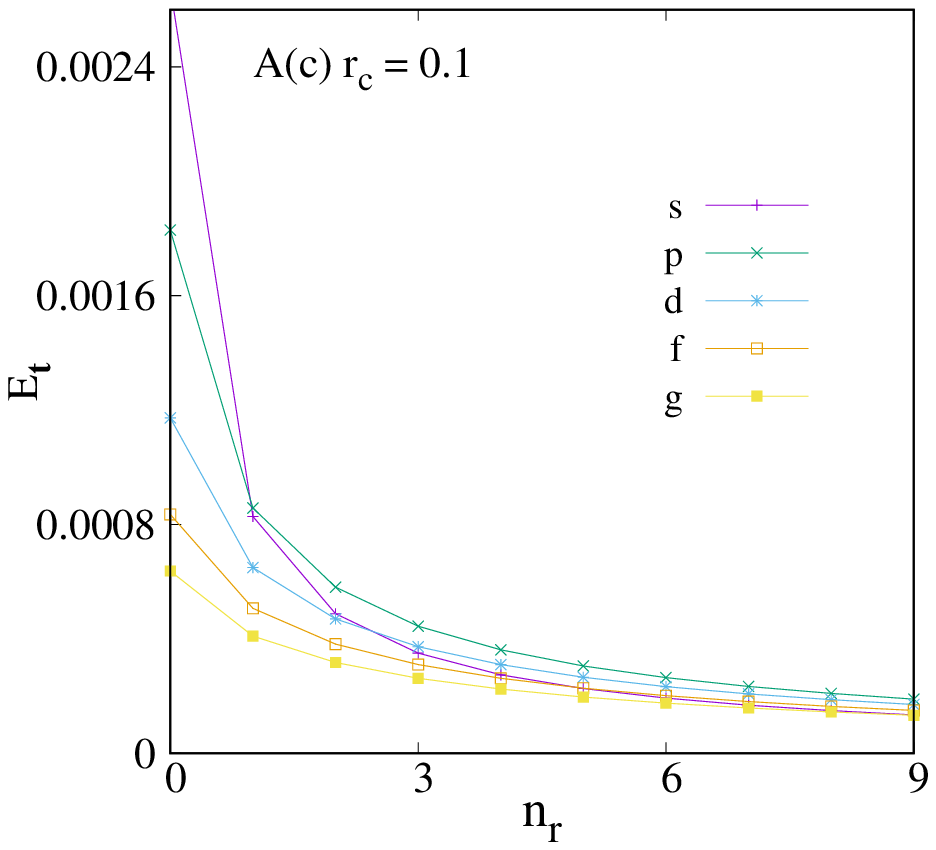}
\end{minipage}%
\caption{Plot of $E_{\rvec}$ (a), $E_{\pvec}$ (a) and $E_{t}$ (c) versus $n_{r}$ (at $\omega=1$) for $s,p,d,f,g$ 
states at five particular $r_{c}$'s of CHO, namely, $0.1,2.5,3,5,\infty$ in panels (A)-(E). $S_{t}$'s 
for all these states obey the lower bound given in Eq.~(19). For more details, consult text.} 
\end{figure}
         
\section{Future and Outlook}
Information theoretic measures like R,~S,~E are pursued for CHO in both $r,~p$ spaces, along with their composite measures. At first,
in order to explore the composite effects of $\omega$ and $r_c$, the Hamiltonian is transformed into a dimensionless form.
This established that, CHO behaves as an interim model between the PISB and IHO. Later, the role of $r_c$ on these 
measures were investigated keeping $\omega$ fixed at $1$. Amongst several interesting features, one notices that, at very low 
$r_c$, $R_{\rvec}^{\alpha}$, $S_{\rvec}$ fall and $E_{\rvec}$ grows as $n_r$ advances, which is in sharp contrast to that found in IHO.  
Furthermore, $r_c$ and $\eta$ produce opposite effects on IE measures. The effect of nonzero $m$ and a penetrable cavity on these measures 
may lead to some other interesting features, which may be pursued later.      

\section{Acknowledgement}
Financial support from DST SERB, New Delhi, India (sanction order: EMR/2014/000838) is gratefully acknowledged. NM thanks DST SERB, 
New Delhi, India, for a National-post-doctoral fellowship (sanction order: PDF/2016/000014/CS).

\end{document}